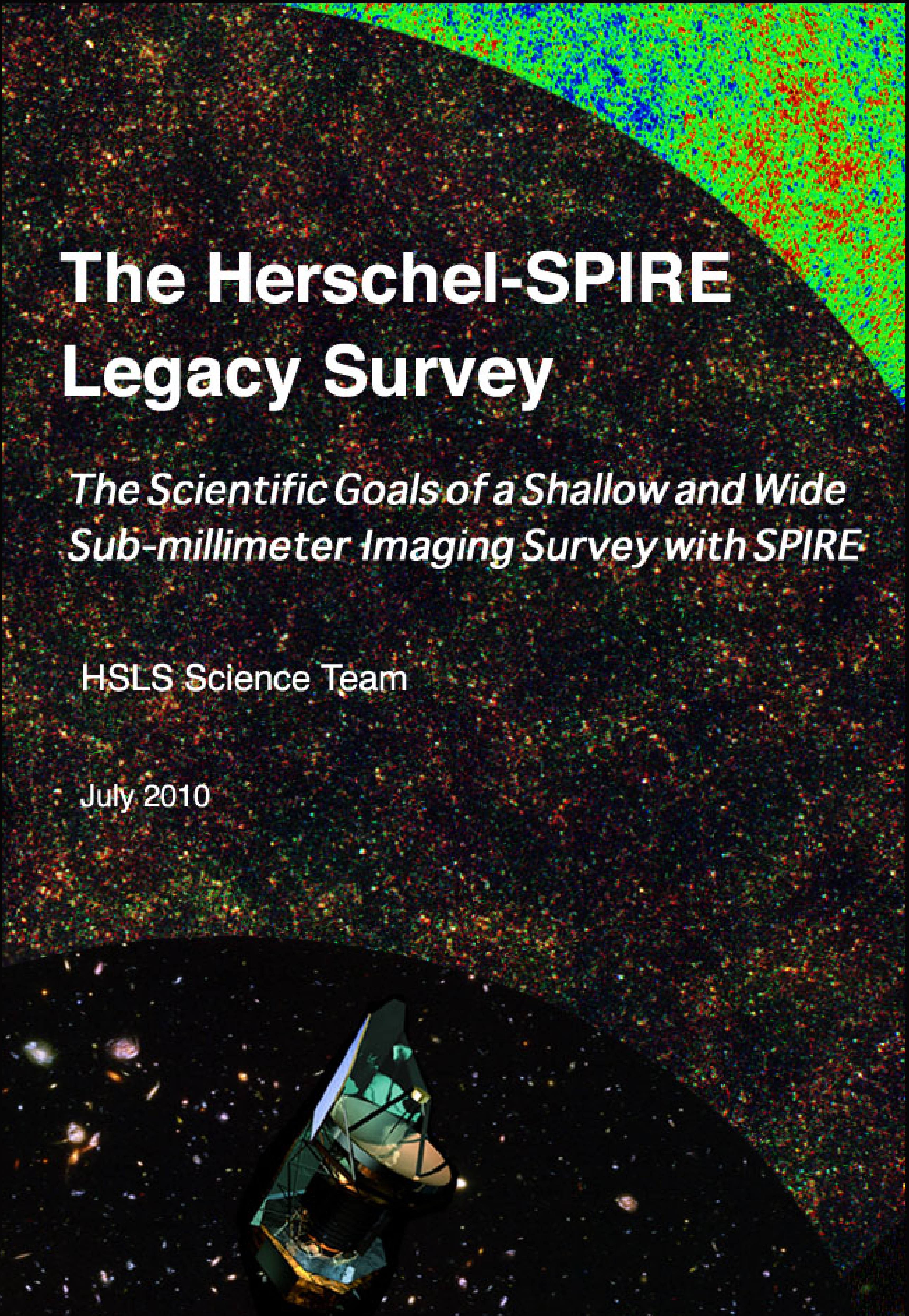

# The Herschel-SPIRE Legacy Survey

*The Scientific Goals of a Shallow and Wide Sub-millimeter Imaging Survey with SPIRE*

HSLS Science Team

July 2010

# The Herschel-SPIRE Legacy Survey (HSLS)

The Scientific Goals of a Shallow and Wide Submillimeter Imaging Survey with SPIRE

Members of the HSLS Study Team

Asantha Cooray[1], Steve Eales[2], Scott Chapman[14], David L. Clements[4], Olivier Doré[5], Duncan Farrah[6], Matt J. Jarvis[8] Manoj Kaplinghat[1], Mattia Negrello[3], Alessandro Melchiorri[9], Hiranya Peiris[11], Alexandra Pope[12], Mario G. Santos[13], Stephen Serjeant[3], Mark Thompson[8], Glenn White[3,15], Alexandre Amblard[1,7], Manda Banerji[14], Pier-Stefano Corasaniti[16], Sudeep Das[17], Francesco de Bernardis[9], Gianfranco de Zotti[78], Tommaso Giannantonio[18,88], Joaquin Gonzalez-Nuevo Gonzalez[19], Ali Ahmad Khostovan[1], Ketron Mitchell-Wynne[1], Paolo Serra[1], Yong-Seon Song[20], Joaquin Vieira[86], Lingyu Wang[6], Michael Zemcov[86,5] Filipe Abdalla[11], Jose Afonso[56], Nabila Aghanim[49], Paola Andreani[21], Itziar Aretxaga[84], Robbie Auld[2], Maarten Baes[85], Andrew Baker[58], Denis Barkats[68], R. Belen Barreiro[60], Nicola Bartolo[89], Elizabeth Barton[1], Sudhanshu Barway[23], Elia Stefano Battistelli[9], Carlton Baugh[25], Alexander Beelen[49], Karim Benabed[49], Andrew Blain[86], Joss Bland-Hawthorn[46], James J. Bock[5,86], J. Richard Bond[40], Julian Borrill[22,41], Colin Borys[67], Alessandro Boselli[51], François R. Bouchet[79], Carrie Bridge[86], Fabrizio Brighenti[96], Veronique Buat[51], David Buote[1], Denis Burgarella[51], Robert Bussmann[39], Erminia Calabrese[9], Christopher Cantalupo[22], Raymond Carlberg[38], Carla Sofia Carvalho[63], Caitlin Casey[100], Antonio Cava[47,77], Jordi Cepa[47], Edward Chapin[37], Ranga Ram Chary[86], Xuelei Chen[87], Natalie Christopher[35], Sergio Colafrancesco[62,78], Shaun Cole[25], Peter Coles[2], Alexander Conley[93], Luca Conversi[74], Jeff Cooke[1], Steven Crawford[23], Catherine Cress[91], Elisabete da Cunha[95], Gavin Dalton[15,35], Luigi Danese[19], Helmut Dannerbauer[80], Jonathan Davies[2], Paolo de Bernardis[9], Roland de Putter[70], Mark Devlin[33], Jose M. Diego[60], Hervé Dole[49], Marian Douspis[49], Joanna Dunkley[35], James Dunlop[54], Loretta Dunne[31], Rolando Dünner[105], Simon Dye[2], George Efstathiou[102], Eiichi Egami[39], Taotao Fang[1], Patrizia Ferrero[47,77], Alberto Franceschini[71], Christopher C. Frazer[1], David Frayer[43], Carlos Frenk[25], Ken Ganga[53], Raphaël Gavazzi[79], Jason Glenn[93], Yan Gong[87], Eduardo Gonzalez-Solares[14], Matt Griffin[2], Qi Guo[25], Mark Gurwell[76], Amir Hajian[40], Mark Halpern[37], Duncan Hanson[14], Martin Hardcastle[8] Evanthia Hatziminaoglou[21], Alan Heavens[54], Sébastien Heinis[51], Diego Herranz[60], Matt Hilton[63], Shirley Ho[22], Benne W. Holwerda[55], Rosalind Hopwood[3], Jonathan Horner[99], Kevin Huffenberger[64], David H. Hughes[84], John P. Hughes[58], Edo Ibar[83], Rob Ivison[54], Neal Jackson[48], Andrew Jaffe[4], Timothy Jenness[45], Gilles Joncas[75], Shahab Joudaki[1], Sugata Kaviraj[4], Sam Kim[1], Lindsay King[14], Theodore Kisner[22], Johan Knapen[47,77], Alexei Kniazev[23], Eiichiro Komatsu[34], Leon Koopmans[82], Chao-Lin Kuo[24], Cedric Lacey[25], Ofer Lahav[11], Anthony N. Lasenby[102], Andy Lawrence[54] Myung Gyoon Lee[69], Lerothodi L. Leeuw[72,73], Louis R. Levenson[86], Geraint Lewis[46], Nicola Loaring[23], Marcos López-Caniego[60], Steve Maddox[31], Tobias Marriage[106], Gaelen Marsden[37], Enrique Martinez-Gonzalez[60], Silvia Masi[9], Sabino Matarrese[89], William G. Mathews[104], Shuji Matsuura[30], Richard McMahon[14], Yannick Mellier[79], Felipe Menanteau[58], Michał J. Michałowski[54], Marius Millea[98], Bahram Mobasher[101], Subhanjoy Mohanty[4], Ludovic Montier[65], Kavilan Moodley[63], Gerald H. Moriarty-Schieven[42], Angela Mortier[4], Dipak Munshi[2], Eric Murphy[26], Kirpal Nandra[97], Paolo Natoli[10], Hien Nguyen[5], Seb Oliver[6], Alain Omont[79], Lyman Page[50], Mathew Page[66], Roberta Paladini[26], Stefania Pandolfi[9], Enzo Pascale[2], Guillaume Patanchon[53], John Peacock[54], Chris Pearson[15,29], Ismael Perez-Fournon[47,77], Pablo G. Pérez-González[52], Francesco Piacentini[9], Elena Pierpaoli[59], Michael Pohlen[2], Etienne Pointecouteau[65], Gianluca Polenta[62], Jason Rawlings[66], Erik D. Reese[33], Emma Rigby[36], Patrick Roche[35], Giulia Rodighiero[71], Encarni Romero-Colmenero[23], Isaac Roseboom[6], Michael Rowan-Robinson[4], Miguel Sánchez-Portal[74], Fabian Schmidt[86], Michael Schneider[25], Bernhard Schulz[67,86], Douglas Scott[37], Chris Sedgwick[3], Neelima Sehgal[94], Nick Seymour[66], Blake D. Sherwin[50], Jo Short[2], David Shupe[67], Jonathan Sievers[40], Ramin Skibba[44], Joseph Smidt[1], Anthony Smith[6], Daniel J. B. Smith[31], Matthew W. L. Smith[2], David Spergel[27], Suzanne Staggs[50], Jason Stevens[8], Eric Switzer[81], Toshinobu Takagi[30], Tsutomu Takeuchi[103], Pasquale Temi[7], Markos Trichas[76], Corrado Trigilio[78], Katherine Tugwell[66], Grazia Umana[78], William Vacca[92], Mattia Vaccari[78], Petri Vaisanen[23], Ivan Valtchanov[74], Kurt van der Heyden[55], Paul P. van der Werf[90], Eelco van Kampen[21], Ludovic van Waerbeke[37], Simona Vegetti[57], Marcella Veneziani[9], Licia Verde[61], Aprajita Verma[35], Patricio Vielva[60], Marco P. Viero[86], Baltasar Vila Vilaro[68], Julie Wardlow[32], Grant Wilson[12], Edward L. Wright[28], C. Kevin Xu[67,86], Min S. Yun[12]

# Contents

## Executive Summary

During the next decade several wide optical and radio surveys will map most of the sky, detect $\sim 10^9$ galaxies, and provide a scientific goldmine for extragalactic astronomers and cosmologists. At the same time, surveys of the cosmic microwave background (CMB) are being carried out over areas ranging from thousands of deg$^2$ to the whole sky (Planck), leading to new information about the Universe since recombination through secondary anisotropies. To fully exploit these large datasets, however, it is vital to have a survey of a similar area in the far-IR/submm waveband. *Herschel* is the last chance to achieve this, since SPICA (launch date ∼2020) is primarily a spectroscopic mission and will not have coverage at $\lambda > 200\,\mu$m, and there are no other missions planned for the foreseeable future. We propose the *Herschel*-SPIRE Legacy Survey (HSLS), a project using SPIRE that will cover 4000 deg$^2$, one tenth of the whole sky and one fifth of the extragalactic sky at $|b| > 30°$.

### The key goals of Herschel-SPIRE Legacy Survey (HSLS) are to:

- Produce a catalog of 2.5 to 3 million galaxies down to 26, 27 and 33 mJy (50% completeness; $\sim 5\sigma$ confusion noise) at 250, 350 and 500 $\mu$m, respectively, in the southern hemisphere (3000 deg$^2$) and an expanded area centered on SDSS Stripe-82 (1000 deg$^2$); these areas have wide multi-wavelength coverage and are easily accessible from ALMA for detailed studies over the next several decades. One third of these sources are expected to be at $z < 1$, with most of the rest at $z \sim 2$ and $\sim$ 2000 sources at $z > 5$.

- Remove point source confusion in secondary anisotropy studies with Planck and ground-based CMB data, improving Planck cluster detection (finding 50% more clusters); detect the integrated Sachs-Wolfe signal at $z \sim 2$, by cross-correlating HSLS with Planck, for a powerful test of modified gravity theories; detect at better than $20\sigma$ large-scale structure responsible for lensing of the CMB.

- Find at least 1200 strongly lensed sub-mm sources with $S_{500} > 100$ mJy, leading to a 2% test of General Relativity and a 10% measurement on the number count amplitude and slope below the source confusion. Approximately 20 of these will be "golden lenses", with two background sources lensed by the same galaxy, making possible cosmological tests that are independent of the mass of the lens. These $\sim$ 1200 sources will be fundamental for followup studies of the properties of ULIRGs from $z \sim 1$ to 6 using ALMA.

- Measure clustering of bright sources in several broad sub-mm photometric redshift bins and combine the results with Planck CMB spectra to improve measurements of cosmological parameters by factors of 2 to 3, including a cosmological measurement of neutrino masses; measure clustering of unresolved fluctuations in the cosmic far-IR background (FIRB) with a signal-to-noise ratio of $10^3$; possibly allowing new studies on cosmic magnification of FIRB fluctuations.

- Identify $\sim$ 200 proto-cluster regions at $z \sim 2$ and perform an unbiased study of the environmental dependence of star formation from protoclusters at $z \sim 2$ to virialized clusters at $z = 0$.

- Perform an unbiased survey for star formation and dust at high Galactic latitude and make a census of debris disks and dust around AGB stars and white dwarfs.

# 1 Introduction

‘

Approximately half of the energy radiated by stars and galaxies since the time of recombination has been reprocessed by dust and is now detected on Earth in the far-IR and sub-mm wavebands ([145]). Therefore, to completely understand the origin and evolution of galaxies it is crucial to follow the energy by determining the properties of the sources making up this far-IR and sub-mm background (FIRB). Until Herschel, our knowledge of these sources was largely limited to the far-IR properties of galaxies in the nearby Universe, provided by IRAS in the eighties, and the properties of dusty galaxies in the distant (early) Universe provided by Spitzer, SCUBA and BLAST. The surveys of the distant Universe were inevitably limited by statistics, and both Spitzer and SCUBA operated at wavelengths where the strength of the background is much less than at its peak ($\lambda \simeq 200\mu m$). *Herschel*, of course, has changed this completely, with a number of ground-breaking projects to study dust and dust-obscured star formation in both the nearby and distant Universe. The existing *Herschel* surveys range from very deep surveys of tiny regions of sky (PEP, H-GOODS) to shallower surveys of larger regions (HerMES and H-ATLAS). The largest contiguous region of the extragalactic sky for which time has been allocated for a *Herschel* survey is 250 deg$^2$.

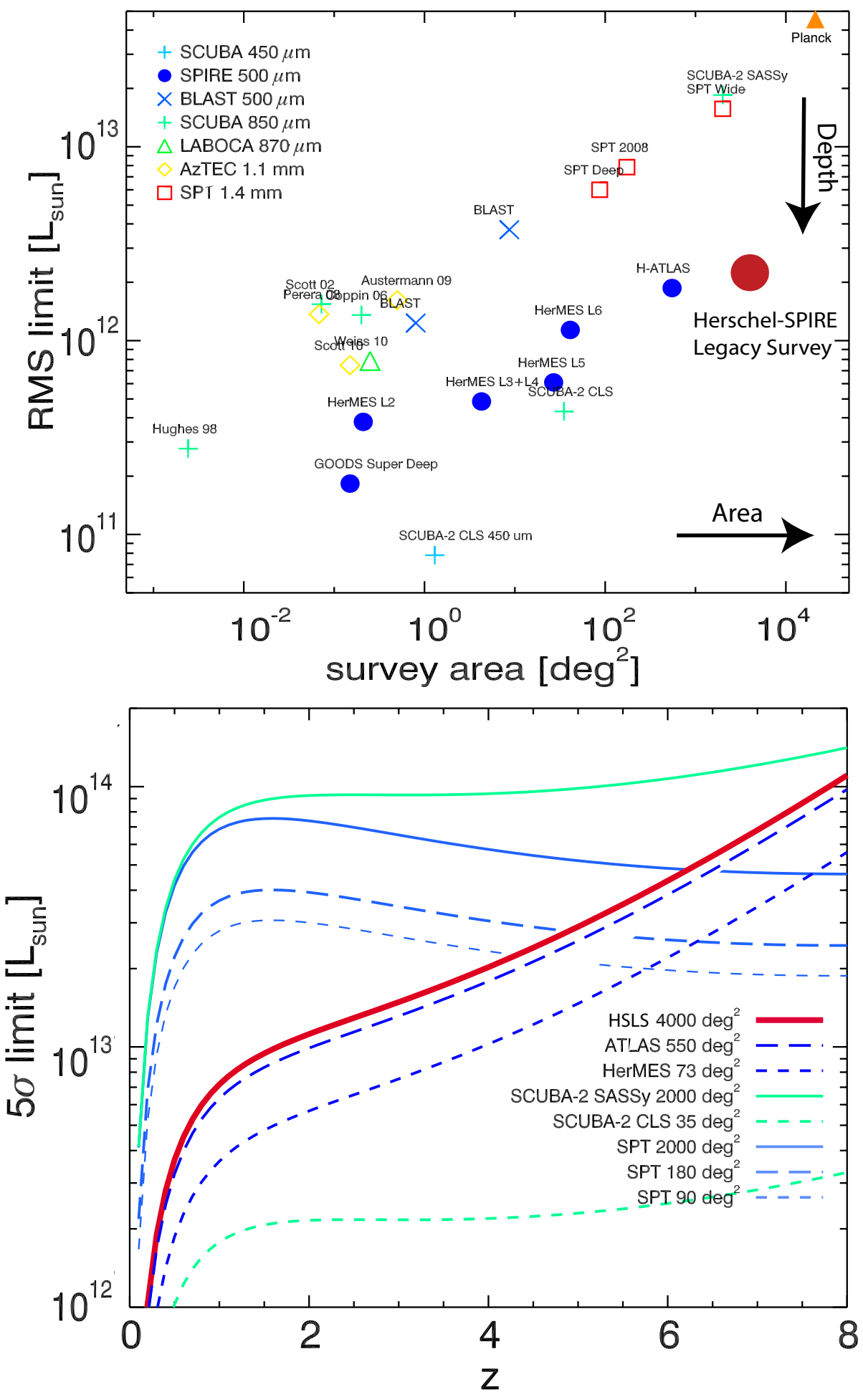

Figure 1: *Top:* The depth vs. survey area in deg$^2$ for a variety of extragalactic surveys at sub-mm wavelengths, from 250 $\mu$m to 1.4 mm. For easy comparison, the rms depth is given in terms of the equivalent far-IR luminosity with a sub-mm galaxy SED at $z = 2$ with $T_d = 35$ K and greybody spectral index $\beta = 1.5$. *Herschel* surveys in general are an order of magnitude more sensitive than surveys from the ground covering similar areas. H-ATLAS covers a total of 550 deg$^2$ with several fields spread over the sky with the largest contiguous area of about 220 deg$^2$. Confusion-noise limited *Herschel*-SPIRE surveys are sensitive to $10^{12}$ $L_\odot$ at $z = 2$. *Bottom:* The $5\sigma$ detection luminosity as a function of redshift for a variety of sub-mm surveys at a wide range of wavelengths.

In this white paper we consider the arguments for a much larger *Herschel* survey covering 4000 deg$^2$, one tenth of the sky or one fifth of the extragalactic sky ($|b| > 30°$). This survey would be similar in area to many of the surveys being carried out in the optical wavebands (Dark Energy Survey, Pan-STARRS) and with the SKA pathfinders (LOFAR, MeerKAT and ASKAP). We show that with a shallow SPIRE-only survey it would be possible to detect $\simeq$six times as many galaxies as will be detected in all the existing *Herschel* surveys in approximately the same observing time as one of the large existing surveys. Apart from the huge legacy value for after the helium runs out, there are many scientific projects that would suddenly become possible with a survey of such a large area of sky. These include projects to study the origin and evolution of galaxies, to investigate the evolution of the dark matter structures (one of the most important and untested predictions of the current paradigm for the formation of structure) and to measure the equation-of-state of dark energy. Many of these are only possible because of the synergy between such a large *Herschel* survey and the surveys of the cosmic microwave background being carried out with Planck, South Pole Telescope (SPT) and the Atacama Cosmology Telescope (ACT).

The largest existing extragalactic programs with *Herschel*-SPIRE, HerMES [304] and H-ATLAS [130], cover in total about 620 deg$^2$. They consist of imaging with PACS and SPIRE of areas with multi-wavelength data on hand already. Together these two surveys are expected to detect $\simeq 5 \times 10^5$ sources. These surveys were designed before the launch of *Herschel* with the pre-launch assumptions about the instrumental performance. With the benefit of post-launch observations, we now realize that even the shallowest survey that one could implement with SPIRE alone, using single-scans at $60''$/sec, would be dominated by the confusion [300] and not by instrumental noise. Due to the remarkable stability of SPIRE [177], which allows long scans to minimize turn-around overheads, we can survey $\simeq$7 times the area of the largest existing survey (H-ATLAS) in roughly the same observing time and with only a small degradation in sensitivity. At far-IR wavelengths, the IRAS all-sky survey [299] in 1983 led to a catalog of about 250,000 sources, mostly bright starbursts in the local universe that contain $\sim 40$ K warm dust and have SEDs which peak at short sub-mm wavelengths of 60 and 100 $\mu$m [116]. Though it has now been more than 25 years, IRAS sources are still followed-up and studied by new generations of astronomers. Citations to primary IRAS papers are now constant at the level of 400/year, with a total of 40,000 citations. We believe that a SPIRE-only survey of a large area of sky would provide a similar legacy from *Herschel* to the astronomical community.

The *Herschel*-SPIRE Legacy Survey (HSLS) aims to cover 4000 deg$^2$, requiring 780 hours of observations with *Herschel.* HSLS will generate a catalog of sub-mm galaxies an order of magnitude larger than IRAS, roughly 6 times the number expected from existing surveys [86, 304], with a substantial fraction of the sources at $z > 1$. Studies made with single-scan maps show that we will recover a point source catalog that is 50% complete down to 26.5, 26.7 and 32.6 mJy at 250, 350, 500 $\mu$m, respectively, close to the $5\sigma$ source confusion noise of 29, 31, 34 mJy at these three wavelengths [300]. HSLS will be an extraordinary achievement with an immense legacy value not only for *Herschel*, but for sub-mm astronomy in general. Figs. 1 and 2 put HSLS into context with planned and existing surveys over a range of wavelengths.

To enable maximum use of HSLS maps and catalogs over the next decade and for first science studies with ALMA, 3000 deg$^2$ will be in the southern hemisphere, while $\sim$1000 deg$^2$ will overlap with SDSS Stripe-82 in the ecliptic plane. This latter area has deep optical imaging and spectroscopy data already from SDSS and BOSS, near-IR from VISTA-VHS, radio data from the VLA, and mm-wave data from ACT. An overlapping wide survey with Hypersuprime Camera on Subaru is also planned. It is likely that the large source catalog produced by the proposed *Herschel*-SPIRE Legacy Survey will be followed up by several generations of astronomers, and will provide the basis for scientific studies with not only ground-based instruments such as ALMA, but also coming and next-generation space-based observatories such as JWST, SPICA, and Euclid.

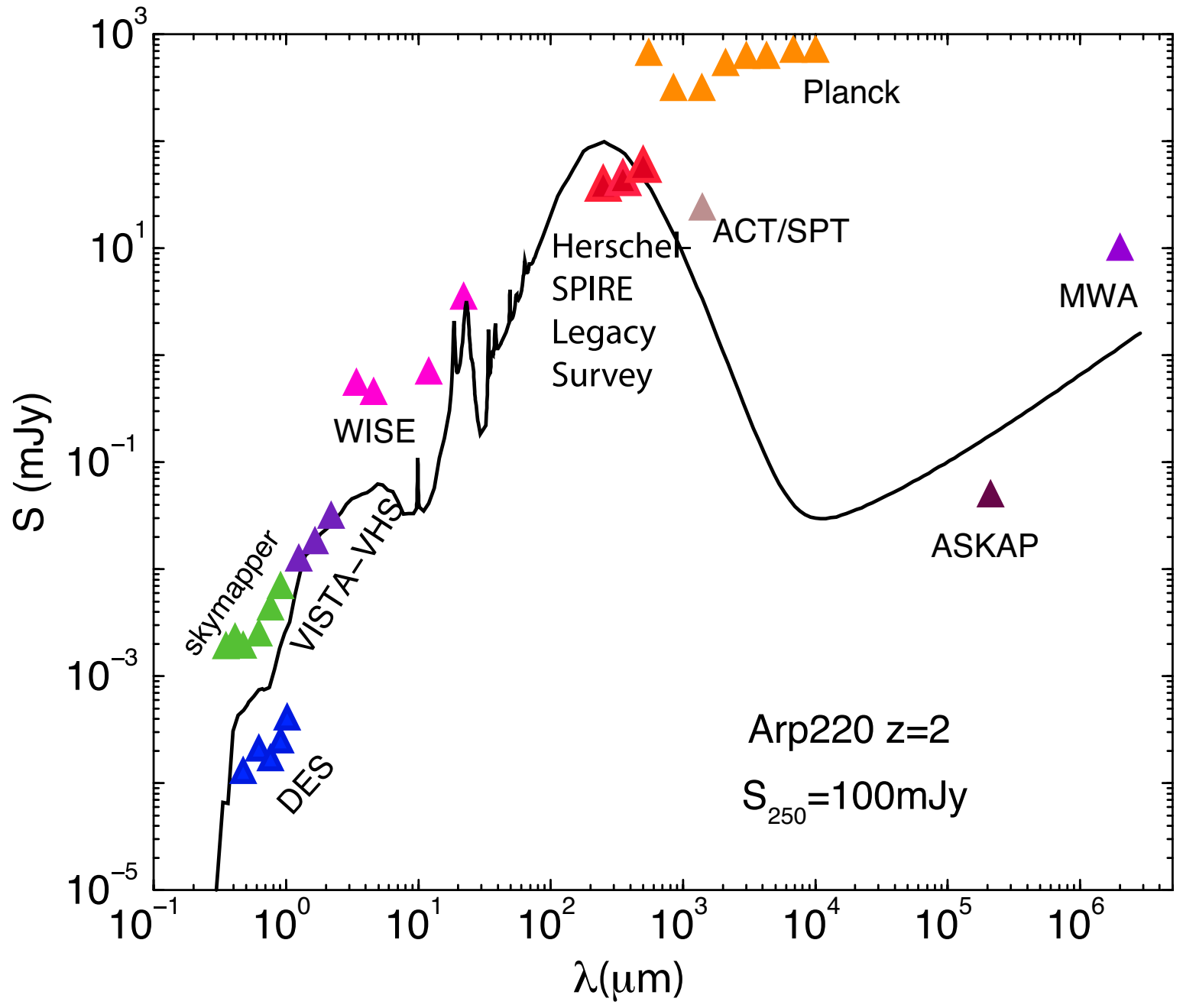

Figure 2: The multi-wavelength coverage of the southern sky over an area of order 4000 deg$^2$ with $5\sigma$ point source detection limit from a wide variety of surveys. For comparison we show Arp220's SED placed at $z = 2$ and normalized to have a 250 $\mu$m flux of 100 mJy. Red triangles are the *Herschel*-SPIRE Legacy Survey.

Beyond the large number of sources, HSLS maps will also contain information on the unresolved background sources responsible for the confusion noise, though such sources are not individually detected. These faint, unresolved galaxies are likely to be further out in redshift, and the cosmological community can deploy a large suite of statistical analysis methods to extract information out of these maps similar to cosmological studies involving Planck and other CMB maps.

Beyond the large legacy value, HSLS will make possible three immediate science programs:

(1) **The Origin and Evolution of Galaxies:** Approximately half the energy emitted by the galaxy population is now detected as a far-IR-submillimetre background (FIRB) [128, 153, 326]. The angular resolution of SPIRE means that even the deepest surveys at 250 $\mu$m, where the FIRB is close to its peak, have only resolved about 15% of the background, and so we still have no direct way of studying the sources of a large fraction of the energy emitted since recombination. Another practical limitation of the existing surveys is that the deep surveys find few galaxies at $z > 3$, the epoch in which the first galaxies and sub-galactic units were formed – the epoch which will be probed by JWST.

We can solve the first two problems because of one of the most exciting early *Herschel* results. Since the discovery of gravitational lensing 30 years ago, theorists have suggested many uses for large samples of lenses, but observers have struggled to discover a method of producing useful samples. The early results from H-ATLAS imply that almost all of the bright 500-$\mu$m sources are lensed systems or low-$z$ spirals. Based on our experience of finding lensed sources in H-ATLAS, we expect $\sim$ 1200 bright lensed sources with $S_{500}$ > 100 mJy. Existing optical and near-IR data can be used to separate the lensed source sample from low-$z$ ($z$ < 0.1) bright spirals, and to identify the foreground lenses. It is relatively easy to measure the redshifts of the background sources using existing CO spectrometers (and soon ALMA), and the redshifts of the lenses with optical telescopes. This lensed source sample can also be used for a number of experiments to investigate dark matter and dark energy that have been suggested in the literature, but never properly applied due to lack of a large complete sample. Without lensing magnification ($\sim$10–20), many of these sources would be below the SPIRE confusion limit, and thus a straightforward way to study the sources producing the missing 85% of the FIRB will be to find lensed HSLS sources and observe them with ALMA.

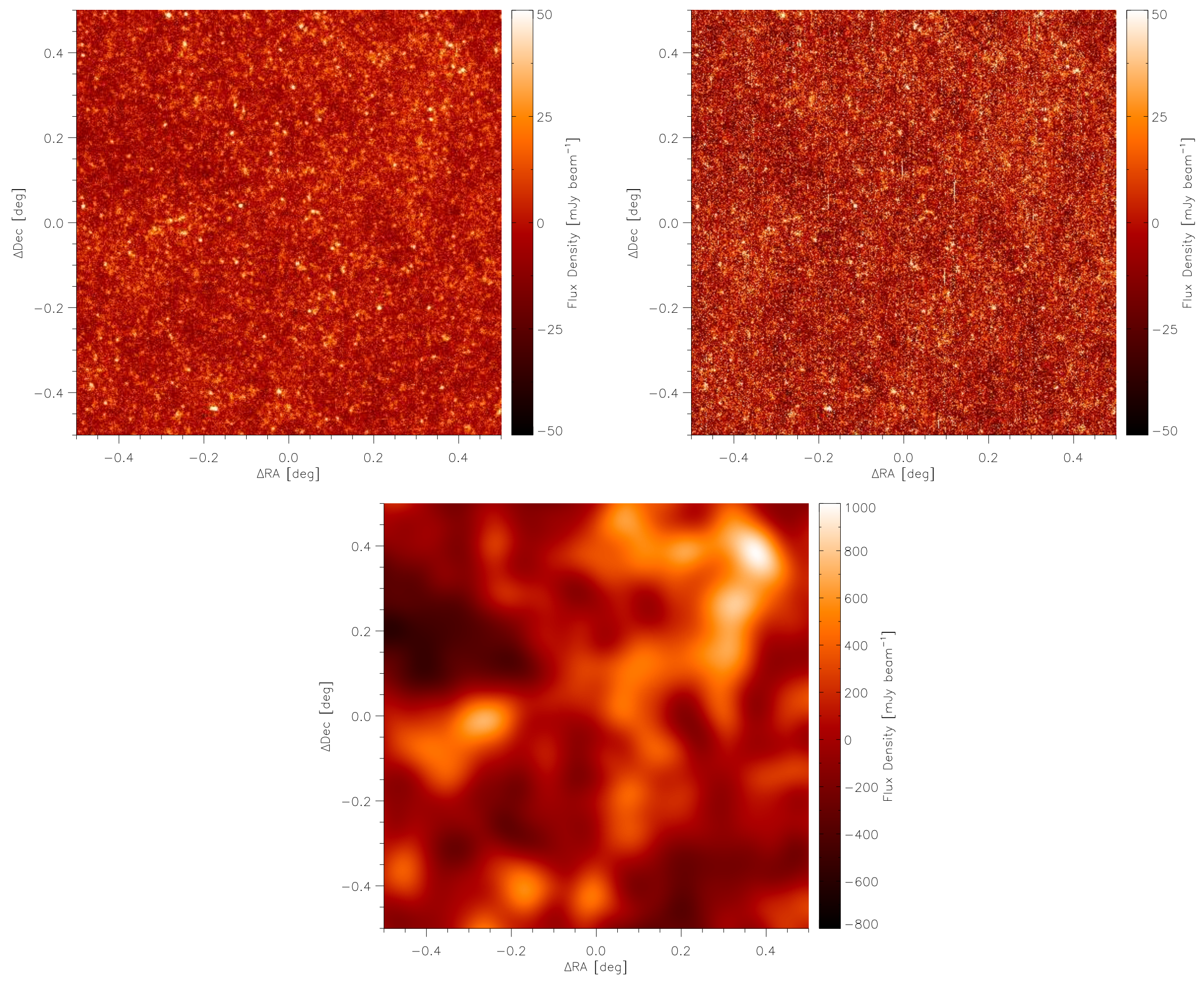

Figure 3: 1 deg$^2$ maps made from SPIRE data in the Lockman-SW field. *Top left*: SPIRE 350 $\mu$m map with 4 scans (2 orthogonal scans in the fast-scan mode). *Top right:* SPIRE 350 $\mu$m map made with a single-scan (HSLS map will be similar to this). *Bottom:* Expected Planck HFI 850 GHz channel map in the same 1 deg$^2$ area; we show the SPIRE map convolved with the Planck beam.

The large comoving volume sampled by HSLS means that we will detect significant numbers of the rarest low-redshift systems, including Ultraluminous Infrared Galaxies (ULIRGs), the most luminous quasars, and 350 very massive (M $> 10^{15}$ M$_\odot$) clusters, with 20 at $z > 1$. Being able to simultaneously study very high-redshift systems and the low redshift universe, thanks to the large comoving volume of HSLS, enables us to follow for the first time the evolution of ULIRGs over most of the history of the universe. It will also allow us to investigate the inter-connected evolution of galaxies and nuclear black holes by measuring (from the submillimetre emission) the evolution of the star-formation rate in quasars. Using empirical models and the results from H-ATLAS, we predict that we will detect $\simeq 1000$ $z > 5$ sub-mm galaxies (SMGs), extending the study of dust-obscured star formation (and the formation of dust) into the epoch that will be studied by JWST.

Environment must play an important role in the process of galaxy formation, the most striking observational evidence being that clusters today have a much higher fraction of early-type galaxies than is found in the field. At $z > 1.5$, we predict that the HSLS will find $\sim 200$ "proto-clusters," unvirialized over-densities extending 10-20 Mpc that will evolve into virialized clusters like those we see around us today. These protoclusters have so far only been found as clusters of sources around high-redshift radio galaxies and quasars. By observing these proto-clusters in other wavebands, we will be able to learn much about the evolution of these structures and their populations. Since the HSLS will also observe very massive clusters out to $z \sim 1$, we will again be able to follow the evolution of dust and dust-obscured star formation over most of the history of the universe.

(2) **The Evolution of Dark Matter:** The paradigm for the evolution of structure in the universe is based on the gravitational coalescence of non-baryonic dark matter, which has a mass six times greater than the mass of the baryonic matter. Yet there has been no way of directly measuring the evolution of the dark-matter structures predicted by this paradigm. We will use the HSLS to test and extend this paradigm.

*Lensing by Galaxies:* Given the existing measurements of the cosmological parameters, we can use the sample of 1200 HSLS lensed sources to directly measure the mass distributions in the lenses. Using follow-up optical and CO spectroscopy and shallow ALMA maps, we will directly measure the number-density of halos as a function of mass and redshift, which can be compared with the theoretical predictions. With longer ALMA observations and using lens reconstruction techniques, we will be able to determine the mass profiles of the lenses. By also incorporating optical observations, we will investigate separately the evolution of the profiles of the baryonic and non-baryonic matter of our foreground lenses out to $z \sim 2$.

*Lensing of the CMB:* The intervening large-scale structure acts as a foreground lens to CMB photons. The large-scale structure at $2 < z < 3$ traced by HSLS lies halfway between the surface of recombination and the observer in comoving angular diameter distance, well matched to the lensing kernel. Cross-correlating the line-of-sight projected density of the structure traced by HSLS with a convergence map reconstructed from CMB lensing provides a measure of the linear bias of the structure. We predict that by combining the HSLS with Planck we will detect this effect with a signal-to-noise ratio greater than 20. This signal-to-noise ratio increases by a factor of 3 to 60 when the HSLS maps are combined with CMB lensing from ACT or SPT, which have $\sim 1'$ beams. The measurements will allow us to constrain the sum of the neutrino masses down to 0.05 eV, leading to an identification of the neutrino mass hierarchy from cosmological structure formation measurements for the first time.

*Lensing of the Far-Infrared Background:* A particularly exciting possibility is to use the HSLS maps as the background to investigate the evolution of dark matter at $z < 2$. As with all wide SPIRE maps, the HSLS maps will contain information about the angular fluctuations ($C_\ell$) of the cosmic far-IR background, making it possible to probe the nature of faint sources that are below the confusion level. The HSLS will be better for this kind of investigation than the existing *Herschel* surveys, because the statistical weight provided by the area beats the extra depth of HeRMES and H-ATLAS; we predict that we will be able to measure this effect with a signal-to-noise ratio of $\simeq 1000$.

*The Integrated Sachs-Wolfe Effect:* This is the differential redshift effect from photons climbing in and out of a time-varying gravitational potential. With Planck+optical surveys, it will be possible to measure the ISW at $z < 1$. The HSLS is the only way to extend measurements of the ISW out to $z \sim 2$. At this redshift the ISW is insensitive to dark energy properties as long as dark energy is similar to a cosmological constant or a fluid with negative pressure. Any unusual signature in the ISW signal will therefore test the basis of the structure formation paradigm, such as the possibility that, on large scales, General Relativity (GR) needs to be replaced by a modified gravity theory that also explains cosmic acceleration today. Planck+HSLS ISW will provide the strongest constraint on GR at the largest cosmological scales probed.

(3) **Dark Energy and Cosmology:** We will be able to use the HSLS for a number of projects to measure cosmological parameters and to investigate the evolution of the equation-of-state of dark energy.

*The Evolution of Clusters:* The number-density of clusters as a function of redshift and mass provides a sensitive test of cosmological parameters. One of the key science goals of all CMB experiments is to provide samples of clusters for cosmological tests, with the clusters being detected via the scattering of photons by the hot gas in the clusters – the Sunyaev-Zel'dovich (SZ) effect. Planck, for example, should detect $\simeq 200$ clusters in the HSLS region, but the $5'$ resolution of Planck means the SZ signal will be contaminated by point sources (lensed sources and cluster galaxies). With HSLS, we will be able to remove these contaminating populations and develop statistical methods for decontaminating the rest of the Planck SZ sample. A decontaminated cluster sample will also open up the possibility of looking for the kinetic SZ effect, which can provide measurements of the peculiar velocities of clusters – another test of the cosmological paradigm.

*A Search for Non-Gaussianity (NG):* The standard cosmological paradigm is based on the assumption that the density fluctuations in the early universe are Gaussian. This is the simplest assumption, but there are many alternative models in which the fluctuations are non-Gaussian. We can use HSLS sources at $z \sim 2$ to carry out a sensitive test of this assumption, because at this epoch most of the structure traced by HSLS sources is still in the linear regime of gravitational evolution. We estimate that the non-Gaussianity parameter, $\mathrm{f_{NL}}$, can be constrained with HSLS to an accuracy of $\pm 11$ at $z \sim 2$ (current constraint $32 \pm 21$), with a potential improvement to $\pm 7$ with exact redshifts. While the measurement of NG is a key goal of Planck and other CMB experiments, our survey will allow a measurement of the $f_{\mathrm{NL}}$ parameter competitive with Planck, but using a different technique, probing different scales and redshifts. If Planck detects a signal, then HSLS will offer a strong independent confirmation. If Planck does not see any NG signal, then HSLS will explore a complementary domain.

*The Dark Energy Equation of State (EOS):* One of the methods that has been suggested for determining the evolution of the EOS of dark energy is to look for the baryon acoustic oscillation wiggles in the matter power spectrum at different redshifts; these are imprinted at the time of recombination, providing a standard ruler. By estimating the redshifts of the HSLS from the SPIRE colours alone ($\sigma_{\rm z} \simeq 0.3$), and combined with Planck, we will measure the EOS parameter $w$ to an accuracy of $\pm$ 0.17 compared to $\pm$ 0.26 with Planck data alone. This will improve to $\pm$ 0.11 as we obtain the redshifts more accurately from DES and other optical/near-IR surveys, with a considerable improvement to $\pm$ 0.05 with exact redshifts for HSLS sources. With ALMA and follow-up of $\sim$ 1000 strongly lensed sources, HSLS will measure the EOS to $\pm$ 0.06, independent of all other cosmological probes.

## 2 The Joint Planck/CMB-Herschel Studies

While the CMB is an exquisite probe of the physics of the early universe, CMB photons also encounter the large-scale structure (LSS) of the universe – especially baryons and clumps of dark matter that define the universe at $z \sim 0$ to 5 – while they are in transit to us. These encounters affect several aspects of the photon properties, such as the frequency or direction of propagation, leading to secondary anisotropies in the CMB. A careful study of CMB maps, complemented by information on the LSS at a variety of redshifts as traced by multi-wavelength sources, can help extract these signals and quantify key cosmological and astrophysical properties of the universe.

- The joint analysis of Planck and HSLS allows a measurement of the Integrated Sachs-Wolfe effect at an unprecedented high redshift ($z \sim 2.0$), providing a potent test of alternative theories of gravity.

- The sources mapped by HSLS correlate strongly with the CMB lensing signal. The cross-correlation signal reaches 20-$\sigma$ for Planck and 16/42-$\sigma$ for ACT/ACTPol. This yields more robust CMB lensing measurements for these experiments, and corresponding improvements in cosmological constraints.

- The extra information on dust emission provided by HSLS will substantially improve the Planck Sunyaev-Zel'dovich (SZ) cluster catalog photometry in the overlap region, allowing 95% completeness for SNR>5 clusters.

- The dust mapping allowed by HSLS enhances the prospects of measuring the kinetic SZ and patchy reionization signals for overlapping CMB experiments like Planck, SPT or ACT.

### 2.1 Integrated Sachs-Wolfe (ISW) Effect

The ISW effect arises from an imbalance between the blueshift a CMB photon suffers while falling into a gravitational potential well and the redshift while climbing out, if the gravitational potential evolves during transit. This effect is important in low matter density universes with $\Omega_m < 1$, where the gravitational potentials decay at low redshift; it contributes anisotropies on and above the scale of the horizon at the time of decay. The CMB temperature field is hence correlated with tracers of gravitational potentials such as galaxies, and measuring this correlation provides important cosmological constraints (e.g. Ref. [309]). The ISW effect has now been shown to be present, with varying levels of significance, in cross-correlations of WMAP data with LSS tracers going back to $z \sim 1.5$. With HSLS, we will extend ISW studies with a detection at $z \sim 2$ using combined HSLS and Planck maps.

**Constraints on modified gravity models:** An ISW measurement at $z \sim 2$ provides a stringent test of modifications to General Relativity (GR). We will now consider how an ISW detection with HSLS will discriminate among different theories of Modified Gravity (MG). The main modification made by MG models on the CMB is a change in the evolution of gravitational potentials during the acceleration epoch. The significance of a detection of this departure from GR is stronger if the redshift of the LSS tracers is matched to the epoch at which the gravitational potentials evolve, thus changing the growth function. *This enables the HSLS to be a powerful probe of MG, as it sits at a "sweet spot" at $z \sim 2$ where the difference in the growth function between GR and MG models peaks.* We shall now consider

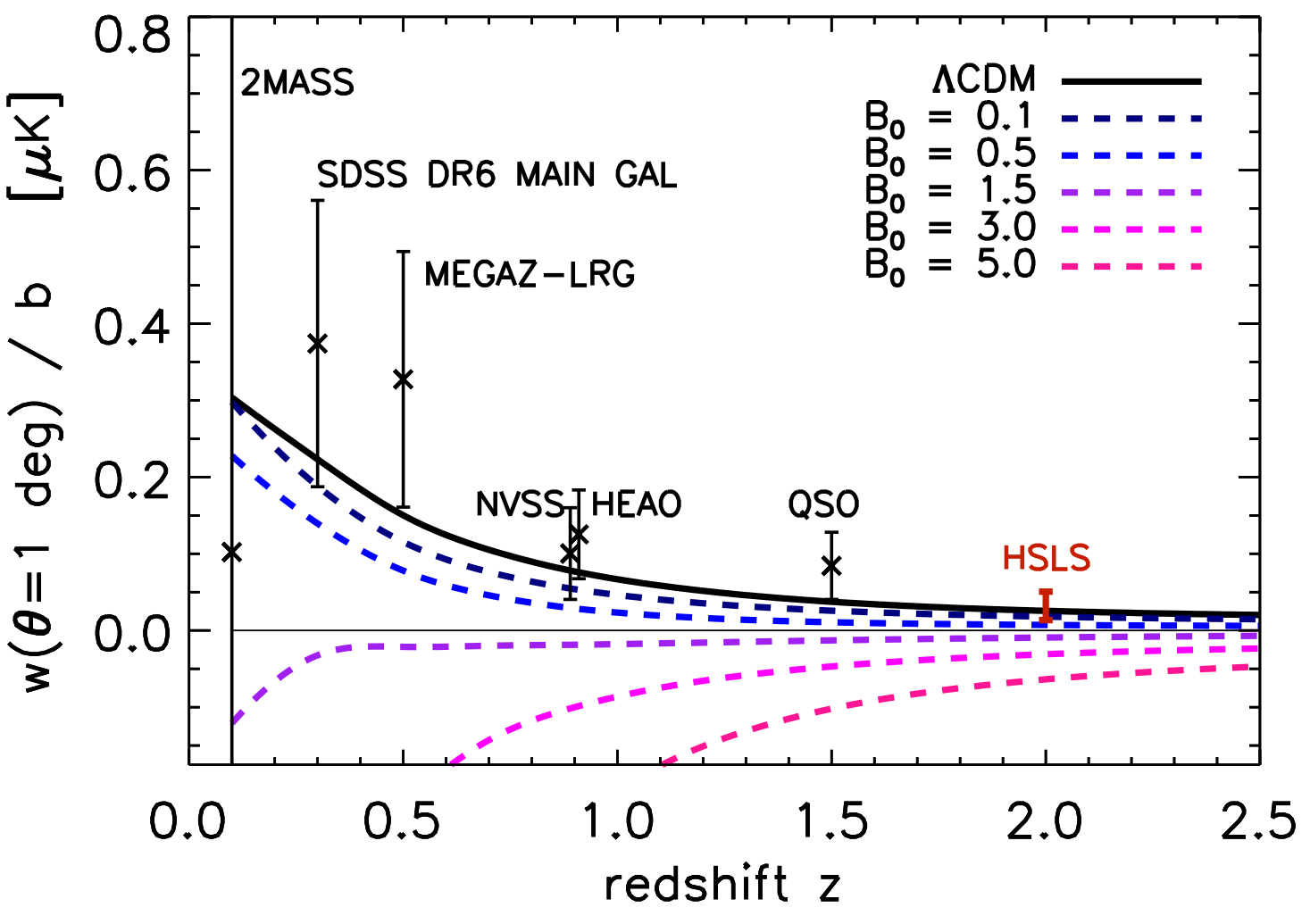

Figure 4: The ISW–source angular correlation function at 1 degree is shown as a function of redshift as measured by a variety of surveys, with the bias divided out. The WMAP7-normalized ΛCDM model (limit $B_0 \to 0$) is shown as a solid curve, and $f(R)$ gravity models with $B_0 = 0.1$, 0.5, 1.5, 3.0 and 5.0 are shown as dashed curves (top to bottom), where $B_0$ is the $f(R)$ Compton wavelength parameter. The forecast for the cross-correlation of ISW–HSLS sources (red) is shown, with current measurements at lower redshifts presented for comparison with black error bars (all at 1-$\sigma$).

several specific MG models that can be constrained via this effect using HSLS ISW.

In Figure 4, we show the ISW-source angular correlation function at 1 degree, $w(\theta = 1^\circ)$, for the redshift range of $0 < z < 3$, comparing current measurements of this quantity with a forecast for the ISW–HSLS sources correlation at $z \sim 2$, assuming sky coverage of 4000 deg$^2$. The WMAP7-normalized ΛCDM model (limit $B_0 \to 0$) is shown as a solid curve, and a family of $f(R)$ gravity models with Compton length scale $B_0 = 0.1$, 0.5, 1.5, 3.0 and 5.0 (see Ref. [381] for details) are shown as colored dashed curves. With a cosmological constant, gravitational potentials decay during the acceleration epoch. For $f(R)$ models, the enhancement of the growth rate below the Compton scale can change the decay into growth. CMB photons then become colder along directions associated with overdense regions. This reversal changes the sign of the ISW correlation, as seen in Figure 4. The theoretical curves are calculated for a set of redshift points of median redshifts $z_i = (0.1, 0.3, ..., 2.9)$, assuming ideal galaxy catalogues of "broad" redshift distributions

$$n_i(z) = \frac{1.5}{\Gamma(2)} \frac{z^2}{z_i^3} e^{-(z/z_i)^{1.5}}, \tag{1}$$

following Ref. [381].

We show the observational data compiled by Ref. [170], obtained through correlating the WMAP CMB maps with six different galaxy catalogues ranging between $0.1 < z < 1.5$: the infrared 2MASS survey; the SDSS main galaxy sample; the MegaZ data set of SDSS luminous red galaxies (LRGs); the NVSS radio galaxy survey; the HEAO X-ray catalogue; SDSS quasars. For a uniform comparison, the linear bias – assumed constant for each catalogue, and fitted from the auto-correlation functions – is removed. We can see that these data are generally compatible with ΛCDM, within the 1-$\sigma$ error bars, although the observed amplitude is generally higher. The 2MASS error bar is arbitrarily large, because these angular scales are heavily contaminated by the Sunyaev-Zel'dovich effect in this survey, thus hindering a significant measurement.

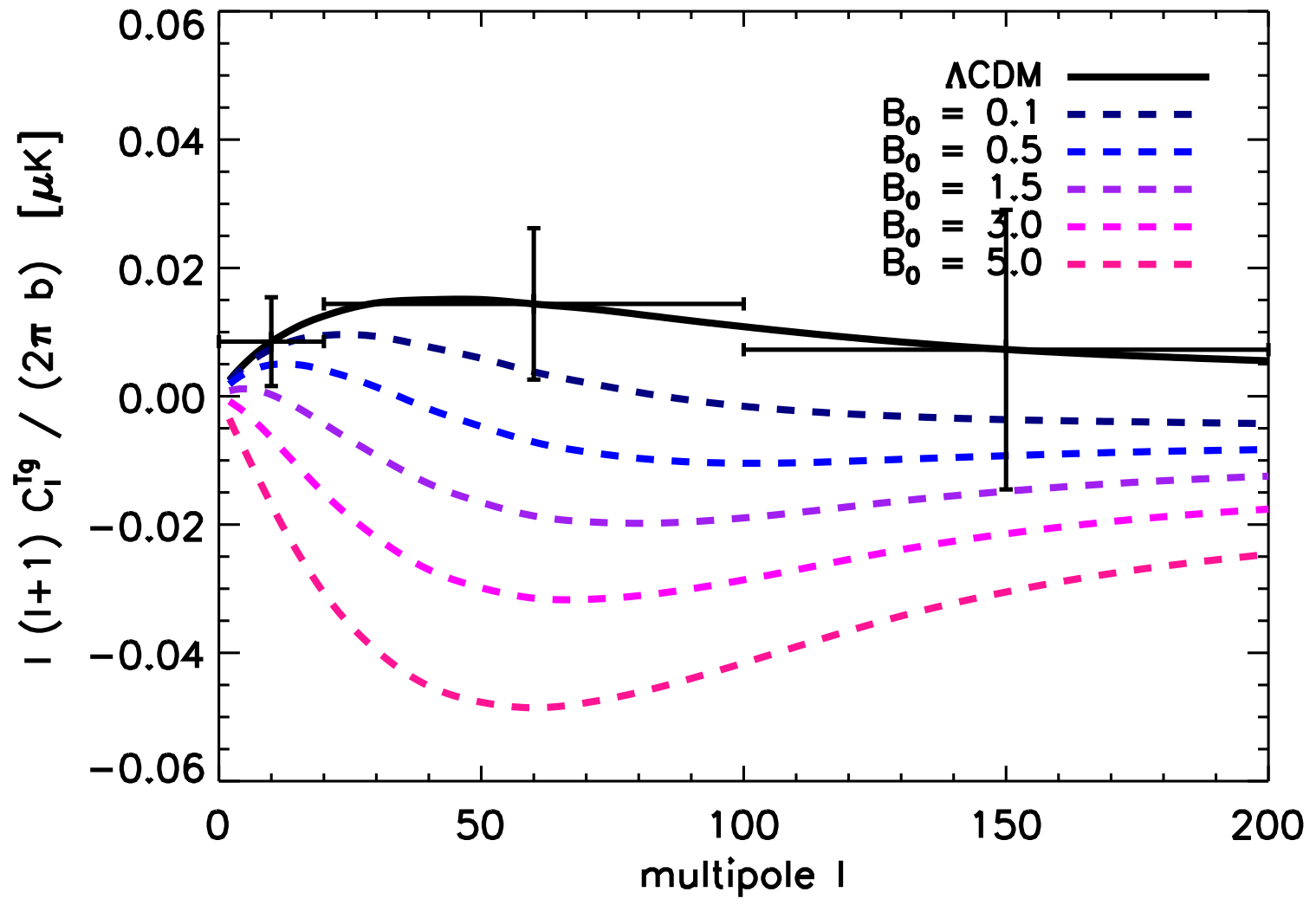

Figure 5: The ISW–source angular power spectrum with the bias divided out, with theoretical curves for ΛCDM and $f(R)$ presented in the same format as in Figure 4. The forecast for the cross-correlation of ISW–HSLS sources in various multipole bins is indicated by black error bars.

The red error bar is the forecast for Planck-HSLS at $z \sim 2$. We see that not only is the HSLS S/N significantly better than with current data, but the measurement will be made at a high redshift which will not be achievable by any other survey till the advent of LSST, enhancing cosmological constraints using the ISW effect.

Figure 4 should be interpreted with the caveat that the S/N which is dispersed over a range of angular scales is assigned solely to the 1 degree angular scale. In Figure 5 we show the detection as a function of multipole, showing the full range of angular scales from which the signal arises at the expense of suppressing redshift information.

Current constraints on $f(R)$ gravity are $B_0 < 1$ (at 95 % CL) when using the CMB plus the condition of not observing any negative ISW [381], which is reduced to $B_0 < 0.4$ when performing a full likelihood analysis of current ISW data [169] with many different systematics. Even stricter bounds are found in the non-linear regime using cluster constraints [348]. Figure 5 shows that $f(R)$ gravity can be constrained by Planck-HSLS ISW *alone* at $B_0 < 1$ (95% CL) in a hitherto-unexplored redshift range.

The Tensor-Vector-Scalar (TeVeS) theory [29] is a metric theory of gravity which was designed to encompass a MOND regime in the low-acceleration limit, and to attempt to explain all cosmological observations without dark matter. While the formation of large-scale structure is possible without dark matter in this model [119, 370], the history of structure formation is very different, which leaves a distinct signature in the ISW effect [347]. The predictions shown in Figure 6 are for the TeVeS model including massive neutrinos described in Ref. [347], which was adjusted to fit the CMB [370]. The HSLS-Planck cross-correlation will be able to constrain the TeVeS model with high significance.

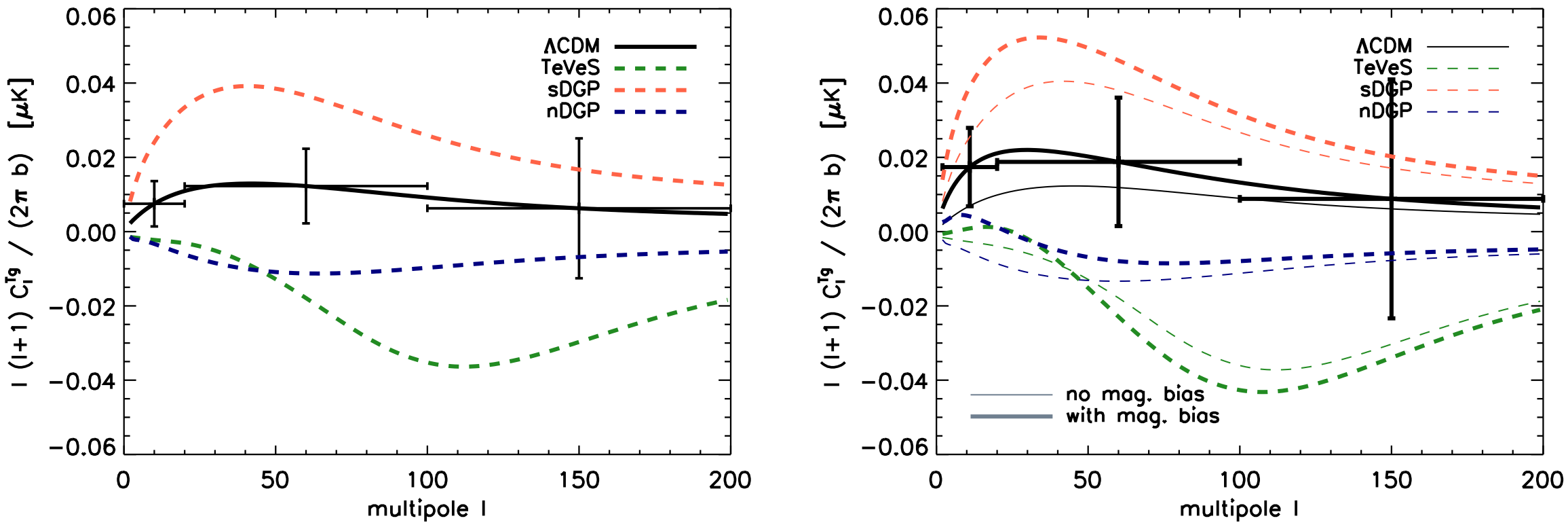

Figure 6: (Left) The ISW–source angular power spectrum with the bias divided out, with theoretical curves for ΛCDM and TeVeS/DGP MG models as indicated. (Right) The same models, showing the effect of accounting for magnification bias, which is significant for HSLS. The forecast for the cross-correlation of ISW–HSLS sources in various multipole bins is indicated by black error bars.

Another popular modified gravity scenario is the Dvali-Gabadadze-Porrati braneworld model [127]. In this model, it is possible to achieve an accelerating Universe without dark energy [110]. In Figure 6 we also show two examples of DGP models: a *self-accelerating* DGP model without any cosmological constant or dark energy; and a *normal-branch* DGP model with a dark energy component adjusted to yield an expansion history identical to ΛCDM [338, 346]. For the self-accelerating model, we choose parameters from Ref. [146] which yield the best fit to current CMB and expansion history (SN, BAO, $H_0$) data. Note that this model is already in $\sim 4\sigma$ conflict with the data [146]. The normal-branch DGP model with dark energy from Ref. [346] can be seen as an effective model for generalized braneworld models whose precise predictions have not been calculated yet. Since the expansion history is identical to ΛCDM, the parameter controlling the modification to gravity, the cross-over scale $r_c$, can be adjusted freely. Here, we set it to $r_c = 500$ Mpc.

In a flux-limited survey, the observed overdensity of galaxies is given by (e.g., Ref. [63]) $\delta_{\rm obs} = \delta_g + 2(S-1)\kappa$, where $\delta_g = b\delta$ is the intrinsic galaxy overdensity (and $b$ the galaxy bias), $S = d\ln N/d\ln f_{\rm min} \approx 2.8$ is the logarithmic slope of the number counts at the flux threshold, and $\kappa$ is the weak lensing convergence (e.g. Eq. (2) in Ref. [349]). Correspondingly, the galaxy-temperature cross-correlation power is given by [244]

$$C^{gT}(\ell) = b\, C^{\delta T}(\ell) + 2(S-1)C^{\kappa T}(\ell). \quad (2)$$

Both contributions are due to the integrated Sachs-Wolfe effect. While the first term only receives contributions at the redshift of the source galaxies, the magnification contribution is sensitive to all structure along the line of sight to the source galaxies. Both terms can be calculated straightforwardly if the evolution of the matter overdensity $\delta(k,z)$ and lensing potential $(\Phi-\Psi)(k,z)$ is known (see Ref. [349] for further details). Thus far, we have only considered the first term, $C^{\delta T}(\ell)$. In the right panel of Figure 6, we show the effect of including the second term $C^{\kappa T}(\ell)$, which accounts for the magnification bias. A bias of 3, a number density of 700 per deg$^2$, and a sky coverage of 4030 deg$^2$ has been assumed here. We see that the effect is significant for HSLS due to the steep flux slope for sub-mm galaxies, and generally leads to an increase in the S/N for differentiating between ΛCDM and the MG models (though the noise due to clustering of the sources will also be increased somewhat by magnification).

**Generic tests of GR** The ISW effect also provides a generic test of GR without reference to alternative MG models. Similar to a weak lensing experiment, the ISW cross correlation will constrain *gravitation slip* $\Sigma$ [169] defined as

$$k^2(\Phi - \Psi) \;=\; 8\pi a^2 G\Sigma(k,a)\rho_m\delta_m, \tag{3}$$

where $\Sigma(k,a) \equiv 1 + \Sigma_0 a^3$. While a Planck prior on the angular size distance to the last scattering surface alone yields no constraints on the gravitational slip, we forecast that a full joint analysis of Planck-HSLS ISW yields a 1-$\sigma$ error of $\sigma(\Sigma_0) = 1.90$. For comparison, a Planck prior on the angular size distance to the last scattering surface jointly with a CMB lensing analysis would yield $\sigma(\Sigma_0) = 0.75$; the combination of full Planck analysis with CMB lensing and HSLS would yield $\sigma(\Sigma_0) = 0.68$. These constraints are obtained using a Fisher matrix formalism, marginalising over the standard cosmological parameter set. While not competitive with Planck lensing, HSLS ISW provides an important cross-check with different systematics.

**Constraints on dark energy:** It is of prime interest to constrain dark energy by exploiting large scale structure probes, such as weak lensing, redshift surveys, and the ISW effect. Among these experiments, the ISW effect gives the weakest constraint on dark energy. However, it gives an independent confirmation of cosmic acceleration, and it is the only probe of the consistency between the early universe and the late-time acceleration. We estimate the constraint on dark energy via the ISW effect using a Fisher matrix analysis. Since we do not want to double count late-time information, we use Planck priors on the angular size distance to the last scattering surface and on $\Omega_m h^2$, and do not include CMB lensing. The Planck distance prior alone does not constrain dark energy, but we forecast that the combination of a full Planck analysis with HSLS will yield a 1-$\sigma$ error on the equation of state parameter for dark energy of $\sigma(w) = 0.52$. For comparison, a Planck analysis including CMB lensing would yield $\sigma(w) = 0.12$.

Our forecasts indicate that cosmic acceleration can be confirmed using the Planck-HSLS ISW effect alone (without relying on any other LSS experiments), and that ISW-HSLS will provide information on cosmic acceleration at high redshift (where the effect, at least in standard $\Lambda$CDM, should be very small).

### 2.2 Gravitational Lensing of the CMB

The intervening large-scale structure acts as a foreground lens to CMB photons, both redistributing power in multipole space and enhancing it at arcminute scales due to large density perturbations such as dark matter halos and other structures that sit on large CMB temperature gradients. The most effective structures for lensing lie halfway between the surface of recombination and the observer in comoving angular diameter distance. In the fiducial $\Lambda$CDM, this is at $z \sim 3.5$, but the growth of structure skews this to somewhat lower redshifts of $z \sim 2.5$. *Sub-mm sources are the best matched tracer population for CMB lensing kernel, especially when compared to radio and optical surveys.* Combined with Planck, we will make a S/N $\sim 20$ detection of the lensing signature.

### 2.2.1 What can we learn from CMB lensing and HSLS ?

While *Herschel* is mapping the sky in the infrared and sub-mm wavebands, a window of observations that is largely unexplored, a new generation of CMB experiments like Planck and its ground-based counterparts, the Atacama Cosmology Telescope (ACT) [157] and the South Pole Telescope [185, 246] are mapping out the CMB anisotropies with unprecedented precision and sensitivity. One of the many exciting outcomes of the high-resolution CMB experiments will be the reconstruction of the projected matter density of the universe, i.e. the convergence, by studying subtle distortions on the CMB due to lensing by intervening large scale structure. Galaxies, being biased tracers of the same large scale structure, are expected to correlate strongly with the CMB-lensing reconstructed density field. The galaxy density map to be derived from the HSLS will be a "golden candidate" for such a cross-correlation study. This is because the strong negative K-corrections for this class of galaxies make their predicted redshift distribution overlap strongly with the CMB lensing kernel. Also, the steep faint-end slope of the number counts makes these galaxies exhibit a strong magnification bias which enhances the cross-correlation signal. As discussed below, the cross-correlation signal-to-noise is expected to be high enough to provide excellent estimates of the bias of these galaxies. Alternatively, if the bias and redshift distribution could be estimated otherwise, then this galaxy-CMB lensing cross-correlation can improve the constraints on cosmological parameters.

Another way in which IR galaxies enter the CMB lensing scenario is as an extra source of background noise. This is another aspect in which a HSLS-like survey could be useful by providing a template for subtracting off this diffuse IR background.

### 2.2.2 CMB Lensing

Large scale structure in the universe deflects CMB photons, making the lensed CMB sky $\tilde{T}(\hat{\mathbf{n}})$ essentially a remapping of the primordial sky $T(\hat{\mathbf{n}})$:

$$\tilde{T}(\hat{\mathbf{n}}) = T(\hat{\mathbf{n}} + \mathbf{d}), \tag{4}$$

where $\mathbf{d}$ is the effective deflection field. The typical deflection in a $\Lambda$CDM universe is about $2.7'$ and is coherent over degree scales on the sky. The statistical distortions of the temperature and polarization patterns of the CMB sky can be used to reconstruct the deflection field, or the closely related quantity convergence, $\kappa = \frac{1}{2}\nabla \cdot \mathbf{d}$, which is related to the line-of-sight density field:

$$\kappa = \frac{3}{2}\Omega_m H_0^2 \int \frac{d\eta}{a(\eta)} \frac{d_A(\eta) d_A(\eta_0 - \eta)}{d_A(\eta_0)} \delta(d_A(\eta)\hat{\eta}, \eta), \tag{5}$$

where $\eta$ is the comoving lookback time, $d_A$ denote comoving angular diameter distances, $\Omega_m$ and $H_0$ are the matter density and Hubble parameters, and $\delta$ is the fractional overdensity of matter.

### 2.2.3 The cross-correlation signal

We assume that the HSLS galaxies are biased tracers of the underlying dark matter, with linear bias $b$. The projected fractional overdensity of galaxies can be written as,

$$\Sigma_g(\hat{\mathbf{n}}) = \int_0^{\eta_0} \left[ bN(\eta)\delta(d_A(\eta)\hat{\eta}, \eta) + 2(\alpha - 1)\kappa_G(\hat{\mathbf{n}}) \right] \tag{6}$$

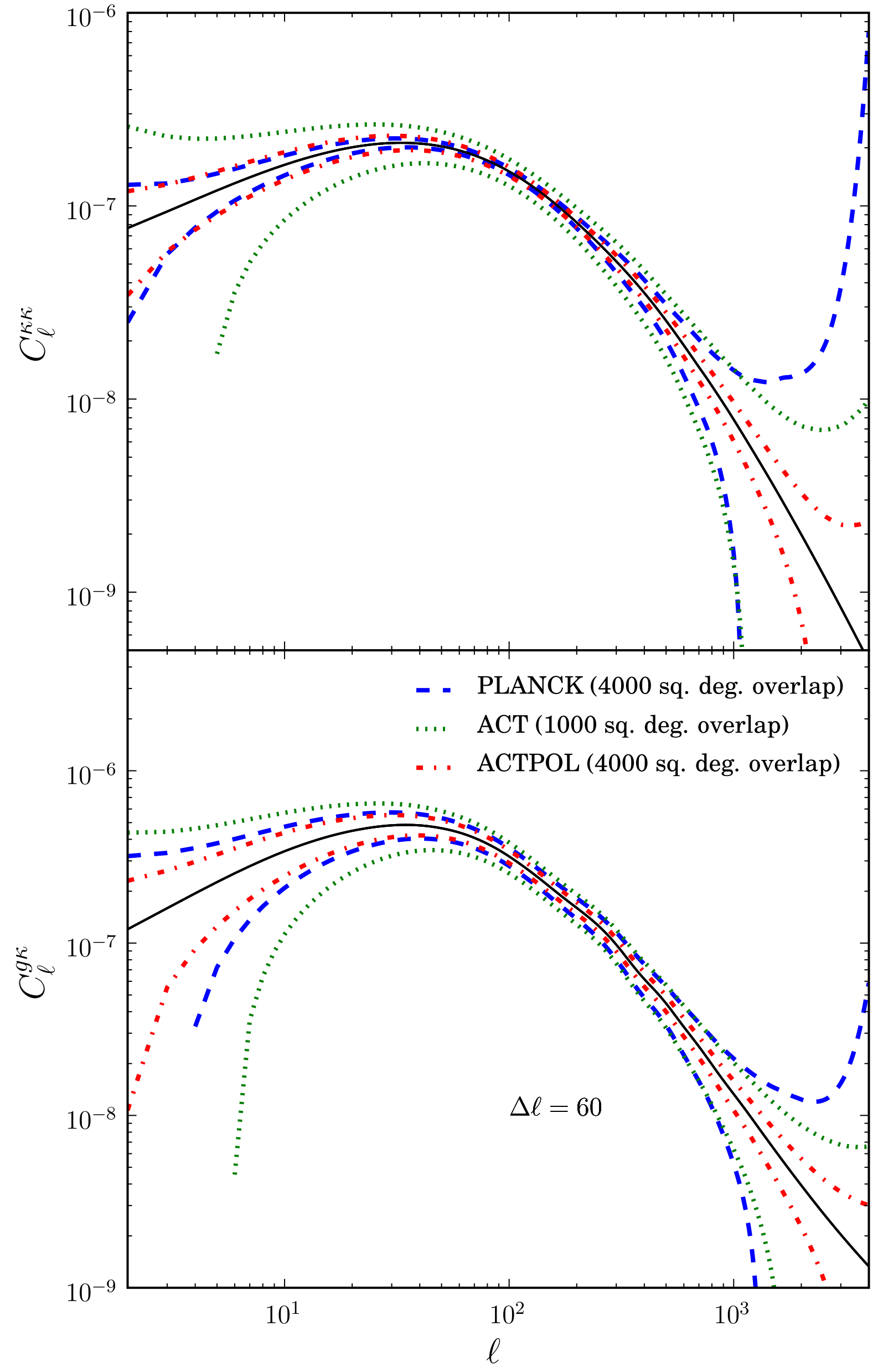

Figure 7: Uncertainties in power spectrum estimates. The top panel shows the uncertainties in the power spectrum of the CMB-lensing reconstructed convergence field, $C_\ell^{\kappa\kappa}$, achievable with Planck, ACT and ACTpol. The bottom panel shows the power spectrum of the cross-correlation signal between HSLS galaxies and CMB-lensing reconstructed convergence for the same experiments. The overlap area between each experiment and HSLS are shown in the legend. It is evident that a precise measurement of the cross-correlation signal is possible with current instruments and should lead to improved cosmological constraints and better understanding of galaxy bias.

| Experiment | S/N | $\Delta b/b(\%)$ |
|---|---|---|
| Planck | 20 | 5.0 |
| ACT | 16 | 6.2 |
| ACTPol | 42 | 2.3 |

Table 1: Estimated signal-to-noise for the cross-correlation signal discussed in the text and the resulting error on galaxy bias.

where $N(\eta)$ is the normalized distribution of galaxies in comoving distance. The second term represents the magnification bias, where $\alpha$ is the logarithmic slope of the source counts at the faint end, and $\kappa_G$ is the convergence for these galaxies treated as lensing sources.

From equations (5) and (6) it is straightforward to compute the cross-spectrum $C_\ell^{\kappa g}$ which is displayed in the bottom panel of Figure 7. The uncertainties on the cross-spectrum in bands of $\Delta\ell$ are also shown, and are estimated as,

$$\Delta C_\ell^{\kappa g} = \frac{1}{\sqrt{\ell \Delta\ell f_{\rm sky}}} \left[ (C_\ell^{\kappa\kappa} + N_\ell^{\kappa\kappa})(C_\ell^{gg} + N_\ell^{gg}) \right]^{1/2}, \quad (7)$$

where $C_\ell^{gg}$ is the galaxy auto-spectrum and $N_\ell^{gg} = 1/\bar{n}_g$ is the galaxy shot noise, $\bar{n}_g$ being the number of galaxies per steradian. We assume HSLS to have a coverage of 4000 deg$^2$ with 700 galaxies per deg$^2$, a linear bias, $b = 3$, number count slope $\alpha = 3$, and a redshift distribution discussed previously. We consider three CMB experiments — Planck, for which we adopt specifications similar to the Planck Blue Book [318], the Atacama Cosmology Telescope (ACT) survey [157], which we take to be a 1000 deg$^2$ survey with 1.4 arcminute resolution at a sensitivity of $\sim 30$ $\mu$K-arcmin, and ACT fitted with a polarization sensitive camera (ACTPol) [301] for which we consider a 4000 deg$^2$ wide survey with 1.4 arcmin resolution with 20 and 28 $\mu$K-arcmin sensitivity for its temperature and polarization maps. The predicted uncertainties and S/N numbers are shown in Figure 7 and Table 1. Also shown are the expected errors on the galaxy bias (which is simply the inverse of the S/N) assuming the underlying cosmology and the redshift distribution are known. The effect of neglecting redshift uncertainties is expected to be small given the width of the CMB lensing kernel.

If the galaxy bias and redshift distribution can be controlled, then combining measurements of the (projected) galaxy overdensity on the sky with CMB lensing information can break parameter degeneracies that are present with CMB lensing data only, and can thus improve cosmological constraints. In particular, constraints on the sum of neutrino masses from CMB lensing are expected to improve significantly with the addition of HSLS data.

#### 2.2.4 IR Contamination in CMB Lensing estimators

To determine the convergence field and power spectrum from a measurement of the CMB temperature, one uses the fact that weak gravitational lensing modifies the statistics of the CMB so that it is no longer a purely Gaussian field – lensing introduces correlations between different Fourier modes of the temperature field. One can thus define a quadratic estimator for the convergence field by using a product of two different Fourier modes of the temperature field [201]. The noise in such a reconstruction arises from mainly three sources – the primary CMB itself, the instrumental and/or atmospheric noise, and secondary effects like the SZ effect and emission from dusty IR galaxies. The third contribution is usually ignored in theoretical calculations, as has been done in the previous two subsections. Here, we make a simple attempt to obtain an idea of the IR background contamination to CMB lensing noise, by assuming that the IR sources are uncorrelated and unlensed (which are both over-simplifications) and basically appear in all relevant equations as an extra source of noise.

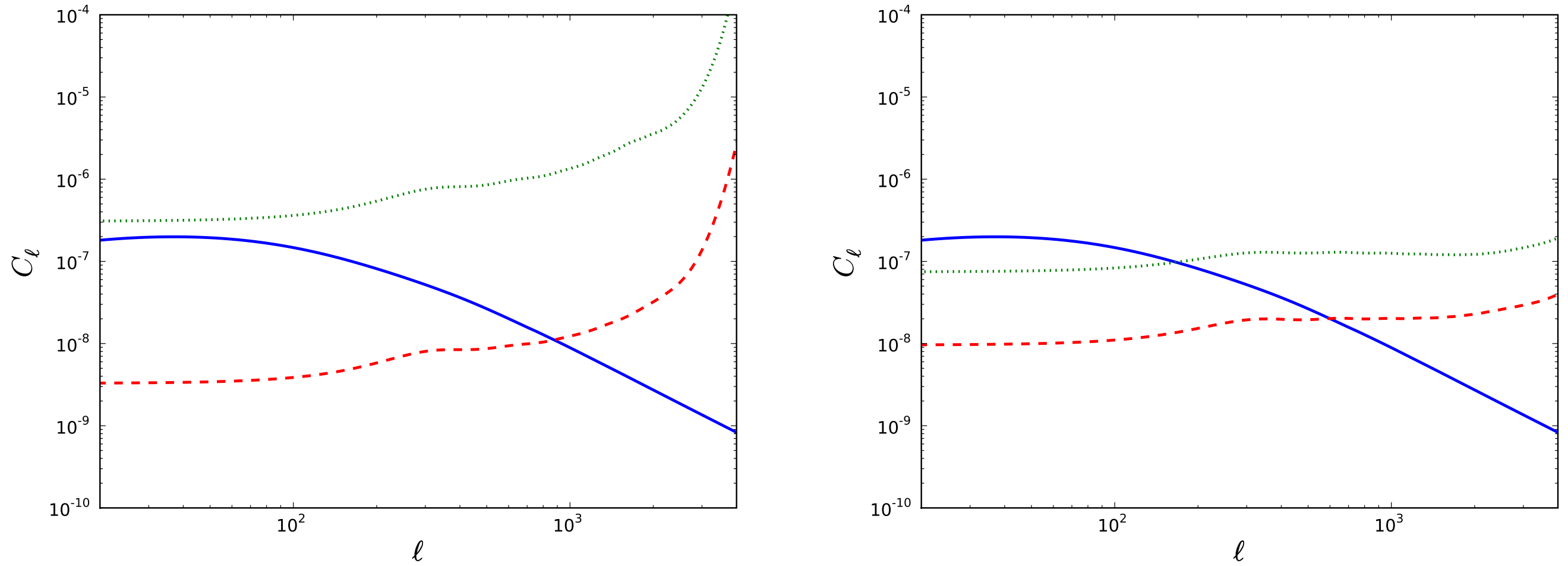

Figure 8: Power spectrum of the signal i.e. convergence (solid line), noise bias without IR effects (dotted line), and the additional IR contribution to the noise bias (dashed line) for a Planck-like (left panel) and an ACTpol-like (right panel) experiment.

We do all calculations at 148 GHz for temperature only, and use the Sehgal et al. (2009) [356] IR galaxy template to create an IR background by subtracting all sources brighter than 5 mJy from it. We also take a very conservative approach in the reconstruction by restricting the maximum multipole in the lensing estimator to $\ell < 3300$. This is done to make sure that we are using a range of multipoles where we are dominated by the CMB. (With information from HSLS, we may be able to push this to higher multipoles, thus increasing S/N.) Next, we perform lensing noise calculations, once with just the CMB and white noise at the level of the experiment considered, and then with the CMB, white noise, and the IR background added together. A comparison of the two noise estimates gives us an idea of the excess noise coming from the unresolved IR background.

The results are shown for the temperature channels of Planck-like and ACTPol-like experiments in Figure 8. It appears that, because lensing reconstruction with Planck will effectively use a lower range of multipoles due to its relatively large beam, IR effects would be small and at a percent level or so. However, the caveat here is that we have not considered the actual frequency bands of Planck, and a more detailed modeling of the IR contribution is necessary. For experiments like ACT and ACTpol which have higher resolution and can use higher multipoles in the lensing reconstruction, the IR background becomes a more tangible $\sim 10\%$ contamination. This excess source of noise can in principle be avoided by subtracting a statistical template based on HSLS observations scaled to ACTpol frequencies.

One has to keep in mind that the IR galaxies are in principle correlated with large scale structure and are themselves lensed. These effects will make the separation of the IR contribution a harder problem than envisioned here, and a promising way forward is to cross correlate the CMB maps with the HSLS maps to pull out the clustered IR signal, and normalize our models to such measurements.

## 2.3 Sunyaev-Zel'dovich Effect

### 2.3.1 Improving Sunyaev-Zel'dovich Cluster Detection in Planck

In the framework of hierarchical structure formation, the population of clusters is a powerful way to probe the matter and density content of the Universe, as well as providing strong constraints on the physical processes driving the formation and the evolution of structure [399]. Massive halos in the Universe contain huge amounts of hot ionised gas (i.e. $T \sim 10^{6-8}$K) which contribute about 10–15% of their total mass. This gas interacts by inverse Compton scattering with the CMB, distorting its spectrum in the direction of clusters. This so called Sunyaev-Zel'dovich effect (SZE) is proportional to the pressure integrated along the line of sight, and detected from centimetre to sub-millimetre wavelengths [37]. The detection of new clusters via the SZE is one of the major objectives of the Planck mission, which has already surveyed the whole sky [318]. Simulations using the predicted noise properties and based on the Planck Sky Model for modeling the astrophysical contamination predict that Planck will detect around 1000 clusters with SNR>5 during its 14 month nominal mission duration. This number can be twice as large for the extended mission.

The hundreds of SZE detections expected from the Planck survey will constitute a unique, complete, nearly mass limited, all sky sample with a few tens of clusters at redshifts higher than 0.6. Despite the wide frequency coverage of its high frequency instrument (i.e. HFI, from 100 to 857 GHz), Planck is designed to have a moderate spatial resolution (5'at 857 GHz, 10' at 100 GHz). Combining Planck with relevant external data provides a way to improve and optimise the SZE information through a better handling of the astrophysical background.

The Galactic and extragalactic dust emission is a well-known issue for SZE measurements [2]. Combining far infra-red data with sub-millimetre and millimetre measurements is a powerful way to tackle this issue; this has been specifically demonstrated for the combination of Planck and *Herschel* measurements [319]. Therefore, the combination of SPIRE measurements with Planck will provide a powerful and efficient way to remove dust emission (from Galactic cirrus, extragalactic sources or the CIB) and significantly improve the SZE photometry. This will help to push down the detection limit for clusters in the Planck survey. Such a combination will improve the catalogue of Planck clusters, and may allow us to attain, in the regions common to Planck and *Herschel* , a completeness of better than 95% at SNR $> 5$.

In practice, the two shared frequencies between Planck and SPIRE (350 and 500 $\mu$m) and the high spatial resolution of SPIRE enable a search for the structure of the SZE signal within the potential well of clusters. Indeed, it has also recently been demonstrated that the SZE could be radially traced in clusters from SPIRE measurements [411]. This should provide constraints on the actual thermal pressure supporting the intra-cluster medium by comparing with up-to-date results from X-ray observations of clusters and numerical simulations of structure formation [15]. Moreover, it will open a new window on the study of the population of galaxies and their evolution in dense environments. Indeed, the IR lu-minosity in the direction of clusters are comparable to and even exceed the X-ray emission [172]. Characterising this emission on the basis of the IR emission from cluster galaxies and correlating it with the overall properties of clusters (from the SZE and/or the X-rays) will allow us to address the question of how and when the SF quenching arises in massive halos across time, and consequently how the galaxies' properties evolve with time, or even try to provide further constraints on the energy exchange between galaxies and the intra-cluster medium. Finally, tentative measurements of the cluster radial velocities and bulk flows will be made on a sample of about a hundred clusters in the overlapping regions. All this issues will be addressed for a sample of over a few dozen to a few hundred clusters, spanning from $z = 0$ to $z \sim 1$, found in the Planck-HSLS overlap.

It is important to note that a wide non-targeted survey overlapping with Planck is the proper way to build a fair sample and thus to define the selection function, especially for rare objects. Smaller surveys with *Herschel*-SPIRE have already resolved $\sim 15\,\%$ of the sub-mm background [304], with clustering of both resolved [91] and unresolved fluctuations detected [10]. By surveying an area large enough to be statistically representative of the whole sky, HSLS will refine our knowledge of the effect of bright sources and correlated sub-mm structure on measurements of the Wien part of the CMB. Planck can take advantage of such measurements, in conjunction with its highest frequency channels, to obtain a truly comprehensive view of the effect of the sub-mm background on measurements of the CMB.

Gravitational lensing of faint background galaxies by galaxy clusters causes a large fraction of clusters to have bright sub-mm sources $< 1\,\mathrm{arcmin}$ from their center of mass [412]. Even with many bands, instruments with beam sizes significantly larger than this can suffer from potentially large biases when measuring the spectrum of the inferred SZE result due to such contamination [2]. Fortunately, SPIRE's characteristics allow identification and removal of such contaminating sources far below that achievable with Planck data alone. Using SPIRE's $250\,\mu$m channel as a tracer of the sub-mm background has proven so effective that SPIRE alone has been able to image the Wien part of the SZ effect in a single cluster with signal to noise ratio $> 8$ at $500\,\mu$m [411]. Using SPIRE maps, it should be possible to identify and subtract all strongly lensed sub-mm contaminants in Planck clusters in the region of overlap, and also to determine statistical models for the expected contamination in clusters in the rest of the sky.

Finally, via the kinetic SZ effect, Planck-HSLS will provide the strongest constraints yet on the peculiar velocities of single clusters, and statistical detections of ensemble peculiar velocities in small cluster samples. Additionally, the presence of relativistic electrons in the intra-cluster medium causes changes in the scattering processes which produce the SZE (these are termed the 'relativistic corrections' to the SZE) which are largest at the peak and Wien parts of the SZE increment at $\lambda < 1\,\mathrm{mm}$. Measurements of hundreds of clusters using the combination of Planck's shortest wavelength channels and SPIRE will allow a measurement of these relativistic corrections as a function of cluster temperature for the first time. *Herschel*-Planck multi-frequency SZ spectra will also measure $T_{\mathrm{CMB}}(\mathbf{x}, z)$, providing a fundamental test of our cosmological framework.

### 2.4 Extracting the kinetic SZ and patchy reionization signal

The same hot gas that gives rise to the thermal SZE emission has a coherent motion on large scales. This bulk motion gives rise to a Doppler shift on the CMB photons, known as the *kinetic* SZ effect (kSZ) [37]. Like the CMB, this secondary source of anisotropies has a black body spectrum; unlike the thermal SZE, it cannot easily be separated from the primordial CMB. Also, unlike the SZE, it receives a substantial contribution from gas with temperatures as low as $10^4$K, making it sensitive to less evolved structures. The signal is made up of two parts. The first contribution is from collapsed, virialized, low redshift clusters, typically referred to as kSZ. The second contribution is thought to come from reionization patches, thus known as *patchy reionization*, which should give a non-negligible contribution.

A first detection of the kSZ signal and/or its patchy reionization component will contain valuable information about the missing baryons, the large-scale velocity flows, and the interplay between baryon momentum, dark matter and the physics of reionization.

However, as illustrated in the ACT forecast in Figure 10, the kSZ signal (dotted green line) is dominated by thermal SZ, radio sources and dusty galaxies. As discussed in the SZE section above, using HSLS to isolate and remove dusty galaxies is a very promising tool to clean the power spectra measured in the overlap region. This is shown in the bottom panels of Figure 10, where an HSLS-based dusty-galaxy template has been removed from the data. Our current naive estimates suggest that this exercise will be challenging for ACT and Planck alone, but the joint use of these experiments with HSLS would substantially increase our ability to probe deeper into these secondary anisotropy signals. And while this simple approach shows great promise, in the meantime, more sophisticated and potent multi-frequency methods in the context of Planck and ACT are being developed. Note that although kSZ templates are in principle hard to construct, the method outlined by [199] will allow the creation of reasonable templates using optical surveys. This will be particularly relevant for Stripe-82 observations.

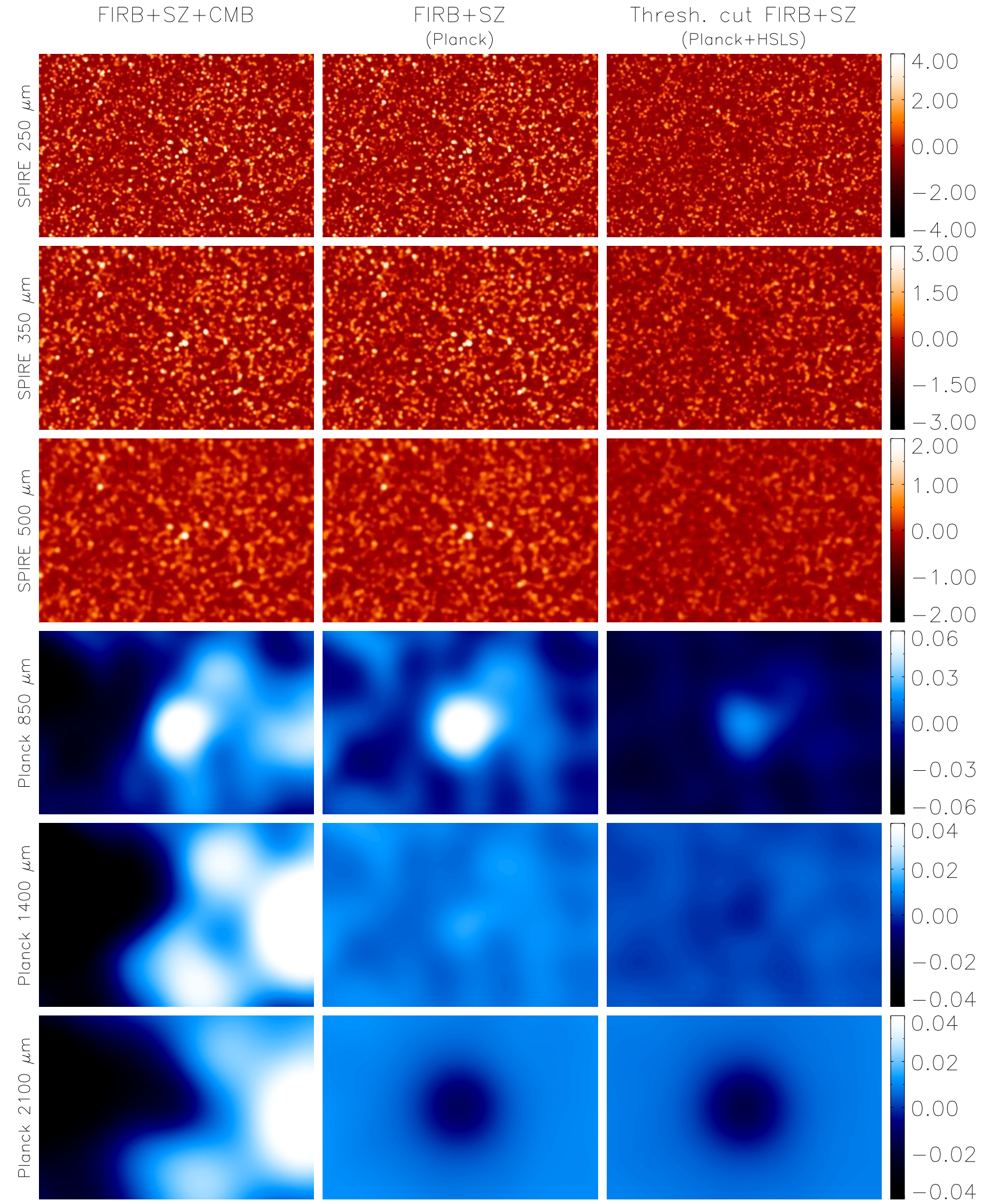

Figure 9: An example of using *Herschel*-SPIRE (red) and Planck-HFI (blue) for spectral component separation for measurements of the Sunyaev-Zel'dovich effect. Color stretches are in MJy/sr and a $\sim 1 \times 10^{15}\,\mathrm{M}_\odot$ cluster is located in the center of the image. The left-most column shows a simulated cluster field as measured by SPIRE at $\{250, 350, 500\}\,\mu$m and HFI at $\{850, 1400, 2100\}\,\mu$m including emission from the CMB, lensed background sub-mm galaxies, and the Sunyaev-Zel'dovich effect. Planck's spectral measurements make it possible to remove contamination from the CMB, leading to the middle column of images. The SPIRE survey proposed here allows measurement and cataloging of the bright sources in and around the cluster; if sources brighter than the survey limit are cut from the simulated maps, the right-most column is obtained. Note in particular the change in the $850\,\mu$m and $1400\,\mu$m channels; as FIRB contaminants are still fairly bright at these frequencies they have a large effect on single-band determinations of the SZ effect amplitude. In particular, since unresolved sub-mm sources contribute a large fraction of the flux at the 850 $\mu$m waveband, cutting the brightest sources with HSLS increases the contrast of the SZ effect peak with the background flux level. This simulation shows that, for measurements where accurate spectral determinations of the SZ effect are key, it is of paramount importance to measure and remove both the lensed images of background sub-mm galaxies and the primary CMB fluctuations which confuse the measurement.

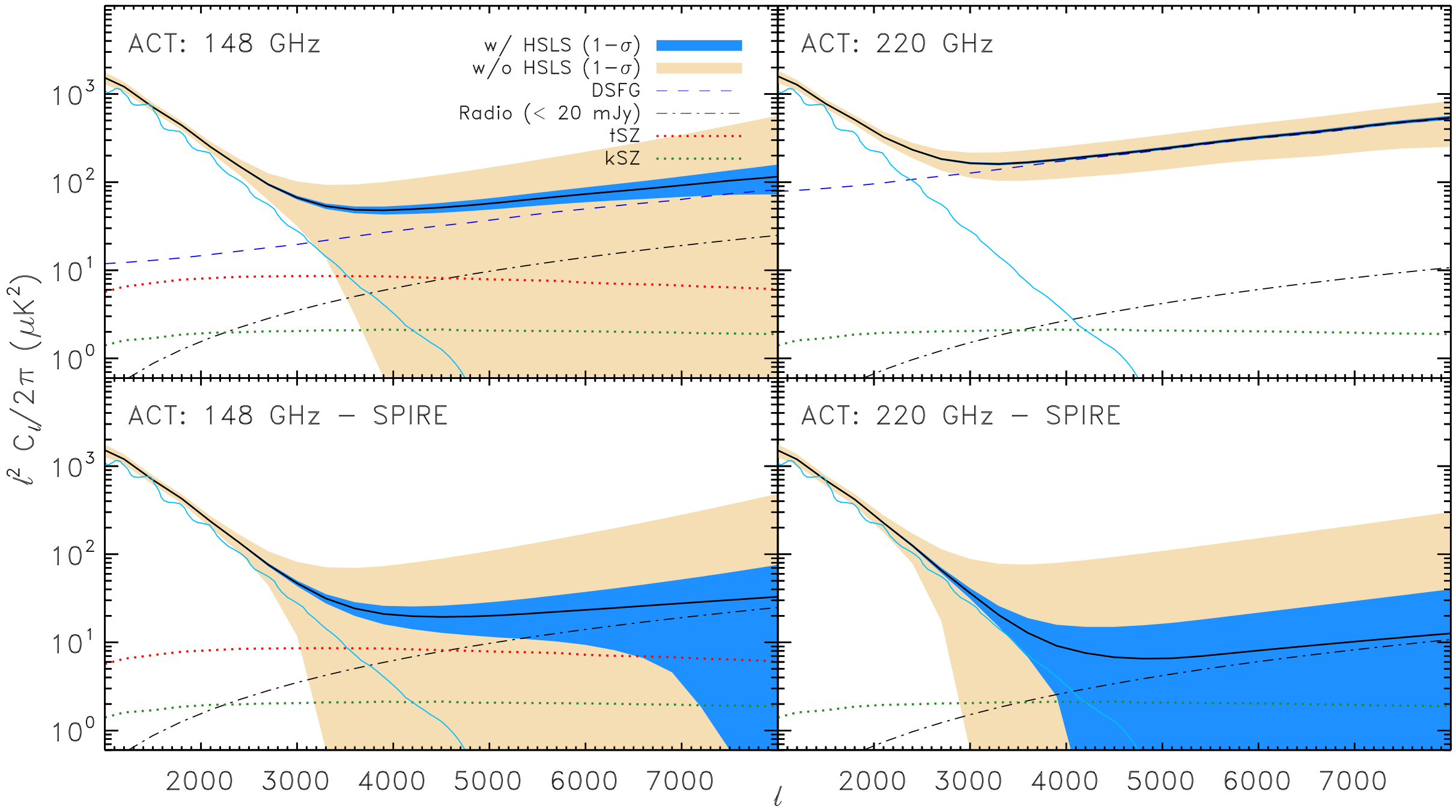

Figure 10: The CMB as measured by ACT, before (top) and after (bottom) removal of power originating from dusty star-forming galaxies, estimated for the $\sim 1400$ deg$^2$ region where the HSLS and ACT surveys overlap. The HSLS is needed to measure the exact shape of the power from dusty star-forming galaxies, both linear and non-linear terms, to high precision. The non-linear term dominates at scales $\ell \sim 3500$, and is similar in shape and amplitude to the thermal SZ (tSZ) component, making its removal critical for estimating the tSZ to high precision. After removal, the dominant residual contaminant comes from shot-noise due to radio sources $< 20$ mJy $(5\sigma)$.

# 3 Cosmological Studies with HSLS

The top level science goals of cosmological studies with HSLS are:

- Galaxy clustering measurements, when combined with Planck, leading to a factor of 2 to 3 improved measurement of the dark energy equation of state and the sum of the neutrino masses when compared to Plank measurement alone; combine Planck and HSLS to identify the neutrino mass hierarchy with cosmological measurements.
- Clustering measurements of unresolved fluctuations with a total signal-to-noise ratio greater than $10^3$ with the possibility for novel studies on lensing of FIRB fluctuations.
- Constraints on the non-Gaussianity parameter $f_{\rm NL}$ at $z \sim 2$ with an accuracy from $\pm 12$ (sub-mm photometric redshifts) to $\pm 7$ (exact spectroscopic redshifts) from observations of the scale-dependent galaxy bias; an independent measurement of $f_{\rm NL}$ with rare, massive clusters at $z > 1$.
- A measurement of the cosmic magnification with significance better than $60\sigma$, leading to complimentary measurements of cosmological parameters.

## 3.1 Source Clustering and BAO Cosmology out to $z \sim 2$

In recent years it has been shown that a powerful probe of dark energy is the measurement of Baryonic Acoustic Oscillations (BAO) [137, 310, 311] features in the galaxy clustering power spectrum. Oscillations in the primordial photon-baryon plasma leave an imprint in the matter distribution. Since the frequency of these oscillations is related to the size of the sound horizon at the epoch of recombination, which is well constrained by CMB measurements, it is possible to use the measurements of acoustic oscillations at different redshifts as a standard ruler. This offers the opportunity to determine the rate of the cosmic expansion and distance measurements as a function of the redshift. It has been found that a number of theoretical systematics including non-linear growth, non-linear bias, and non-linear redshift distortions can be efficiently modeled-out to minimize their contributions to uncertainties in the analysis of BAO measurements [136, 358, 359].

Here we use the Fisher matrix formalism of [360] to forecast the achievable constraints on cosmological parameters from HSLS making use of the large source sample expected over its 4000 $\rm deg^2$ area at $z \sim 2$.

We recall that the observed galaxy power spectrum is given by (see [360]):

$$P(k, \mu, z) = \frac{D^2_{\rm A,r}(z) H(z)}{D^2_{\rm A}(z) H_{\rm r}(z)} b^2 (1 + \beta\mu^2)^2 \left(\frac{G(z)}{G(0)}\right)^2 P^0(z = 0, k)\,, \tag{8}$$

where $D_{\rm A}$ is the angular diameter distance, and the first factor accounts for the fact that the reference cosmology (subscript $r$) could differ from the "true" cosmology. $b$ is the galaxy bias factor, here approximated as $b = \Omega_{\rm m}(z)^{0.6}/\beta(z)$ following linear theory. The term $(1+\beta\mu^2)^2$ describes the linear redshift distortion. The linear growth factor $G(z)/G(0)$ is given by the position-independent ratio $\delta(z)/\delta(0)$ between the linear density contrast at redshift $z$ and at $z = 0$ and $P^0(z = 0, k)$ is the linear matter power spectrum at $z = 0$. The quantity $\mu$ is the cosine of the angle between the unit vector along the line-of-sight $\hat{r}$ and the wave vector $\vec{k}$, $\mu = \vec{k} \cdot \hat{r}/k$.

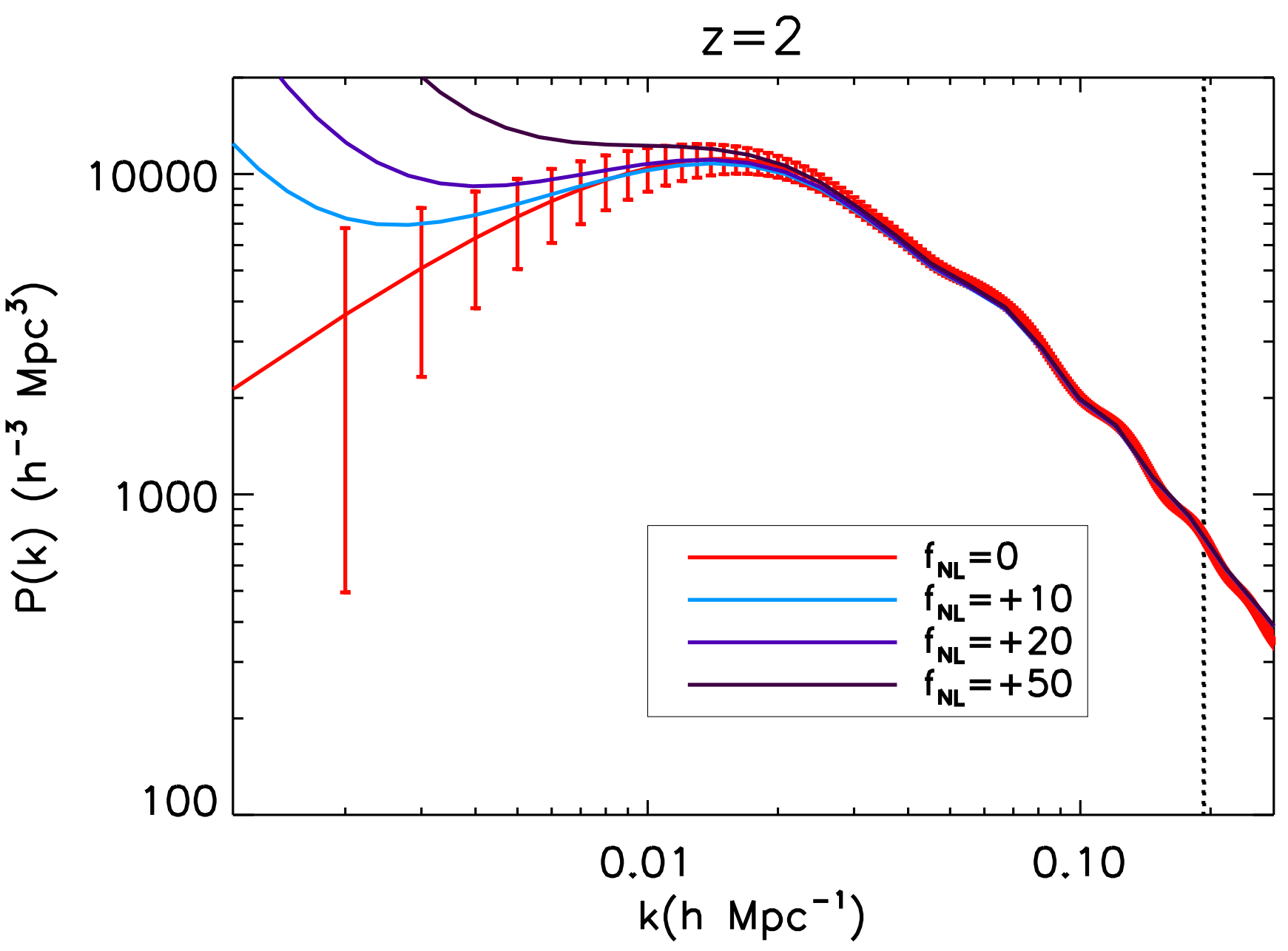

Figure 11: The dark matter halo power spectrum at redshift $z = 2$ with expected error bars from HSLS for the legacy case with $\sigma_z = 0$. We mark the non-linear scale with a vertical line at $z = 2$. For reference, we show several $P(k)$ spectra with different values of the non-Gaussianity parameter, $f_{\rm NL}$.

The Fisher matrix that incorporates the uncertainty on the parameters and their correlations given a theoretical model and an experimental setup can be written as:

$$F_{ij} = \int_{-1}^{1} \int_{k_{\rm min}}^{k_{\rm max}} \frac{\partial \ln P(k,\mu)}{\partial p_i} \frac{\partial \ln P(k,\mu)}{\partial p_j} V_{\rm eff}(k,\mu) \frac{k^2}{8\pi^2} {\rm d}k{\rm d}\mu \tag{9}$$

The integral over $k$ that appears in (9) is performed only up to a $k_{\rm max}$, to exclude non-linear scales. The value of $k_{\rm max}$ is redshift-dependent and we calculate it using the same criterion as in Ref. [360], i.e. requiring $\sigma(R) = 0.5$, where $\sigma$ is the rms fluctuation on the scale $R = \pi/2k$. We choose $k_{\rm min}$ to be $k_{\rm min} = 2\pi/V_{\rm survey}^{1/3} = 0.001h{\rm Mpc}^{-1}$. In describing HSLS we assume a source sample of 2 million sources over 4000 deg$^2$ over the redshift range of 1.8 to 3.; this assumes an optimistic redshift distribution with most of the sources distributed around $z \sim$2. Such an assumption, however, is consistent with existing data of small sub-mm source samples [9, 75, 321]. In Figure 11 we show the expected error bars on the matter power spectrum from the HSLS in the legacy case, where HSLS is followed-up by a spectroscopy campaign leading to $\sigma_z = 0$. With sub-mm data alone, as discussed in Section 5, HSLS redshifts will be uncertain at the level of $\sigma_z = 0.3$. In estimating the cosmological parameter constraints with photometric redshifts only, we account for the degraded radial distance information through a Gaussian suppression of the matter power spectrum for wave vectors parallel to the line of sight [360]:

$$\tilde{P}(\vec{k}) = P(\vec{k}) \exp\left(-k_{\parallel}^2 \sigma_{\rm r}^2\right), \tag{10}$$

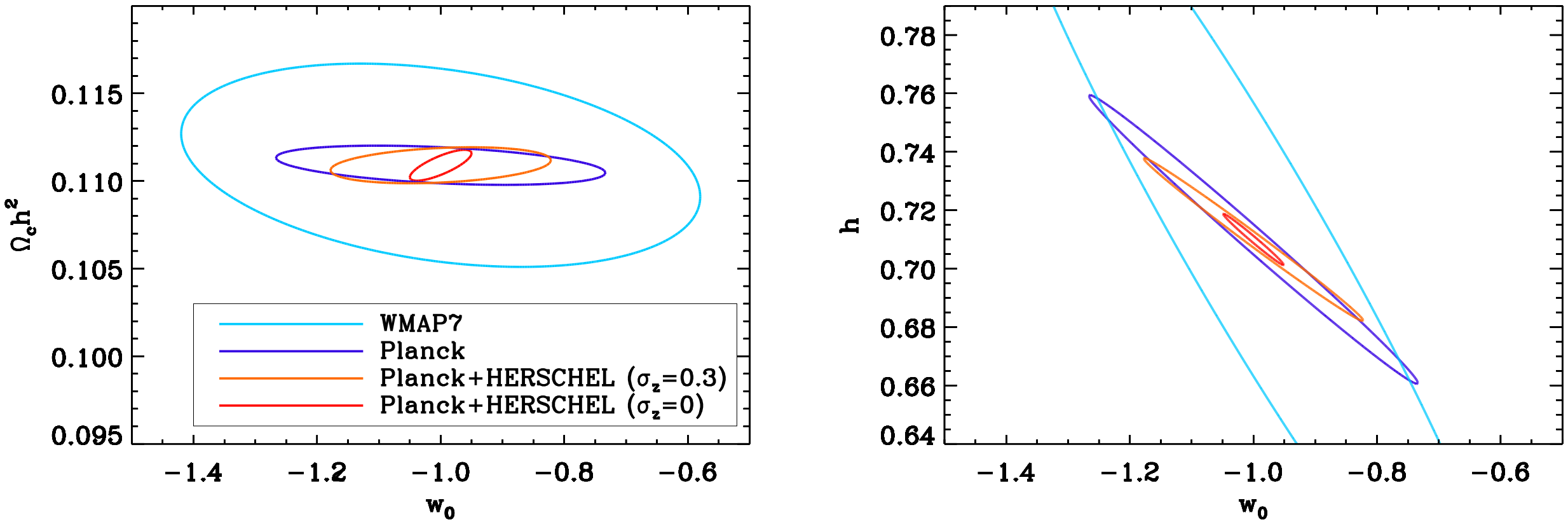

Figure 12: Constraints on the parameters of the $\Lambda$CDM+$w_0$ model from WMAP alone and Planck alone and from Planck combined with HSLS data for two different redshift uncertainties.

with $\sigma_r = c\sigma_z/H(z)$.

We consider the standard $\Lambda$CDM model described by 6 parameters: the baryon density $\Omega_b h^2$, the cold dark matter density $\Omega_c h^2$, the reduced Hubble constant $h$, the scalar spectral index of primordial fluctuations $n_s$ with an amplitude of $\ln[10^{10}A_s]$, and the optical depth $\tau$. We set the target model to the WMAP seven years values [226], $\Omega_b h^2 = 0.02258$, $\Omega_c h^2 = 0.1109$, $h = 0.71$, $n_s = 0.963$, $\tau = 0.088$, $\ln[10^1 0 A_s] = 3.2$. In addition to the standard $\Lambda$CDM parameters we included other, redshift-dependent parameters, such as the angular diameter distance $D_A(z)$, the Hubble parameter $H(z)$, the linear growth factor $G(z)$ and the linear redshift distortion $\beta(z)$. We stress that in this parameterization these parameters are treated as free. We split the survey into three redshift bins equally spaced in the range $1.8 < z < 3$ uniformly distributing galaxies between the bins. With this parameterization each redshift bin has a total of 10 parameters, 6 common to all bins and 4 different for each bin. We note that for photometric redshifts ($\sigma_z = 0.3$) it may be preferable to use a smaller number of redshift bins. Our results, however, are unchanged even using only two broader redshift bins when compared to the case presented here.

| | | $\sigma_{D_A}/D_A(\%)$ | $\sigma_H/H(\%)$ | $\sigma_G/G(\%)$ |
|---|---|---|---|---|
| $\sigma_z = 0.3$ | $z = 2.2$ | 8.6 | $> 100$ | $> 100$ |
| | $z = 2.4$ | 5.5 | $> 100$ | $> 100$ |
| | $z = 2.8$ | 5.8 | $> 100$ | $> 100$ |
| $\sigma_z = 0.1$ | $z = 2.2$ | 6 | $> 100$ | $> 100$ |
| | $z = 2.4$ | 3.6 | 75 | $\sim 100$ |
| | $z = 2.8$ | 3.8 | 65 | 76 |
| $\sigma_z = 0.0$ | $z = 2.2$ | 1.7 | 1.9 | 5.8 |
| | $z = 2.4$ | 1.4 | 1.6 | 4.7 |
| | $z = 2.8$ | 1.6 | 1.8 | 6.5 |

Table 2: Constraints from Planck and HSLS on angular distances, Hubble parameter and growth factor as a function of uncertainty on redshift. Fiducial model is given by $w_0 = -1$, $w_1 = 0$ and $\gamma = 0.55$.

| | $\sigma_{w_0}$ | $\sigma_{w_1}$ | $\sigma_\gamma$ |
|---|---|---|---|
| $\sigma_z = 0.0$ | 0.05 | – | – |
| | 0.08 | 0.14 | – |
| | 0.05 | – | 0.02 |
| | 0.09 | 0.15 | 0.04 |
| Planck only | 0.29 | 1.8 | – |
| | 0.26 | – | – |

Table 3: Constraints from Planck and HSLS on dark energy parameters for the case $\sigma_z = 0$ (the lines correspond to fixing some of the parameters). Fiducial model is given by $w_0 = -1$, $w_1 = 0$ and $\gamma = 0.55$. In the bottom lines we show constraints from Planck alone by comparison. The improvement from the combination with HSLS is of 70% on $\sigma_{w_0}$ and of a factor $\sim 10$ on $w_1$.

| | FoM |
|---|---|
| Planck | 2.4 |
| Planck+HSLS ($\sigma_z = 0$) | 139 |
| Planck+EUCLID | 4381 |

Table 4: Dark Energy Task Force Figure-of-merit (FoM) for $w_0$-$w_1$, for the combination of different probes. The combination Planck+HSLS, with $\sigma_z = 0$, gives an improvement by a factor of $\sim 60$ over Planck alone.

The information on the dark energy equation of state is fully contained in the $D_{\rm A}(z)$ and $H(z)$. We can then project the results onto the parameter space of dark energy testing both a constant dark energy equation of state $w_0$ and a varying equation of state of the form:

$$w(z) = w_0 + w_1 z \tag{11}$$

Moreover, by measuring the growth factor $G$ as a function of redshift one can place constraints on the growth index $\gamma$, which in flat models and on sub-horizon scales is related to $G(z)$ through [236]:

$$\frac{\partial \ln G}{\partial \ln a} = \Omega_{\rm m}(a)^\gamma \tag{12}$$

The growth index is a key quantity for the discrimination of cosmological models and in particular to constrain non-standard models such as modified gravity models, since any model that modifies the Poisson equation will show a different growth index.

Together with the HSLS data we consider the information on the 6 cosmological parameters coming from Planck [318]. We include power spectra for the temperature (TT), temperature-polarization (TE) and E-mode polarization (EE) in the 70, 100, 143, and 217 GHz channels with $\Delta T/T = (3.6, 2.5, 2.2, 4.8) \times 10^{-6}$, $\Delta P/T = (5.1, 4.0, 4.2, 9.8) \times 10^{-6}$ and FWHM of $(114, 9.5, 7.1, 5.0)'$ in the four channels, respectively. We take $f_{\rm sky} = 0.65$ and $\ell_{\rm max} = 1500$.

We forecast constraints first assuming that the redshifts of the sources are known and then showing the degradation in the constraints with redshift errors of $\sigma_z = 0.1$ and $\sigma_z = 0.3$. It is possible that with a modest calibration of the sub-mm color-color diagram with 100 or so exact redshifts, we will be able to derive sub-mm photometric redshifts with an uncertainty of $\sigma_z \simeq 0.3$ (see Section 5.2.2). We note that the combination of HSLS and near-IR surveys such as VISTA-VHS or deep optical surveys such as DES will likely improve the photometric redshift uncertainty to $\sigma_z \simeq 0.1$. Finally, a dedicated sub-mm facility that employs a multi-source instrument for CO-line spectroscopy could obtain exact redshifts for our whole sample.

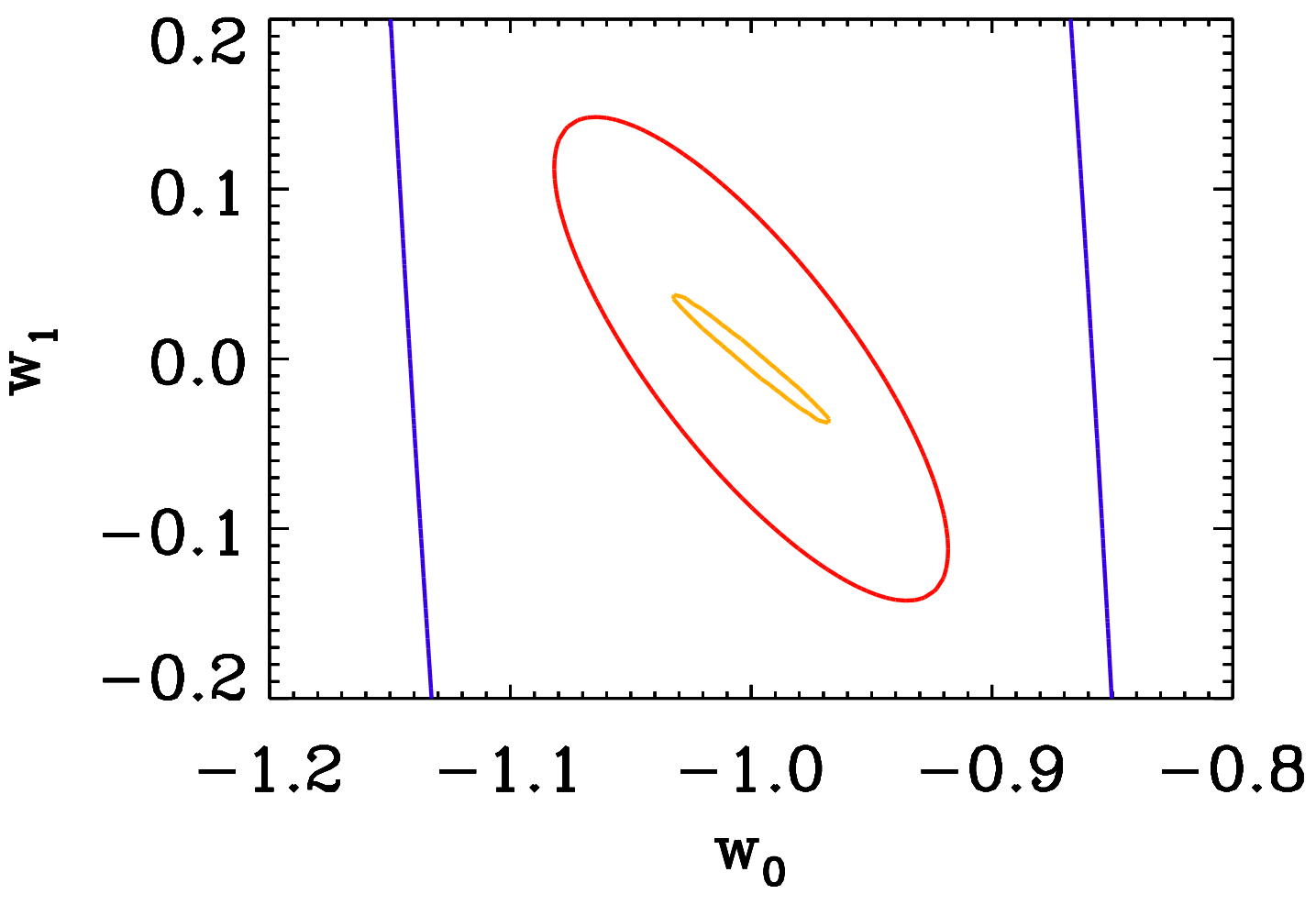

Figure 13: Constraints at $1\sigma$ C.L. on $w_0$ and $w_1$ from the combination of HSLS with Planck for the case $\sigma_z = 0$ and fixing the growth index to $\gamma = 0.55$ (red contour) compared to constraints from Planck alone (blue) and constraints from Planck combined with EUCLID survey (orange).

The results of our forecast are presented in Table 2 through 5. Table 2 shows that when the redshifts of the sources are known, the combination of BAOs measurements from HSLS with Planck could achieve an accuracy of $1\% - 2\%$ on the angular and radial distances and of $5\% - 7\%$ on the growth factor. We also note that non-linearities are less relevant at $z \sim 2$. We show constraints on the $h$-$w_0$ and $\Omega_c h^2$-$w_0$ planes in Figure 12 from the combination of HSLS with Planck, both for the case of a photometric redshifts with $\sigma_z = 0.3$ and for a spectroscopic follow-up of HSLS, compared to the constraints from WMAP or Planck alone, assuming a constant equation of state ($w_1$ fixed to 0). The HSLS improves constraints on $h$ and $w_0$ even when using sub-mm photometric redshifts. For the case $\sigma_z = 0.3$ and $\sigma_z = 0.1$ the constraints on a constant equation of state are improved respectively by 30% and over 50% when compared to Planck alone. A varying equation of state remains however unconstrained for $\sigma_z > 0$.

Figure 13 shows constraints on $w_0$ and $w_1$ from the combination Planck+HSLS for the case $\sigma_z = 0$ and fixing $\gamma$ to its fiducial value $\gamma = 0.55$, compared to constraints achievable from Planck data and from Planck combined with a survey like Euclid [330]. In Table 3 we summarize the constraints achievable on $w_0$, $w_1$ and $\gamma$ for the case $\sigma_z = 0$ marginalizing over different combinations of parameters. As shown the achievable uncertainty in the growth index when redshifts are known precisely is of the order of 4% while it degrades significantly for larger $\sigma_z$. For the case with $\sigma_z = 0.3$ the errors degrade to $\sigma_{w_0} = 0.17$ fixing $w_1$, while in this case, for a varying equation of state, there would be no improvement with respect to Planck alone. Table 4 shows the improvement on $w_0$-$w_1$ in terms of a Figure-of-Merit defined as [5] FoM $= 1/(\sigma_{w_p}\sigma_{w_a})$ where $w_p$ is the minimum uncertainty value of $w$, calculated at the pivot redshift $z_p$. For $\sigma_z = 0$, the improvement is of a factor of $\sim 60$ using the combination

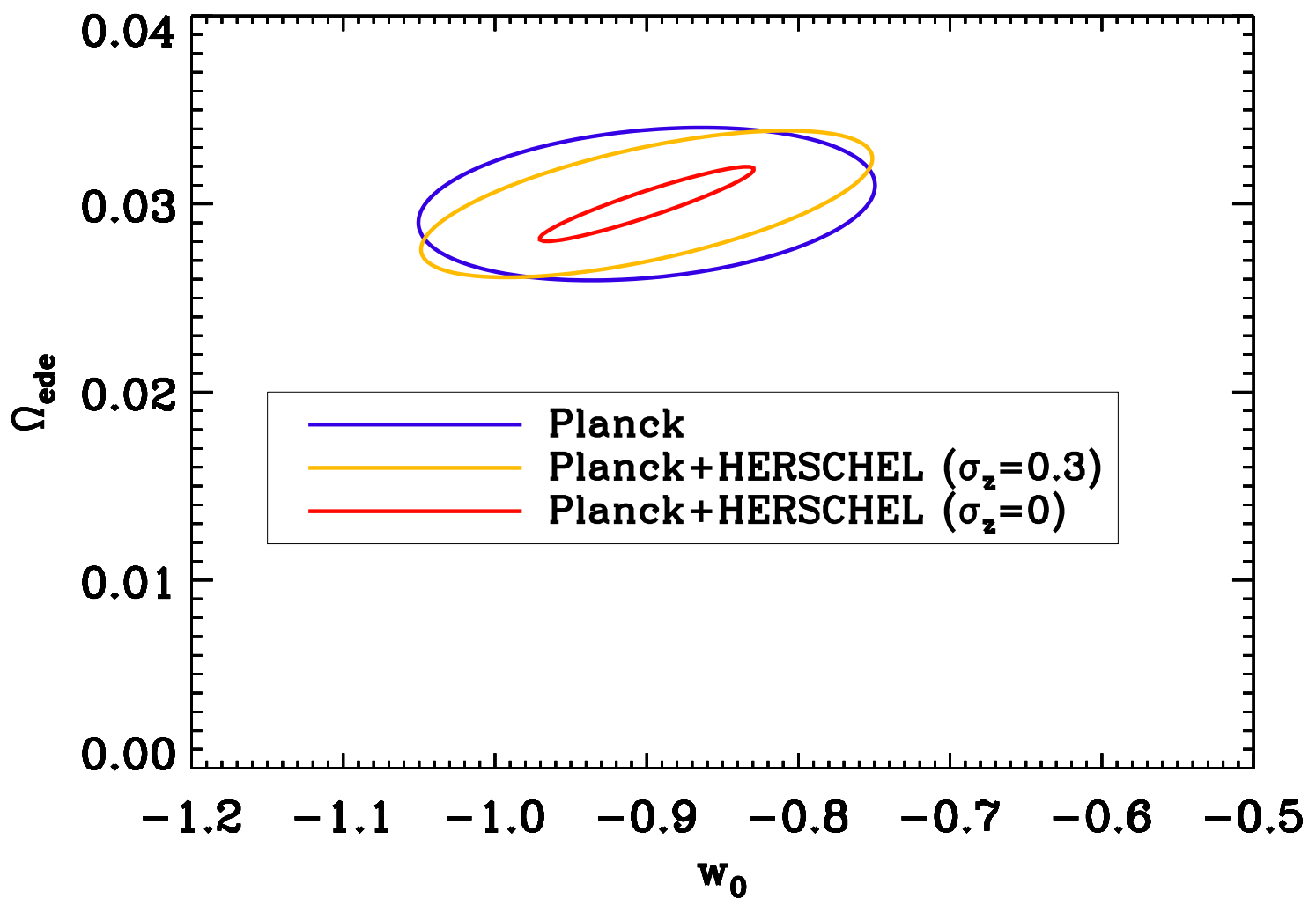

Figure 14: Constraints on $w_0$ and $\Omega_{\rm ede}$ for Planck+HSLS.

Planck+HSLS.

We also forecast constraints achievable on the early dark energy model that allows a significant amount of dark energy at early times. We parameterize the general dark energy density as in [122]:

$$\Omega_{\rm DE}(a) = \frac{\Omega_{\rm DE} - \Omega_{\rm e}(1 - a^{-3w_0})}{\Omega_{\rm DE} + \Omega_{\rm m} a^{3w_0}} + \Omega_{\rm e}(1 - a^{-3w_0}) \tag{13}$$

where $\Omega_{\rm DE}$ is the present dark energy density and $\Omega_{\rm e}$ is the asymptotic value of early dark energy density. We stress that in this parameterization $w_0$ has the usual meaning of present dark energy equation of state, and the only free parameters are $\Omega_{\rm e}$ and $w_0$. Early dark energy changes the Hubble parameter and angular diameter distances and from BAOs measurements one can place constraints on $w_0$ and $\Omega_{\rm e}$. We show results in Figure 14. HSLS data can improve Planck constraints if we use spectroscopic redshifts.

Finally we forecast the constraints achievable from galaxy power spectrum measurements on the sum of the neutrino mass. Neutrino mass is known to have a peculiar effect on the clustering of galaxies through free streaming. While CMB is only slightly sensitive to neutrino mass the combination of CMB information with large scale structure observations can efficiently break many degeneracies and put strong constraints on neutrino mass. We test an 8 parameters model, with the same cosmological parameters stated above plus neutrino mass $\sum m_\nu$ and the number of effective relativistic species $N_{\rm eff}$. We assume $\sum m_\nu = 0.055$ eV and $N_{\rm eff} = 3.04$ as target model for the total mass. Table 5 summarizes expected constraints on the sum of the neutrino masses from Planck and HSLS. Even with photometric redshifts at an accuracy of 0.3, the combination of HSLS with Planck is sensitive to a sum of the neutrino masses of $\sum m_\nu \sim 0.05$ eV at the 68% confidence level.

One of the main issues concerning neutrino masses is the question of distinguishing between two possible mass hierarchies, depending on the sign of the squared mass difference $|\Delta m^2_{31}|$ (see for example [261]). Given the squared mass differences value $|\Delta m^2_{31}|$ and $|\Delta m^2_{21}|$, the total neutrino mass in each hierarchy depends on the mass of the lightest neutrino. This is shown in Figure 15 together with the $1\sigma$ upper limits on the total mass from WMAP, Planck (including lensing extraction) and Planck data combined with HSLS data. The inclusion of HSLS clustering information would put an upper limit of $\sum m_\nu < 0.10$ eV at 68% C.L. and for $\sigma_z = 0.3$. Note that this is close to the upper limit required to rule out the inverted hierarchy (see Figure 15). We note that these constraints are comparable with those achievable from clustering of the DES survey.
The $1\sigma$ error on the number of relativistic neutrinos $N_{\rm eff}$ is also improved, being $\sigma_{N_{\rm eff}} = 0.14$ from Planck alone and reduced to $\sigma_{N_{\rm eff}} = 0.11$ from Planck combined with HSLS data.

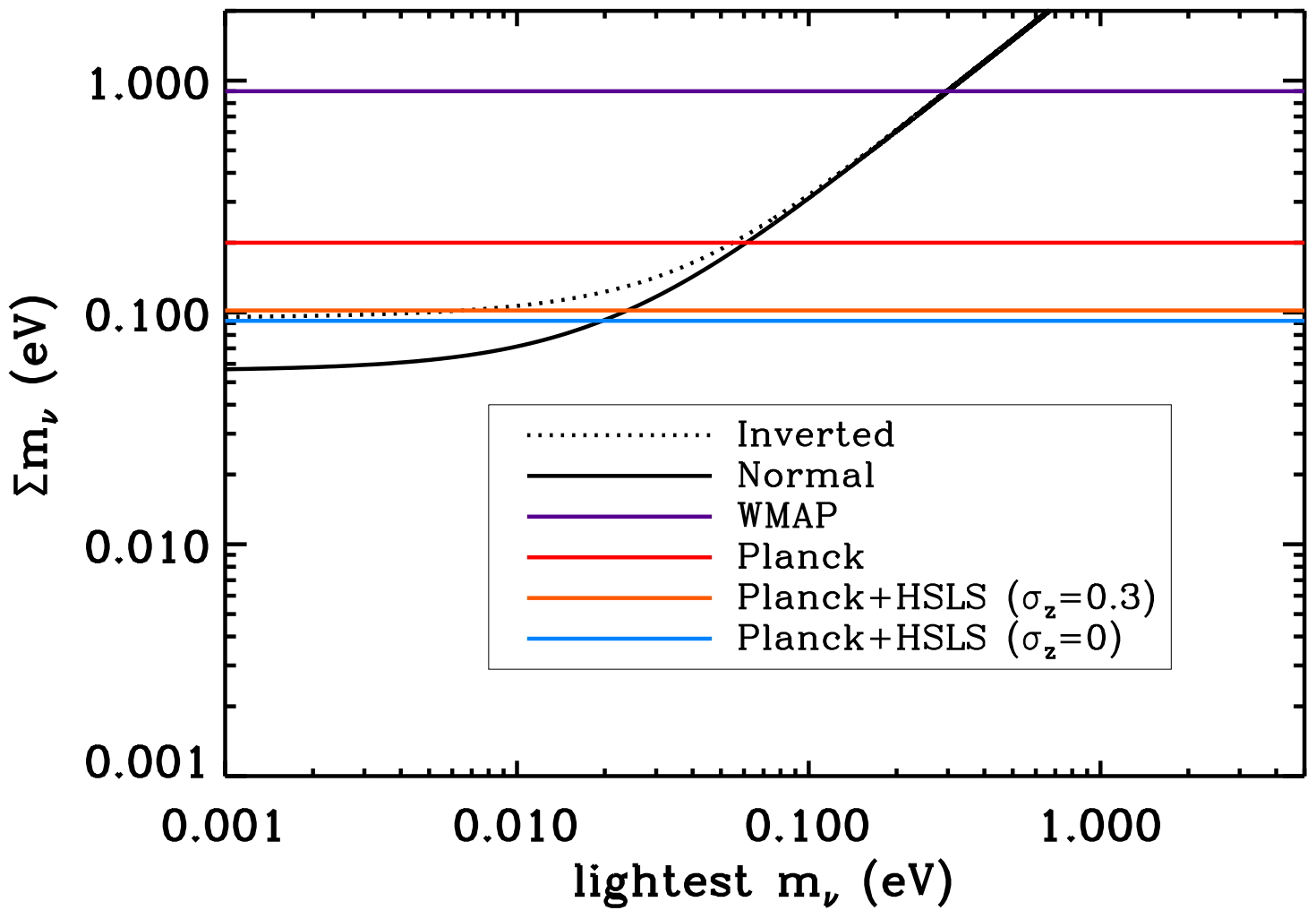

Figure 15: $1\sigma$ upper limits on the neutrino mass from Planck and WMAP in comparison with constraints achievable from Planck combined with HSLS. Black curves show the two possible hierarchies allowed by current flavor oscillation data. The inclusion of HSLS data could lead to a detection of the neutrino mass for a target model $\sum m_\nu = 0.055$ eV.

| Planck | Planck+HSLS |
|---|---|
| 0.13 | 0.036 ($\sigma_z = 0.0$) |
| | 0.040 ($\sigma_z = 0.1$) |
| | 0.047 ($\sigma_z = 0.3$) |

Table 5: $1\sigma$ error in eV on the total neutrino mass $\sum m_\nu$ for Planck, including CMB lensing information (without it would be 0.41 eV), and Planck combined with HSLS for different redshift uncertainties.

### 3.2 Primordial Non-Gaussianity

According to the standard cosmological scenario, quantum-mechanical fluctuations in the scalar field driving inflation lead to primordial density perturbations responsible for the large scale structures we observe today [4, 383]. Although the simplest assumption is that these fluctuations were Gaussian distributed [21], there are several inflationary models [242, 342] involving the existence of a primordial non-Gaussianity. A detection or exclusion of non-Gaussianities would hence be of fundamental interest for the understanding of the physics governing the early Universe. A current standard way to parameterize primordial non-

Gaussianities is to introduce a quadratic correction to the potential [162, 225]:

$$\Phi = \phi + f_{\rm NL}(\phi^2 - \langle\phi^2\rangle)\,, \tag{14}$$

where $\Phi$ is the primordial potential and $\phi$ is a Gaussian random field. Other types of non-Gaussianity are also possible, see Ref. [227] and references therein, such as the so called equilateral and folded types.

A promising method to constraining non-Gaussianity from large-scale structure [101, 270, 371] exploits the fact that the clustering of dark matter halos, where HSLS sources are hosted, on large scales is modified by a non-zero $f_{\rm NL}$. In particular a positive (negative) $f_{\rm NL}$ introduces a scale-dependent boost (suppression) of the halo power spectrum proportional to $f_{\rm NL}1/k^2$ on large scales ($k < 0.03$ h/Mpc) which evolves roughly as $(1+z)$. This effect goes under the name of non-Gaussian halo bias. Below, we forecast constraints attainable on $f_{\rm NL}$ from measurements of the galaxy clustering using HSLS [90]. Note that these constraints can be complemented by using counts of rare, massive clusters (see Section 3.5).

| | Planck | Planck+HSLS ($\sigma_z = 0$); $f_{\rm NL}$ = | | | | |
|---|---|---|---|---|---|---|
| | | 0 | +50 | +25 | +10 | 0 |
| $f_{\rm NL}$ | – | – | 7.7 | 7.2 | 7.4 | 6.25 |
| $\Omega_{\rm b}h^2$ | 0.00013 | 0.00010 (−30%) | 0.00010 | 0.00010 | 0.00010 | 0.00010 |
| $\Omega_{\rm c}h^2$ | 0.0010 | 0.00020 (−80%) | 0.00030 | 0.00030 | 0.00028 | 0.00029 |
| $h$ | 0.0052 | 0.0010 (−81%) | 0.0013 | 0.0013 | 0.0013 | 0.0013 |
| $n_{\rm s}$ | 0.0035 | 0.0033 (−5.7%) | 0.0034 | 0.0034 | 0.0034 | 0.0034 |
| $\tau$ | 0.0029 | 0.0019 (−34%) | 0.0020 | 0.0021 | 0.0020 | 0.0019 |
| $log[10^{10}A_{\rm s}]$ | 0.012 | 0.010 (−17%) | 0.011 | 0.012 | 0.010 | 0.011 |

| | Planck+HSLS ($\sigma_z = 0.3$); $f_{\rm NL}$ = | | | | |
|---|---|---|---|---|---|
| | 0 | +50 | +25 | +10 | 0 |
| $f_{\rm NL}$ | – | 12 | 11 | 12 | 11 |
| $\Omega_{\rm b}h^2$ | 0.00010 (−30%) | 0.00010 | 0.00010 | 0.00010 | 0.00010 |
| $\Omega_{\rm c}h^2$ | 0.00025 (−75%) | 0.00059 | 0.00058 | 0.00058 | 0.00059 |
| $h$ | 0.0014 (−73%) | 0.0029 | 0.0030 | 0.0029 | 0.0030 |
| $n_{\rm s}$ | 0.0034 (−3%) | 0.0034 | 0.0034 | 0.0034 | 0.0034 |
| $\tau$ | 0.0020 (−31%) | 0.0022 | 0.0022 | 0.0022 | 0.0021 |
| $log[10^{10}A_{\rm s}]$ | 0.011 (−8.3%) | 0.011 | 0.012 | 0.011 | 0.011 |

Table 6: $1\sigma$ Fisher matrix errors on $f_{\rm NL}$ and cosmological parameters from the combination of HSLS and Planck measurements both for $\sigma_z = 0.3$ and $\sigma_z = 0$. The first column with $f_{\rm NL} = 0$ case assumes that the $f_{\rm NL}$ parameter is fixed and shows in parenthesis the reduction on the $1\sigma$ error with respect to Planck alone. As one can see there are strong improvements on $\Omega_{\rm c}h^2$ and $h$, and good improvements on $\Omega_{\rm b}h^2$ and $\tau$ regardless of the uncertainty on redshifts.

Following the same procedures presented in [107] we forecast the constraints achievable on $f_{\rm NL}$ again implementing a Fisher matrix calculation. Here we calculate the scale dependent correction to the galaxy bias assuming that the observed sources are old merged objects. For

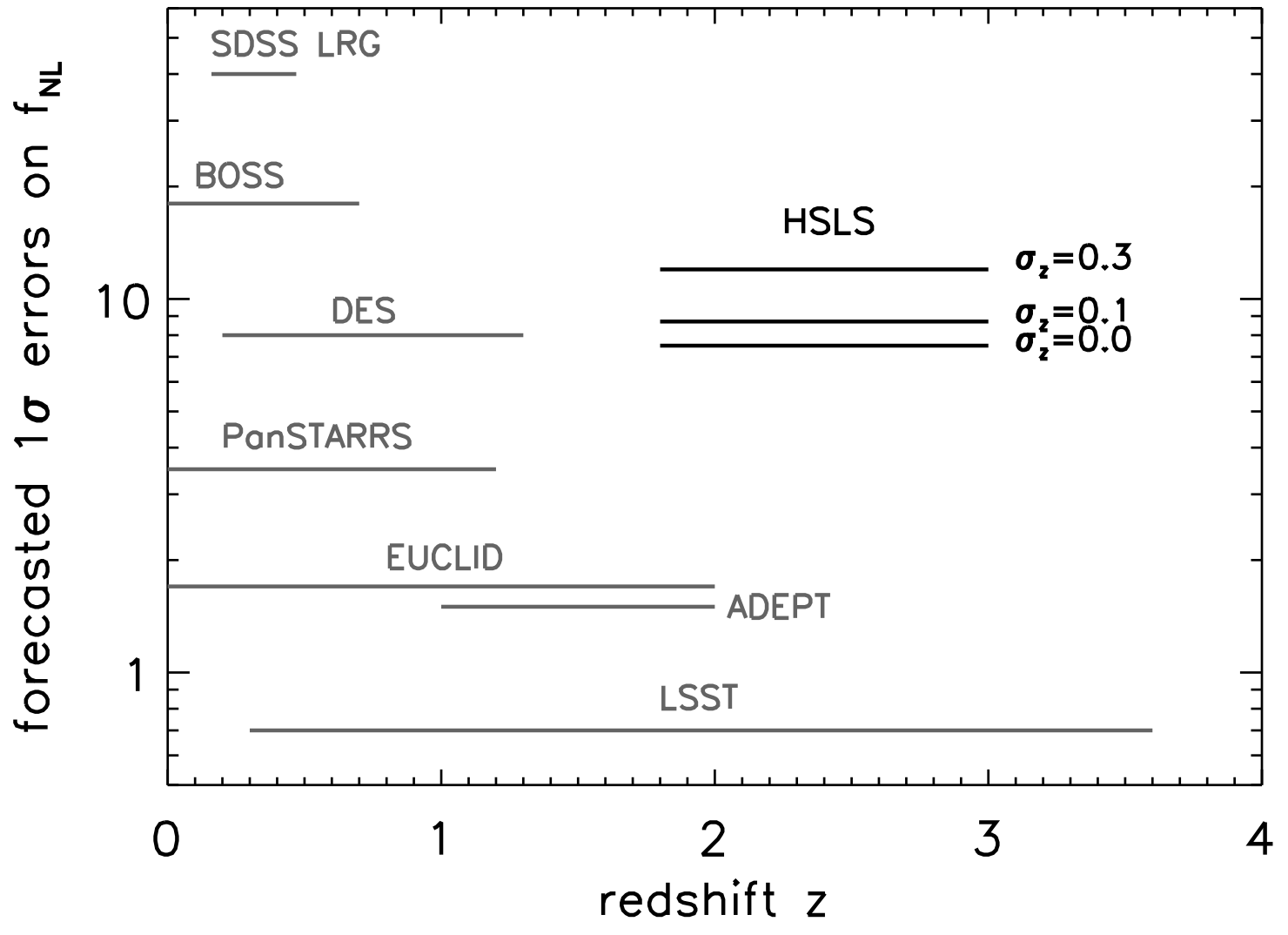

Figure 16: Summary of constraints achievable with future probes on $f_{\rm NL}$ [68] using the non-Gaussian halo bias. HSLS constraints are comparable to DES for $\sigma_z = 0.3$ and could slightly improve DES constraints for lower $\sigma_z$.

a galaxy survey the Fisher matrix can be written as [387]:

$$F_{ij} = \frac{1}{4\pi^2}\int_{k_{\rm min}}^{k_{\rm max}} \frac{\partial \ln P_{\rm gal}}{\partial\theta_i}\frac{\partial \ln P_{\rm gal}}{\partial\theta_j} V_{\rm eff}(k) k^2 {\rm d}k \tag{15}$$

where $V_{\rm eff}$ is the effective volume of the survey, given by:

$$V_{\rm eff}(k,\mu) = \int \left[\frac{n_g(\vec{r})P_{\rm gal}(k)}{n_g(\vec{r})P_{\rm gal}(k)+1}\right]^2 {\rm d}\vec{r} = \left[\frac{n_g P_{\rm gal}(k)}{n_g P_{\rm gal}(k)+1}\right]^2 V_{\rm survey} \tag{16}$$

and in the last equality the comoving number density of galaxies $n_{\rm g}$ is assumed to be constant. The minimum $k$ of the integral in the equation (15) is important, since the $f_{\rm NL}$ correction is more sensitive to small $k$ (see Figure 11 for the effect of $f_{\rm NL}$ on the matter power spectrum).

Again, we use a simple $\Lambda$CDM+$f_{\rm NL}$ model with 6 parameters and we set the target model to the WMAP seven years values [226] but we now include $f_{\rm NL}$ as an extra parameter considering four target values of $f_{\rm NL}$, namely $f_{\rm NL} = 0, +10, +25, +50$. Planck data are used only to constrain the six standard $\Lambda$CDM parameters since the constraints achieved by this experiment will be largely independent of the value of $f_{\rm NL}$. Planck will provide an independent constraint on $f_{\rm NL}$ from higher order statistics of CMB maps, but we do not include this information in the following analysis. Planck data are therefore just used to break the degeneracies between parameters as, for example, the degeneracy between $f_{\rm NL}$ and $h$ and between $f_{\rm NL}$ and $\Omega_{\rm c}h^2$.

Results are shown in Table 6. The combination of galaxy clustering measurements between $1.8 < z < 3$ from *Herschel* with Planck CMB measurements could detect primordial non-Gaussianities of the order of $f_{\rm NL} \simeq 7$ at 68% C.L. when the redshifts of the sources are known, while the detection threshold rises to $f_{\rm NL} \simeq 12$ for $\sigma_z = 0.3$. While the measure of NG has become a key goal of Planck and other CMB experiments, our survey will allow a measurement of the $f_{\rm NL}$ parameter competitive with Planck but using a different technique, probing different scales and redshifts. As such, if Planck detects a signal, then HSLS will offer a strong independent confirmation. If Planck does not see any NG signal, then HSLS will explore a complementary domain.

As we can see from the Table 6, the error on $f_{\rm NL}$ does not have a significant dependence on the target values assumed for $f_{\rm NL}$. The table also shows that HSLS can significantly improve constraints on some fundamental cosmological parameters such as $\Omega_{\rm c}h^2$, $h$, $\Omega_{\rm b}h^2$ and $\tau$ compared to constraints from Planck data alone and even with photometric redshifts ($\sigma_z = 0.3$). Figure 16 shows a summary of constraints achievable with future survey. Photometric HSLS with $\sigma_z = 0.3$ will be comparable to DES thanks to the large redshift range and volume. Reducing the uncertainty in redshift could even slightly improve constraints over the DES survey.

Moreover, if Planck were to detect non-Gaussianity close to the local type with $f_{NL}$ at the level of $\sim 10$, but this does not show up in the non-Gaussian halo bias signal, including HSLS data, then this would indicate that the CMB bispectrum is given by secondary effects in the CMB [186, 263, 363]. If CMB detects non-Gaussianity but is not of the local type, then HSLS clustering can help discriminate between equilateral and enfolded shapes since in the case of non-Gaussian halo bias equilateral type non-Gaussianity should be highly suppressed but not the enfolded type [227]. Thus even a non-detection of the non-Gaussianity in HSLS, in combination with Planck, will have an important discriminative power to distinguish between a variety of models beyond just the value of $f_{\rm NL}$.

### 3.3 Cosmic magnification of sub-mm sources

Large-scale structure at low redshifts magnifies sources at higher redshifts as a result of gravitational light deflection. On the one hand, fewer sources will be observed because lensing stretches the solid angle and dilutes the surface density of sources. Conversely, the effective flux limit is lowered as a result of magnification which leads to a deeper survey. Whether there is an increase or decrease in the observed number density of sources depends on the power-law slope of the background source number counts - an effect known as the magnification bias. At sub-mm wavelengths the magnification bias is expected to be large and positive, resulting in an increase in the number of sources compared to the case without lensing [41, 43, 298]. This has been successfully utilized in the past to search for faint sources lensed by clusters [52, 125, 209, 210].

Cosmic magnification also induces an apparent angular cross-correlation between two source populations with disjoint spatial distributions. It can thus be measured by cross-correlating non-overlapping foreground and background samples. When combined with number counts, such a cross-correlation study can provide constraints on cosmological parameters (e.g. $\Omega_{\rm m}$, $\sigma_8$) and galaxy bias, a key ingredient in galaxy formation and evolution models [250]. As the lensing-induced cross-correlation also probes the dark matter distribution, it provides an independent cross-check of the cosmic shear measurements, which depend on the fundamental assumption that galaxy ellipticities are intrinsically uncorrelated. Most of the previous investigations using foreground galaxies selected in the optical or infrared and background quasars have produced controversial or inconclusive results [23, 24, 357]. The best detection to date is presented in Scranton et al. [355] where cosmic magnification is detected at an $8\sigma$ significance level using 13 million galaxies and $\sim 200{,}000$ quasars from the

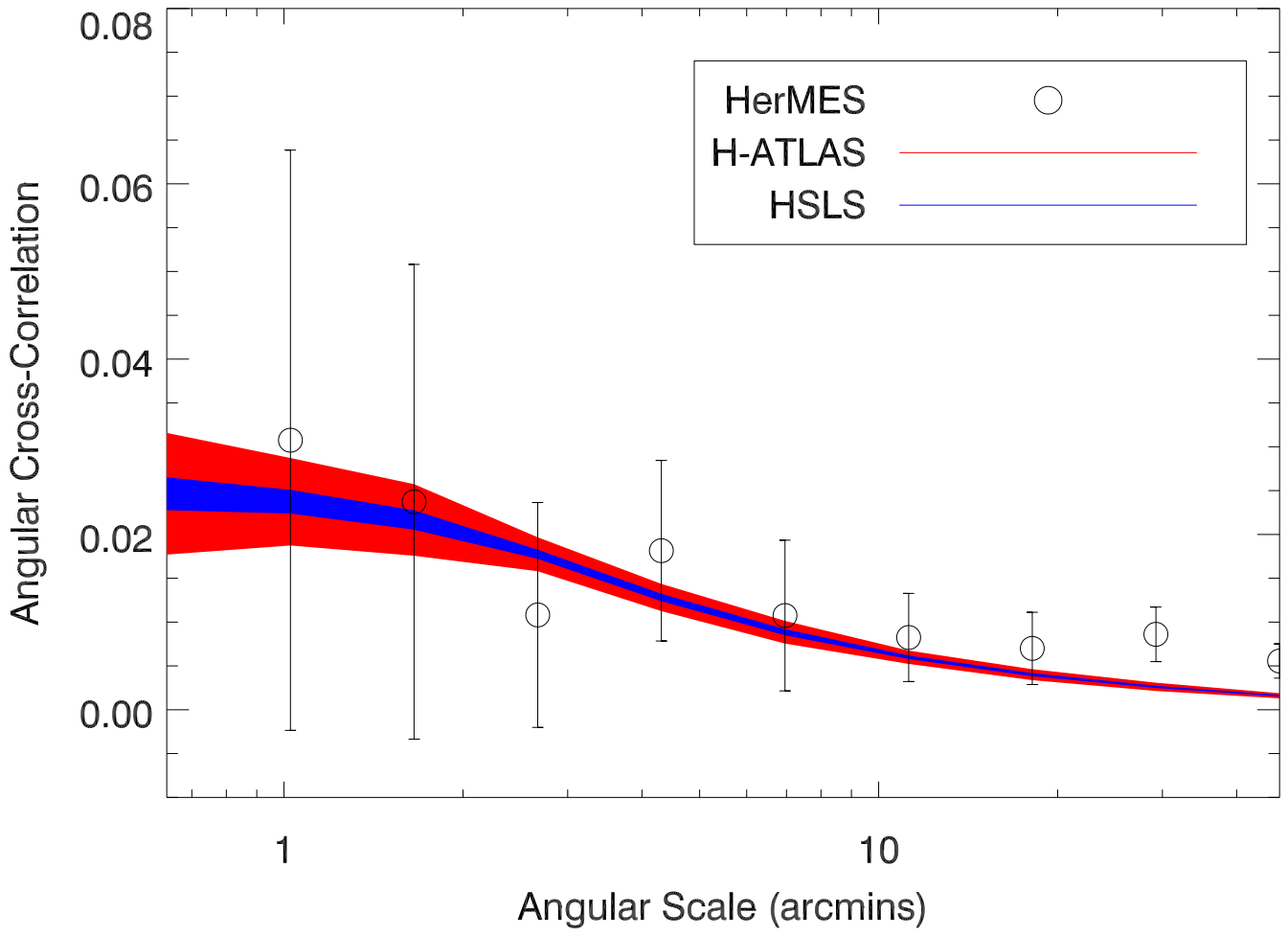

Figure 17: The angular cross-correlations between foreground SDSS/DES galaxies peaking at $z \sim 0.25$ and background sub-mm sources peaking at $z \sim 2$ as a function of the correlation angle. The data points with error show the measurements with existing HerMES data (a $\sim 3.5\,\sigma$ measurement), while the error bands in red and blue show the expected uncertainty for this measurement with H-ATLAS and HSLS, respectively.

Sloan Digital Sky Survey (SDSS).

The amplitude of the lensing-induced cross-correlation is determined by several factors: the dark matter power spectrum and growth function, the power-law slope of the background source number counts and the bias of the foreground sources. At sub-mm wavelengths, the power-law slope of the cumulative number count is exceptionally steep, >2.5 for sources in the flux range 0.02−0.5 Jy at 250, 350 and 500 $\mu m$ [86, 304, 306]. In Scranton et al. [355], the number count slope of the quasar sample is considerably flatter ($\sim 2$ for the brightest ones). In addition, sub-mm sources detected in deep surveys mainly reside in the high-redshift Universe with a median redshift of $z \sim 2$ [9, 13, 77]. The steep number counts together with the large redshift range make sub-mm sources an ideal background sample. So far there have been a few attempts at measuring the lensing-induced cross-correlation between foreground optical galaxies and background sub-mm sources but with conflicting results. Almaini et al. [7] measured the cross-correlation between 39 sub-mm sources detected by SCUBA and optical sources at lower redshifts $\langle z \rangle \sim 0.5$. They found evidence for a significant cross-correlation which might be caused by lensing. Conversely, Blake et al. [45] did not find evidence for cross-correlation due to cosmic magnification using a similar number of sub-mm sources. Using sub-mm sources detected in the HerMES Lockman-SWIRE field during the Science Demonstration Phase as a background population and optical / near-infrared selected sources at low redshifts as a foreground population, Wang et al. (2010) found decisive evidence that the cross-correlation signal measured on angular scales between $\sim 1'$ and $\sim 50'$ (shown in Figure 17) is primarily due to cosmic magnification.

### 3.4 Mapping Cosmic Infrared Background Fluctuations with HSLS

HSLS is capable of detecting Far-Infrared Background (FIRB) fluctuations tracing large-scale structure and galaxy clustering from 100 degrees to $\sim 30$ arcseconds, with an order of magnitude higher signal-to-noise ratio than the measurement expected with Planck HFI down to 5 arcminute angular scales (Figure 18). Background fluctuations probe the physics of galaxy clustering over an ensemble of sources, with the bulk of the signal contribution originating from sources below the *Herschel* point source confusion limit. On large angular scales, background fluctuations measure the linear clustering bias of far-infrared galaxies in dark matter halos, thus probing the dark matter halo mass scale. On small angular scales, fluctuations measure the non-linear clustering within a dark matter halo, and the physics governing how far-infrared galaxies form within a halo. For a fixed survey time over 500 hours, the survey area of HSLS spanning 3000 to 4000 deg$^2$ is optimized for a measurement of clustering of fluctuations, unlike existing surveys that span either 15 deg$^2$ or about 200 deg$^2$ in a contiguous area. *Herschel* measurements will improve on Planck in statistical sensitivity to the overall clustering signal by a factor of 50. Furthermore, *Herschel*, by virtue of its greater angular resolution, will probe the non-linear intra-halo clustering signal (the 1-halo term) which is detectable on angular scales smaller than 5. This is not possible with Planck. The shallow survey map will serve as a rich database with several cosmological applications related to weak lensing. The maps may be analyzed for weak lensing modification to anisotropies, using statistics similar to those developed to study lensing in the CMB, though such a measurement will be challenging as FIRB fluctuations are likely to be non-Gaussian compared to CMB. It is likely to be simple to cross-correlate FIRB maps with foreground large-scale structure as traced by galaxies in DES and VISTA-VHS to extract the cosmic magnification signal and from this measure the integrated dark matter power spectrum out to redshifts of a few on scales from a few to several hundred Mpc.

### 3.4.1 Rationale for a Statistical Approach

The amplitude of the cosmic Far-Infrared Background, measured by COBE [153, 326] gives the redshift-integrated luminosity of galaxy formation throughout cosmic history. The next step is to understand the population and redshift of galaxies producing the FIRB. The standard approach would be to first detect the constituent objects photometrically, and then determine their redshifts with a dedicated and laborious follow-up campaign. Unfortunately, with the limited apertures available in the far-infrared, current and planned surveys can only resolve a fraction of the FIRB into discrete sources. Furthermore, systematic redshift follow-up of even the small (compared to *Herschel*) sample of SCUBA galaxies has shown to be extremely taxing observationally. With *Herschel*, the resolved fraction of the FIRB decreases dramatically from the shorter to longer wavelengths; and unfortunately the longer wavelengths are more sensitive to sources at higher redshift due to the strong K-correction in the far-infrared. For example, SPIRE has only been able to resolve $\sim$15% of the FIRB at 250 $\mu$m in the deep HerMES fields[304]. While surveys resolving discrete sources are indisputably important, they leave unconstrained the majority of objects below the *Herschel* point source detection threshold due to severe confusion noise.

### 3.4.2 Background Fluctuations from Galaxy Clustering

Fortunately, the statistical properties of spatial fluctuations in the FIRB provide a tool to study sources below the point source detection threshold, typically through one-point (Poisson) or two-point (clustering) statistics. Probability of deflection, $P(D)$, perhaps is the more widely known of fluctuation statistics [345]. It is studied by making a histogram of the brightness distribution in a map, and using this information to constrain the number counts of sources below the detection limit. In contrast, the clustering of undetected sources produces fluctuations on larger spatial scales[183, 221]. A first detection of the clustering fluctuations has now been made in the 15 deg$^2$ HerMES fields[10] and found to be brighter than the Poisson fluctuations on spatial scales greater than 10 arcminutes. They are best addressed with a shallow survey strategy.

While $P(D)$ allows one to establish information on number counts below the confusion limit, the detailed measurement of the angular power spectrum of FIRB anisotropies, akin to measurements of the CMB angular power spectrum, are related to the physical distribution of FIRB sources, and thus probe the underlying physics of source clustering. For example, a clustering measurement at large scales probes the average bias parameter and can establish the characteristic dark halo mass scale associated with far-IR sources. Beyond such an average mass, it is also useful to establish detailed statistics on how FIRB sources occupy dark matter halos, which effects fluctuations on small-scales.

The relevant description of galaxy clustering is naturally present in the halo model of large scale structure[90]. In these models, the linear part of clustering (at large angular scales) is determined by the primordial fluctuations while the non-linear part is determined solely by the distribution of dark matter halos, and more importantly, how individual sources (FIRB galaxies in our case) form within these dark matter halos. Thus, if the linear part of clustering can be established, one probes the source bias, namely the relation between how sources are clustered relative to the primordial density field fluctuations, and the shape of the primordial power spectrum. While the former establishes the mass scale, the latter provides information related to cosmological parameters. The non-linear part of source clustering establishes information related to the halo occupation number, mainly the mean and the second moment of the distribution function that describes the number of sources in a given dark matter halo mass.

The halo approach is now routinely used to extract parameters from galaxy clustering at low redshifts and to understand how galaxies are distributed relative to the dark matter field, because relations such as halo occupation are readily available from numerical and semi-analytical models of galaxy formation. Thus, the halo approach allows us to compare the results of HSLS with that of SDSS galaxies, and those at higher redshifts, such as the Lyman-break galaxies, to obtain a consistent picture of galaxy formation and evolution.

Beyond the angular power spectrum, HSLS has adequate area to make measurements of high-order correlations, such as the three-point correlation function or the bispectrum (Figure 19). These higher-order correlations measure the non-linear growth of density perturbations, but could also in principle be used to study non-Gaussianities. Higher-order correlations could also arise when FIRB fluctuations are gravitationally magnified by clustered distribution of the intervening dark matter in the foreground, as we discuss later in Section 3.4.4. In Figure 19 we show the expected errors from HSLS compared to HerMES and H-ATLAS. HSLS provides a factor of 60 to 100 higher signal-to-noise ratio than H-ATLAS and HerMES, respectively, showing the significant potential for detailed statistical studies, such as the configuration-space shapes of the bispectrum, with HSLS. While there is the potential for such higher-order measurements, we note that additional theoretical studies are warranted to fully understand how these higher-order clustering measurements can be used for cosmological parameter estimates.

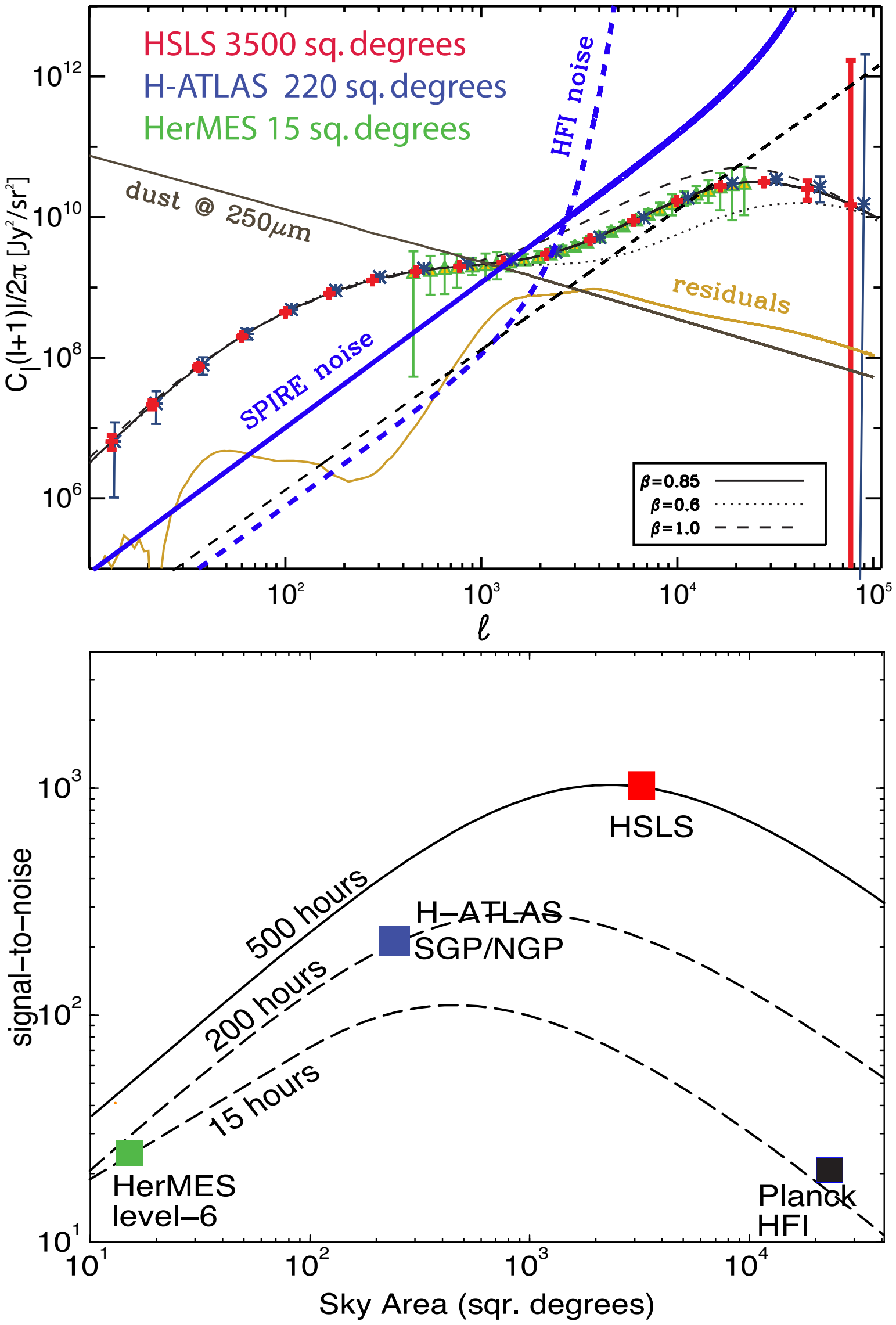

Figure 18: *top:* $C_\ell$'s of FIRB fluctuations. Error bars show three sets of measurements expected from HerMES (15 deg$^2$ Level-6 fields; consistent with measurements reported in Amblard et al. [10]), H-ATLAS (220 sq. degrees), and HSLS (3500 sq. degrees). For reference, we show the Planck-HFI noise that limits measurements to $\ell < 1500$, so that Planck is unable to make a definitive measurement of the 1-halo term of FIRB $C_\ell$'s. The dashed line scaling as $\ell^2$ is the shot-noise. We also show the Galactic dust scaling as $\ell^{-1}$ on the $\ell^2 C_\ell$ plot and the residual foregrounds after multi-frequency cleaning [8]. *bottom:* Cumulative signal-to-noise ratio for a measurement of FIRB $C_\ell$s with *Herschel* and Planck data. The curves assume surveys of fixed integration time showing the maximal signal-to-noise and the area that ought to be covered. For a shallow survey that spends 500 hours, we find the maximum signal-to-noise ratio when the survey spans about $\sim$ 2500 deg$^2$. HSLS is close to such an optimal survey, while both level-6 fields of HerMES, giving the widest coverage, and 220 deg$^2$ of H-ATLAS are sub-optimal. For reference, we also show the expected signal-to-noise ratio of a measurement with Planck at frequencies above 350 GHz. The signal-to-noise ratio with Planck is limited due to the large beam cutting off the measurements at $\ell < 1500$. The beam comparison alone shows that measurements with SPIRE out to 30$''$ should have a factor of $\sim$ 50 larger signal-to-noise ratio in an area of 4000 sq. degrees compared to same for Planck.

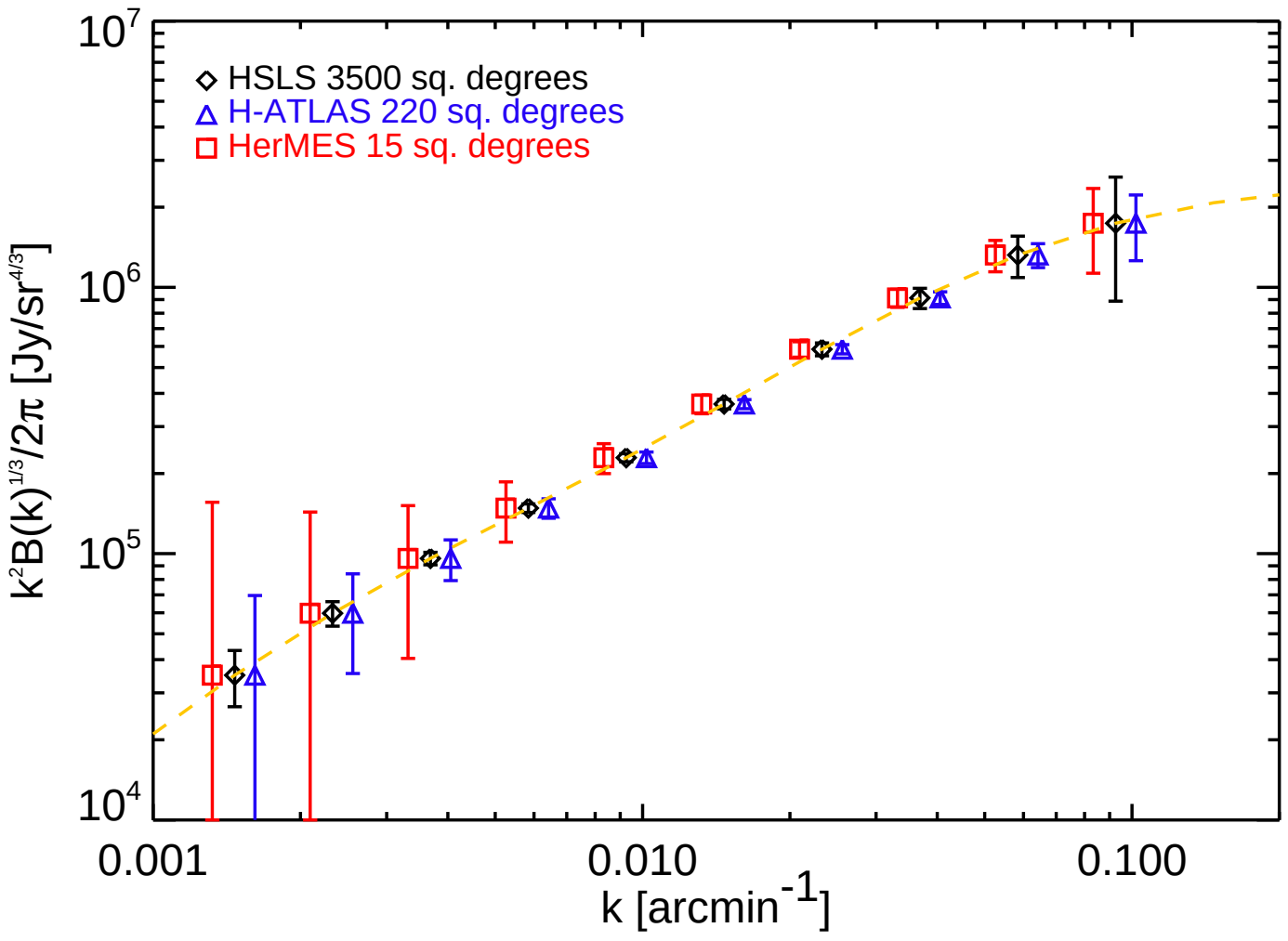

Figure 19: The angular bispectrum of FIRB fluctuations in the equilateral triangle case with $k = k_1 = k_2 = k_3$, where $k$ is the moduli of the 2d wave vector in arcmin$^{-1}$. The line shows an expectation based on the same halo model as Figure 18. The error bars show the measurement uncertainties using HerMES, H-ATLAS, and HSLS.

#### 3.4.3 Detectability of Clustering Fluctuations with Herschel

A comparison of *Herschel*-SPIRE and Planck for probing correlated fluctuations is also shown in Figure 18. Planck, covering the entire sky, has high sensitivity to extended emission but poor spatial information due to its large $5'$ beam, resulting in a loss in sensitivity at higher multipoles. At multipole $\ell > 1000$, HSLS has higher sensitivity and should make a more precise measurement of the far-infrared background correlations overall. At high multipole number, the non-linear 1-halo clustering dominates. This is the term that is heavily dependent on the spatial distribution of FIRB sources within dark matter halos at relevant redshifts and mass scales. With higher angular resolution, *Herschel* is uniquely able to explore the non-linear regime. The Poisson fluctuations from undetected galaxies become the dominant source of confusion at high $\ell$.

The proposed HSLS survey offers higher spatial resolution and point source sensitivity than Planck. Consequently, the shot noise level in these anisotropy studies is reduced by the removal of resolved sources. As a result, HSLS data can better measure and cleanly separate the shot noise signal from the correlation signal. Operationally, by detecting and removing point sources, the HSLS will measure the correlation signal as a function of source cutoff. The clustering can first be measured using detected sources as a function of brightness (or redshift, if known) before moving on to tackle fluctuations below the confusion limit. While Planck can potentially provide information in the linear regime, due to its poor resolution, no information related to non-linear clustering can be obtained. Using *Herschel*-SPIRE, we can completely describe the clustering spectrum, from the linear to the non-linear regime. While Planck allows a bias measurement, due to lack of information on small angular scales, no information related to how FIRB sources occupy dark halos can be obtained.

### 3.4.4 Weak Lensing Applications of Fluctuations

Similar to studies of extract lensing from the CMB, it may be feasible to implement a weak lensing reconstruction on FIRB fluctuations, though the feasibility of such a study has yet to be demonstrated. Such a measurement could make use of the techniques that have now been developed to extract lensing in CMB anisotropy maps [92, 303]. Unlike the case of CMB, where primordial fluctuations are Gaussian and lensing is measured with generated non-Gaussianities, FIRB fluctuations are intrinsically non-Gaussian. Thus, one must pay a careful attention to the non-Gaussian evolution of density perturbations traced by HSLS sources, both resolved and unresolved.

In the case of FIRB fluctuations, the lensing reconstruction could potentially make use of the local anisotropy of the unresolved background induced by large-scale structure modifications under gravitational lensing deflections. The effect can be described as a remapping of the background intensity from one location to another as FIRB photons are deflected due to weak lensing during its journey to us within the large-scale structure mass distribution. The fluctuation intensity towards an arbitrary direction on the sky is $I(\theta) = I(\theta + \delta\theta) \approx I(\theta) + \delta\theta \cdot \nabla I(\theta)$, where $\delta\theta$ is the lensing deflection integrated over the mass distribution between us and the FIRB background and is typically at the level of 0.5 to 1 arcminute for sources at a redshift of 3.

The lensing reconstruction uses statistics of patches separated by this mean deflection angle to directly reconstruct properties of lensing. Though FIRB fluctuations probe a different aspect of weak lensing by the large-scale structure, the deflection angle is still determined by the mass distribution (dominated by dark matter in large scale structure filaments and halos) between the source and us. Thus the data can be used to extract cosmological information on the dark matter distribution. Note that we are not attempting to perform a weak lensing study towards, say, foreground galaxy clusters with background FIRB sources or to look for lensed FIRB sources in the background through a foreground cluster. Rather we use an ensemble of FIRB sources to establish statistical aspects related to the large-scale structure as a whole. This again requires wide-field imaging and while we optimize the shallow survey for FIRB fluctuations as the main science goal, we can equally pursue this secondary science goal of weak lensing.

While this is a possibility, the technique we have described to extract weak lensing information from FIRB fluctuations is yet to be applied due to lack of adequate data. More studies are also needed on the higher-order clustering of FIRB fluctuations. While Planck will provide an all-sky map for a weak lensing study, the lensing information in Planck is intrinsically limited due to poor resolution (one requires CMB maps with $\sim 1'$ resolution for an adequate lensing study). *Herschel* will provide an important and timely opportunity for a weak lensing study, bridging between optical-galaxy and CMB- based lensing measurements.

### 3.4.5 Why Herschel is Essential for Background Fluctuations Studies

*Herschel* offers access to the far-infrared and sub-millimeter spectrum region, and has the capability to map spatially extended low-brightness signals. These observations require understanding of the instrument noise on all spatial scales, so involvement from the instrument team is essential. A survey of this nature uniquely allows us to explore the clustering properties of the majority population contributing to the FIRB, otherwise undetectable with *Herschel* using conventional photometry due to the severe confusion limit. In comparison to Planck, *Herschel* offers higher angular resolution and higher point source sensitivity, allowing *Herschel* to better remove Poisson fluctuations, cover higher multipole space, and access the non-linear clustering signal. *Herschel*'s angular resolution is key to establish weak-lensing statistics. *Herschel* also provides coverage at higher frequencies, allowing multi-color determination of the FIRB fluctuations and cirrus emission. SPIRE in particular is designed to enable the mapping of large regions, using scan map mode, with high fidelity.

### 3.5 Applications related to massive cluster finding

The number and redshift distribution of galaxy clusters can place strong constraints on the cosmological parameters such as the matter density, dark energy density and the properties of galaxy clustering, mainly its normalization $\sigma_8$ [19][398]. The main advantage of using galaxy clusters counts to constraints cosmology relies on the fact that the formation of the dark matter potential wells of clusters involves only gravitational physics and has only a weak dependence on the star formation process and gas physics, that are generally difficult to model [182]. The abundance of clusters is in fact determined only by the geometry of the Universe and by the power spectrum of primordial fluctuations. As a result the redshift evolution of the observed cluster abundance can place strong constraints on the matter density, $\Omega_{\rm m}$, and provide a natural test of the dark energy, both its density and equation of state $w$, that directly affects the linear growth of fluctuations. The effect of the cosmological parameters on the cluster distribution is substantially different than the effect on other observables and as a consequence this kind of observations can be complementary to other cosmological probes such as high redshift type Ia Supernovae (SNe), or cosmic microwave background (CMB) anisotropies. The combination of clusters counts observations with CMB or Supernovae data can hence efficiently break degeneracies between parameters [200].

One of the fundamental quantities in the analysis of cluster counts data is the comoving cluster mass function $dn(z)/dm$, that gives the number density of clusters at a given redshift and mass $m$. The number of clusters as a function of redshift and in a certain mass range $[M_{\rm min}, M_{\rm max}]$ is then given by [181]:

$$\frac{{\rm d}N}{{\rm d}z} = \frac{{\rm d}V}{{\rm d}z} \int_{M_{\rm min}}^{M_{\rm max}} {\rm d}M \frac{{\rm d}n(z)}{{\rm d}M} \tag{17}$$

where ${\rm d}V/{\rm d}z$ is the volume element. Equation 17 clarifies the cosmological dependence of cluster counts: the volume element depends on the geometry of the Universe and its content, while the ${\rm d}n(z)/{\rm d}m$ depends also on the growth function of matter perturbations.

It has also been suggested that counts of massive clusters can place constraints on non-Gaussianity of initial fluctuations, given the connection of the abundance of very massive clusters with the distribution of the primordial matter density field fluctuations [269][243]. Cluster counts may be useful for this aim if non-Gaussianity increases at small scales; in that case it could be difficult to detect it with CMB observations, but it would still leave an imprint on large-scale structures. Moreover the abundance of clusters allows to test even smaller scales with respect to the three point correlation function. Being connected to the primordial density fluctuation field, the abundance of clusters is sensitive to the skewness of the primordial curvature perturbation, i.e. to any primordial non-Gaussianity. Parameterizing the non-Gaussianity as in Section 3.2, it is possible to show that primordial non-Gaussianity affects the mass function $\mathrm{d}n(z)/\mathrm{d}m$, introducing a correction term $R_{\rm NG}$ that depends on $f_{\rm NL}$[71]:

$$\frac{\mathrm{d}n(z, f_{\rm NL})}{\mathrm{d}m} = \frac{\mathrm{d}n(z)}{\mathrm{d}m} R_{\rm NG}(m, z, f_{\rm NL}) \,. \tag{18}$$

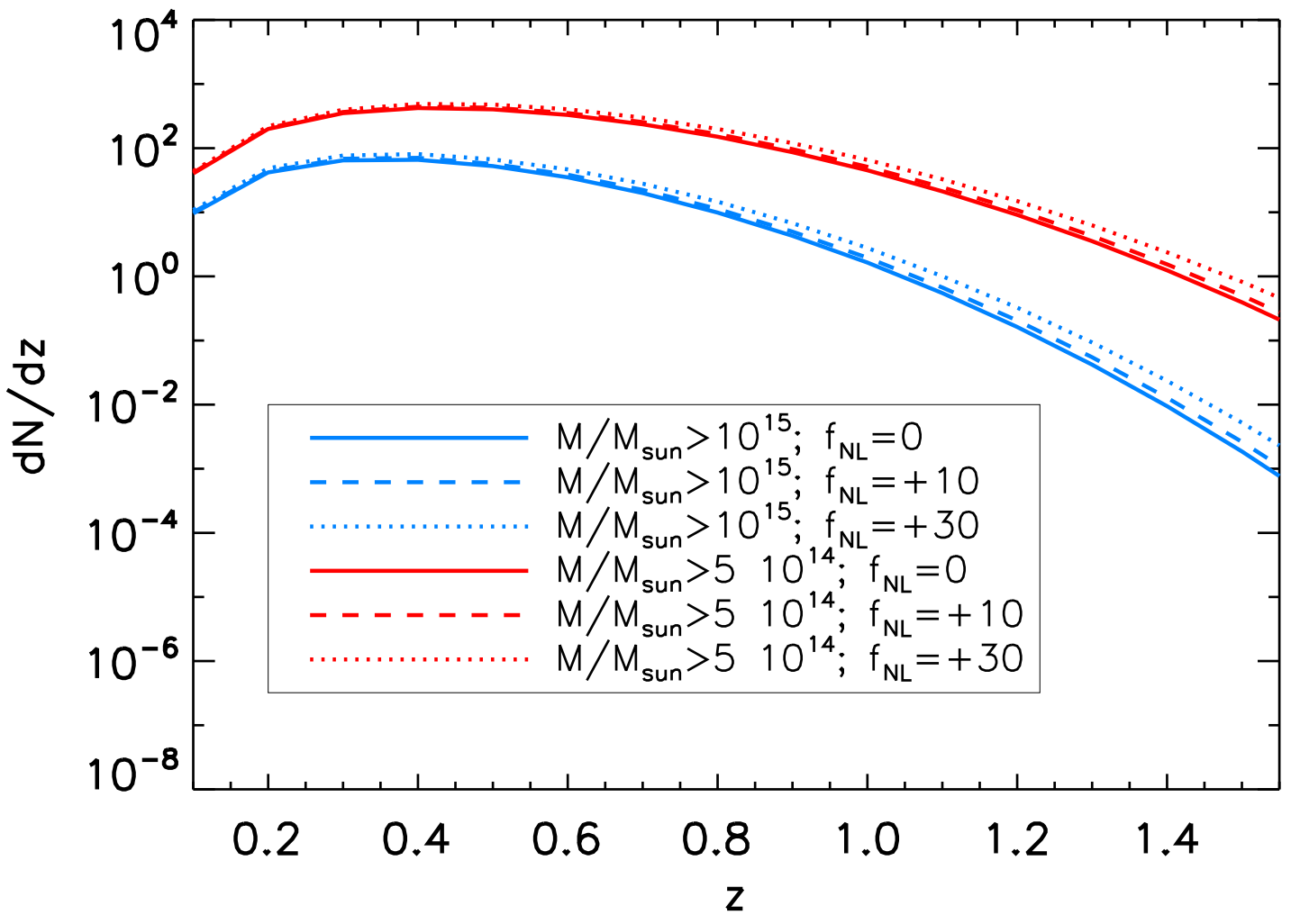

Figure 20: Number of clusters as a function of redshift for two different mass thresholds and different values of $f_{\rm NL}$. Positive values of $f_{\rm NL}$ increase the expected number of clusters and allow to constrain non-Gaussianity through cluster counts.

Generally, large non-Gaussianity would imply a higher probability of finding a very massive cluster [341]. By looking at the number of clusters in a certain redshift interval and in a mass range or to the number of clusters above a mass threshold $M_{\rm lim}$ it is hence possible to constrain non-Gaussianity. In Figure 20 we plot the number of clusters as a function of redshift that are expected to be detected by HSLS above two mass thresholds: massive clusters with $M > 5 \times 10^{14} M_\odot$ and very massive clusters with $M > 10^{15} M_\odot$, for a $\Lambda$CDM model with parameters given by WMAP seven years best fit values and assuming different values for $f_{\rm NL}$. Here we are using the Sheth & Tormen relation [367] for the mass function. By integrating the distribution it is possible to see that HSLS is expected to detect more than $\simeq 300$ very massive clusters ($M > 10^{15} M_\odot$), with about $\simeq 2$ of these clusters at high redshifts ($z > 1$). These numbers increase for positive values of $f_{\rm NL}$, allowing to constrain non Gaussianity using cluster counts. To first approximation it can be assumed that the average predicted number of clusters above $z = 1$ scales roughly linearly with $f_{NL}$. Here we consider the case that 8 clusters with $M > 10^{15} M_\odot$ will be found and we calculate, for Poisson statistics, the ratio of non-Gaussian to Gaussian likelihoods as a function of $f_{NL}$. This is shown in Figure 21: non Gaussianity with $f_{NL} = +30$ would be preferred over $f_{NL} = 0$ by a likelihood factor of $\sim 30$ from this measurement alone if 8 clusters were found at $z > 1$. While HSLS alone cannot find clusters easily, except for the few 500 $\mu$m SZ detections and overdensities of lensed sources, HSLS will help improve the Planck SZ cluster detection. Furthermore HSLS combined with a survey like DES in optical or ROSITA in X-ray will lead to an identification of the massive clusters at $z > 1$.

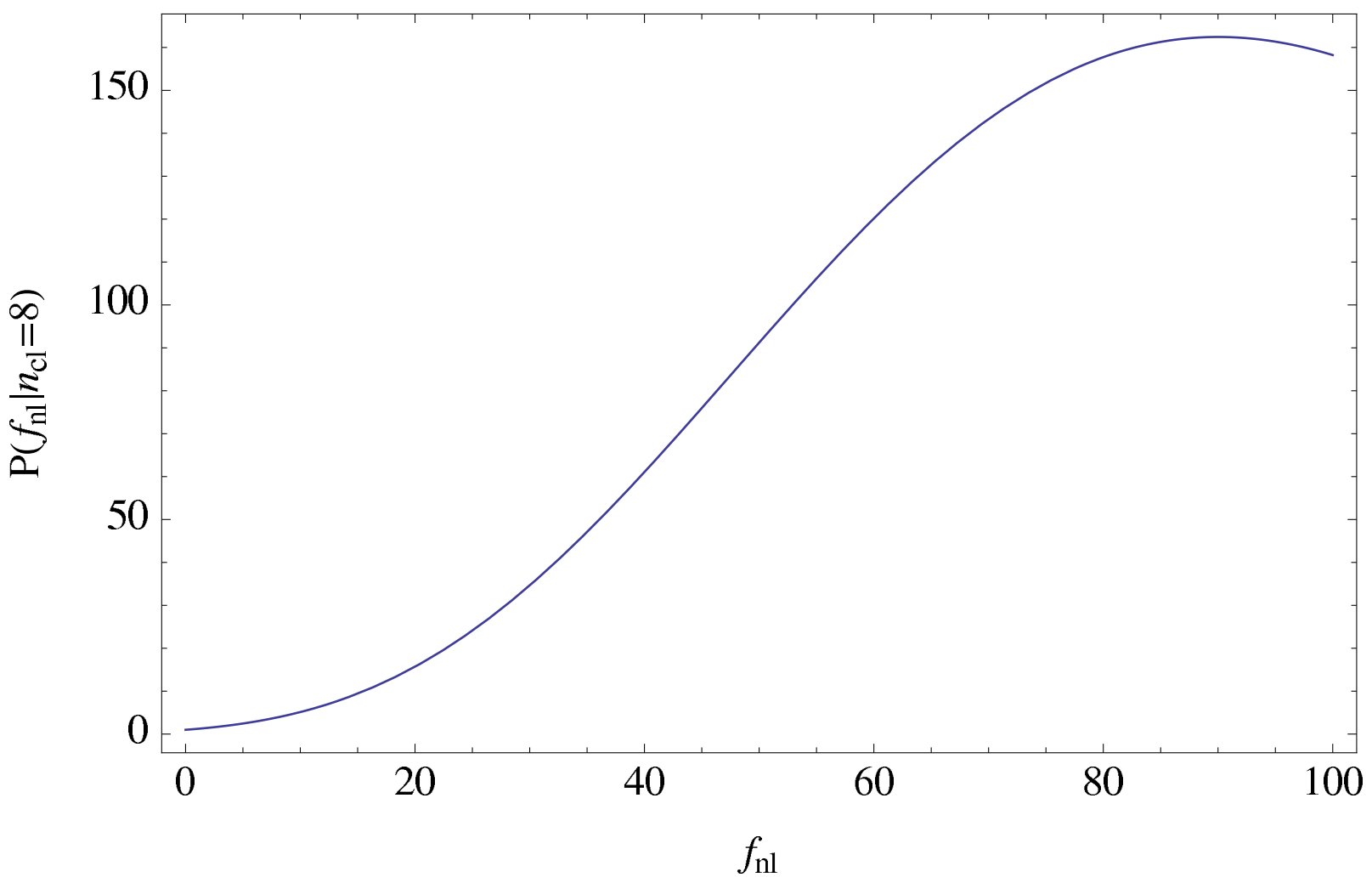

Figure 21: Ratio of the Non-Gaussian to Gaussian likelihoods as a function of $f_{NL}$ for the case that 8, $M > 10^{15} M_{\odot}$, clusters will be found. Non-Gaussianity with $f_{NL} = +30$ would be favored by a likelihood ratio $\sim 30$.

### 3.6 IGM dust with Herschel-SPIRE Legacy Survey Cross-Correlations

The angular distribution of the high-redshift sources ($z > 1$) observable at 250, 350 and 500 $\mu$m with the *Herschel*-SPIRE Legacy Survey can provide unique information on the amount of dust in the IGM through a differential cross-correlation analysis of the survey dataset with an ancillary sample of low-redshift ($z < 1$) galaxies, such as those in the SDSS Stripe-82 and SEP fields.

The presence of dust along the line of sight in the proximity of foreground galaxies alters the flux of background sources, causing fluctuations about the sample average. If dust is mostly concentrated in the halo surrounding the foreground galaxies and/or within galaxy clusters at low-z, one can expect flux fluctuations to increase as the background sources are at smaller angular separations from foreground objects. At optical wavelengths this effect is opposite to the flux fluctuations induced by cosmic magnification. This is because dust causes extinction of the incoming UV/optical photons, thus degrading the correlation signal. A recent analysis of the QSO-galaxy correlation from SDSS has shown evidence of a color-dependent signal suggesting the existence of a diffuse dust component [278]. Similar results were inferred from the QSO-cluster SDSS catalogues in [79]. The *Herschel* dataset can provide a thorough insight since in the far-infrared dust is seen in emission rather than extinction. Moreover by inducing angular correlations between independent samples, a diffuse dust component in the IGM will contribute to the Far-Infrared Background (FIRB) [3] and its angular fluctuations. HSLS in combination with Planck can provide accurate constraints on the FIRB [8] and consequently on the dust contribution as well. Currently various observations constrain the total amount of dust in the IGM to be $\Omega_{\rm d} < 10^{-5}$ for average grain size of $\sim 0.1$ $\mu$m (see e.g. [206, 316]). The SDSS cross-correlation signal [278] is consistent with $\Omega_{\rm d} \approx 10^{-6}$; if this is confirmed at higher statistical significance it will imply that IGM dust can significantly affect future dark energy searches with large samples of SN Ia luminosity distance measurements [94]. Hence a detailed study of cosmic dust is relevant not only for a better understanding of the complex physics of the IGM, but also for assessing source of systematics in SN datasets. Angular differential cross-correlation measurements of the SN flux fluctuations with foreground sources can also shed light on the level of IGM dust extinction [414] and disentangle its effect from the cosmic magnification [89]. From this point of view, HSLS can add complementary information, provided a large sample of SN Ia will be available in the survey field. One possibility would be to remove the point sources from the 250, 350 and 500 $\mu$m maps and cross-correlated the far-infrared background diffuse emission with SN flux fluctuations.

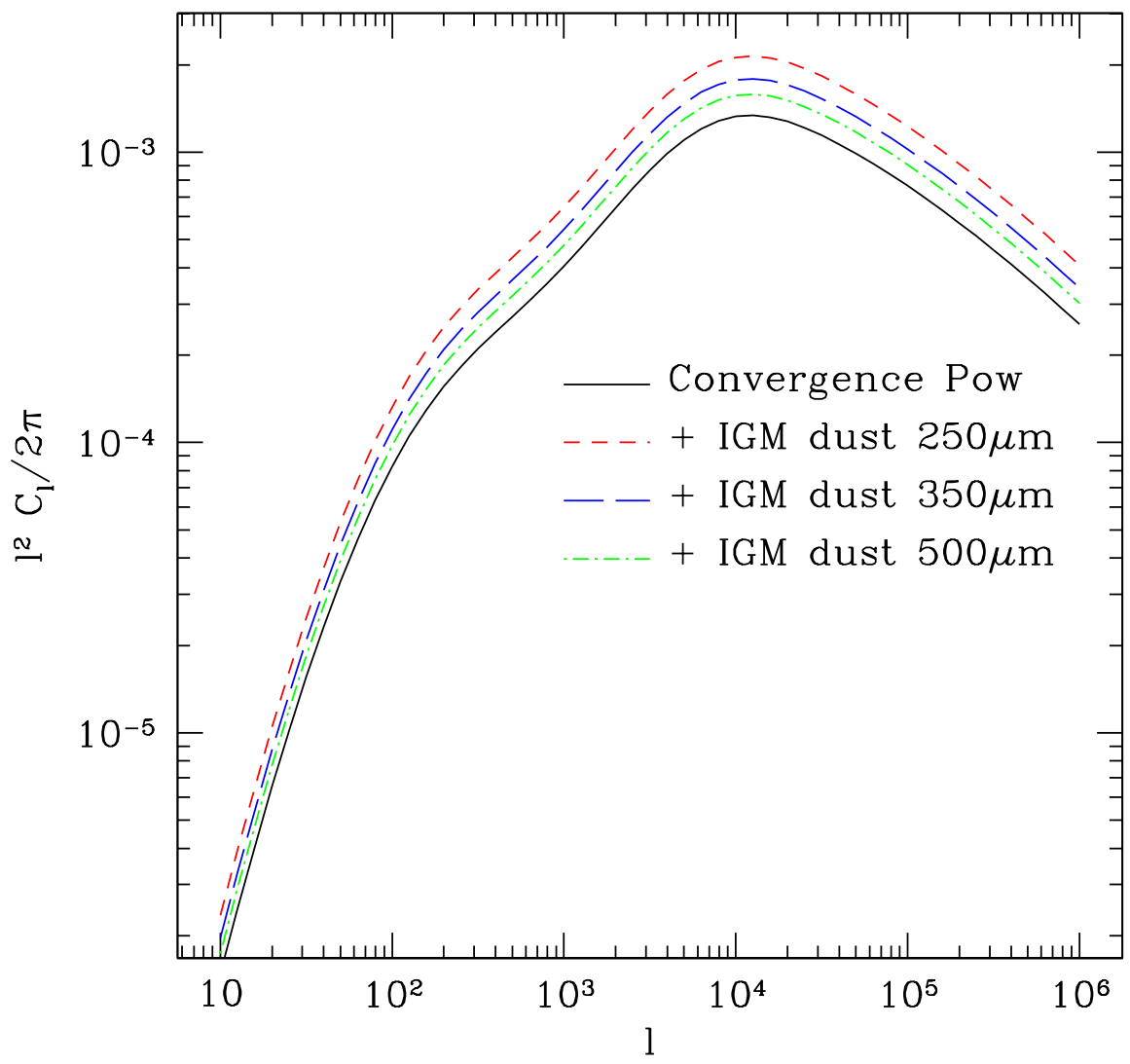

Figure 22: Cross-correlation angular power spectrum between high-z sources expected from the *Herschel*-SPIRE survey and low-z galaxy distribution. The cross-power spectrum signal from cosmic magnification only is shown as black solid line. Accounting for the IGM dust emission increase the signal in a wavelength dependent manner: the various curves show the full cross-power spectrum at 250 $\mu$m (red short dash line), 350 $\mu$m (blue long dash line) and 500 $\mu$m (green dot dash line).

Here we focus on a simple estimate of the angular cross-correlation between high-$z$ *Herschel* sources and low-z galaxies from an optical ancillary data sample. Hereafter we assume a 4000 deg$^2$ sky coverage, and consider an IGM dust model consisting of carbonate and silicate grains (in equal proportions) with an average grain size distribution of 0.1 $\mu$m. We assume the optical and far-infrared properties given in [123] and compute the scattering and emission coefficients using a numerical code which solves the Mie equations.

The flux fluctuation of a high-z source in HSLS can be written as $\delta_F^\lambda(\hat{n}) = \delta\mu + B_\lambda \delta\tau_\lambda$, where $\delta\mu$ is the fluctuation in the cosmic magnification, $B_\lambda$ is the emission intensity of the dust along the line of sight relative to the average flux of the sample and $\delta\tau_\lambda$ is the dust optical depth fluctuation. For a sample of foreground galaxies detected in the optical let $\delta_g$ be the galaxy number density fluctuation. The angular cross-correlation power spectrum between background sources and foreground galaxies then reads as $C_\ell^{bf}(\lambda) = C_\ell^{\mu-\delta_g} + C_\ell^{\delta\tau_\lambda-\delta_g}$, the first term is independent of wavelength since cosmic magnification is a purely gravitational effect, while the second term due to dust emission is wavelength-dependent. Therefore a differential angular cross-correlation analysis would allow us to isolate the dust contribution, i.e. by measuring $\Delta^\ell_{\lambda-\lambda'} = C_\ell^{bf}(\lambda) - C_\ell^{bf}(\lambda')$. With the HSLS sample we can measure three observable quantities: $\Delta^{\ell}_{250-350}$, $\Delta^{\ell}_{250-500}$, $\Delta^{\ell}_{350-500}$.

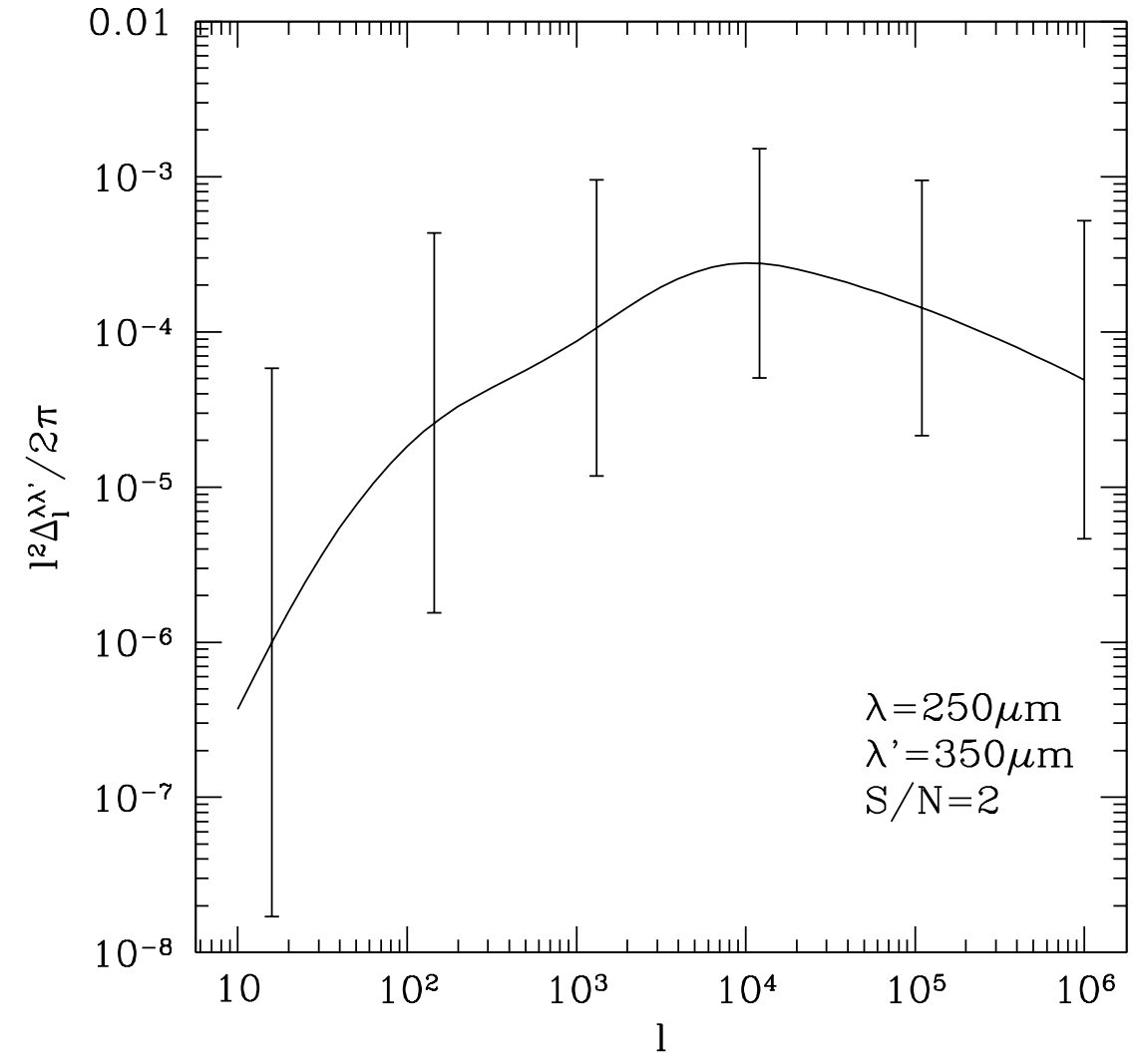

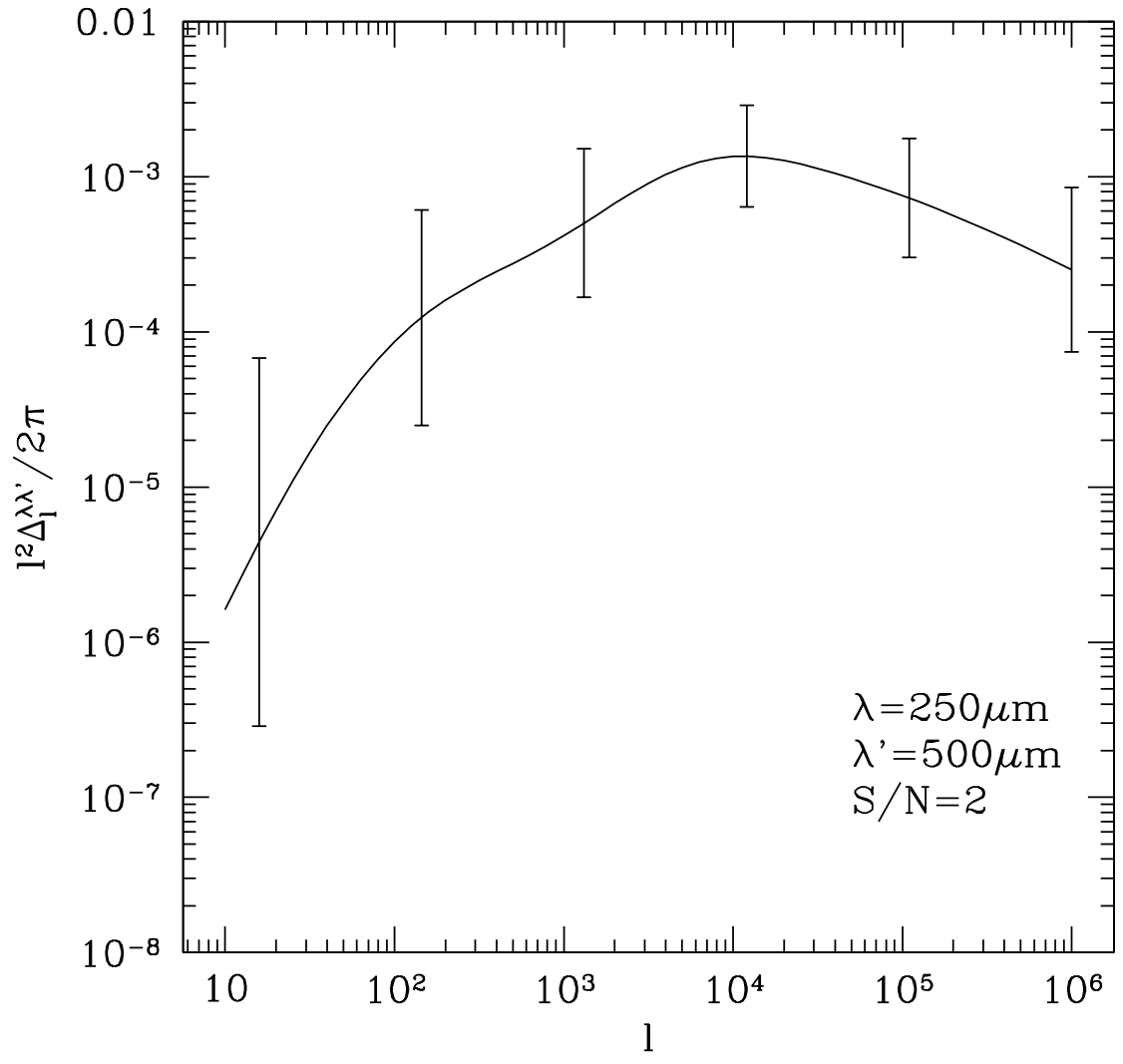

Figure 23: Differential cross-power spectra: $\Delta^{\ell}_{250-350}$ (left panel) and $\Delta^{\ell}_{250-500}$ (right panel). Error bars are the expected uncertainty on the cross-correlation between the expected *Herschel*-SPIRE sources and a low-$z$ galaxy survey distribution.

Assuming that IGM dust is a biased tracer of the matter distribution, $\delta\tau \propto \delta_m$, we have

$$\frac{\ell^2}{2\pi}C_\ell^{\delta\tau_\lambda-\delta_g} = \frac{\pi}{\ell}\int \Delta^2_\delta\left(\frac{l+1/2}{\chi}\right)\frac{d\bar{\tau}_\lambda}{d\chi}\chi \mathrm{d}\chi, \tag{19}$$

where $\chi$ is the comoving distance, $\Delta^2_\delta$ is the non-linear matter power spectrum (in the Limber approximation) which we assume to be in the parameterized form given in [307] and $\bar{\tau}_\lambda$ is the average dust optical depth as function of comoving distance. For simplicity we assume a dust bias parameter $b_d = 1$, and set the other dust model parameters such that the emissivity is about one percent of the average background flux corresponding to an IGM dust density $\Omega_\mathrm{d} \sim 10^{-7}$ (for more details on the dust model characteristics see [94]). In Figure 22 we plot the magnification cross-power spectrum (black solid line) for a $\Lambda$CDM cosmology assuming the expected *Herschel* source redshift distribution and an optical survey with galaxy distribution peaked around $z = 0.5$. The other curves show the effect of including the dust emission at $\lambda = 250$ $\mu$m (short-dash red line), 350 $\mu$m (long-dash blue line) and 500 $\mu$m (dot-dash green line). As we can see, the dust emission increases the cross-correlation signal, with the overall amplitude peaking at 250 $\mu$m and decreasing at longer wavelengths. This is the dust emission for the fiducial model considered here peaking at around 200 $\mu$m (for more details see [3]).

Following [89], we compute the expected cross-power spectrum uncertainties as

$$\Delta C_\ell^{bf} = \sqrt{\frac{1}{(2\ell+1)f_\mathrm{sky}\Delta\ell}}\left[(C_\ell^{bf})^2 + \left(C_\ell^{ff} + 1/N_f\right)\left(C_\ell^{bb} + 1/N_b\right)\right]^{1/2}, \tag{20}$$

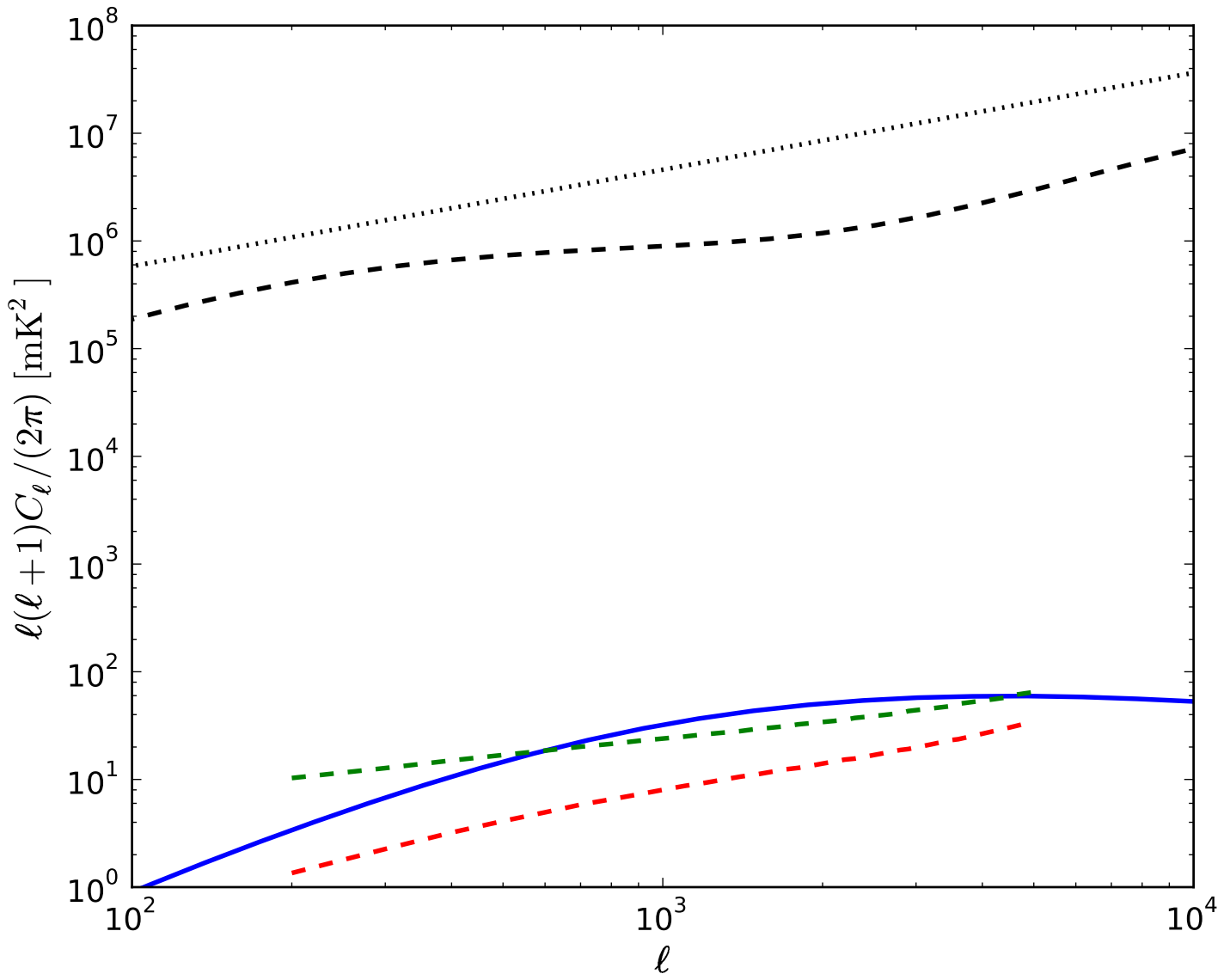

Figure 24: The angular power spectrum of the 21cm signal and expected contamination from point sources. Blue solid line - the expected 21cm signal at 140 MHz ($z \sim 9$, ionization fraction of 0.7, 1 MHz bin size). Dotted black - the model used for the point source contamination. Dashed-dotted green and red lines - residuals (error) after point source cleaning for different assumptions about the smoothness of point sources across frequency (see [339] for details). The dashed black line shows the expected point source clustering at 140 MHz based on *Herschel* fluctuations [10] with a 100 mJy flux cut in the far-IR.

where $N_i$ ($i = f, b$) is the surface number density of objects. Then the signal-to-noise reads as

$$\left(\frac{S}{N}\right)^2 = \sum_\ell \left(\frac{C_\ell^{bf}}{\Delta C_\ell^{bf}}\right)^2 . \tag{21}$$

In Figure 23 we plot the differential cross-power spectra, $\Delta^\ell_{250-350}$ (left panel) and $\Delta^\ell_{250-500}$ (right panel), where the error bars are the expected HSLS uncertainties. The signal-to-noise for each power spectrum is $\approx 2$, and combining all three spectra (including $\Delta^\ell_{350-500}$) gives an expected $2.5\sigma$ detection of the fiducial dust model considered here. On the other hand we find that a model with $\Omega_{\rm d} \sim 10^{-6}$, consistent with the estimate obtained from SDSS correlation in [278], will be detectable with HSLS at $\sim 10\,\sigma$. These forecasts suggest that HSLS can greatly contribute to a better understanding of the dust in the IGM and its cosmological implications.

### 3.7 Applications to 21-cm Measurements

The epoch of Reionization (EoR) is a crucial stage in the history of galaxy and structure formation, signaling the birth of the first luminous objects as structures first evolved beyond the well understood linear regime. The observation of the hyperfine 21cm transition from neutral hydrogen, redshifted to radio frequencies below 200 MHz, promises to give a wealth of information about this epoch and even the preceding period [58, 159, 266, 277, 325, 340].

Several radio-interferometers are now in construction (MWA[1], LOFAR[2], 21CMA[3],PAST[4]) with the aim of probing this signal essentially through statistical measurements using the angular and 3D power spectrum, with more powerful telescopes being planned for the future (SKA[5]).

The expected rms of the signal is $\sim 10$ mK which should be within the reach of the first generation of interferometers. Unfortunately, astrophysical foreground contaminants have fluctuations on the same angular scales up to five orders of magnitude higher than the signal, making its actual extraction a difficult challenge for the proposed experiments, even at the statistical level. However, several studies have shown that the large coherence (smoothness) of the signal across the frequency direction should in principal allow to remove these contaminants to a level below the redshifted 21cm signal [189, 291, 339, 402]. Nevertheless, problems related to the instrument response across frequency [57, 105, 106, 232], as well as the behavior of the huge foreground contaminants [365] might severely complicate this cleaning process. Galactic synchrotron emission accounts for $\sim 70\%$ of the total brightness temperature of these foregrounds while the extra-galactic continuum point sources account for roughly the other 30%. Note however, that depending on the frequency behavior of the point source continuum emission, they might be the most damaging foreground. Figure 24 shows the expected 21cm angular power spectrum at $\nu = 140$ MHz, together with a model for the point source contamination and the expected residuals after cleaning for different assumptions about their frequency smoothness (see [339] for details). As stated, the situation might be even more difficult depending on the instrumental setup.

The proposed HSLS survey covers the area of the sky that will be used by MWA to probe the EoR signal and given that HSLS data products will start to become public by early 2012 (MWA should start full operations in 2011) this will represent an excellent opportunity to analyze the point source contamination in detail. Note that MWA should be able to calibrate out the stronger radio point sources (AGNs), which do not correlate with the IR, leaving a residual source confusion of a few mJy that should then be removed through a cleaning technique. Using the well known far-IR-radio relation [88, 111, 163, 198, 207] we can predict the expected level of contamination from star-forming galaxies at the MWA frequencies using *Herschel* data (see black solid line in Figure 24). Note that this requires a large extrapolation in frequency which might affect the expected level of the signal in the radio. Nevertheless, by comparing the correlation of the HSLS data with MWA before and after foreground removal, we can check how effective the cleaning technique is. This can be particularly important for the MWA, since their lack of long baselines and real time calibration might affect even further the point source cleaning process.

---

[1]http://www.mwatelescope.org
[2]http://lofar.org
[3]http://21cma.bao.ac.cn
[4]http://astro.berkeley.edu/ dbacker/eor
[5]http://www.skatelescope.org

## 4 The Strongly Lensed Sub-mm Sources

The top-level science goals of lensing studies in the HSLS are:

- Statistical power to constrain the dark energy density to 0.01 and the equation of state of dark energy to 0.06, or deviations from General Relativity to about 2%.
- Constrain the lens galaxy dark matter mass profile, ellipticity, substructure fraction and the evolution of the fundamental plane to $z \sim 1$.
- Luminosity function, redshift distribution, star formation and morphology of sub-mm sources.

Systematic searches for gravitational lenses have usually relied on the identification of multiple high-resolution radio and optical images of a lensed source, or the analysis of optical spectra. The Cosmic Lens All-Sky Survey (CLASS) [65, 293] had a low efficiency for selecting lenses, with 22 confirmed lenses out of 16,000 radio lens candidates. The Sloan Lens ACS Survey (SLACS) [47] was efficient, but, because of the preselection based on Sloan spectra, could only identify low redshift ($z \lesssim 0.5$) lenses. An ideal method of selecting lensed galaxies would be more efficient than CLASS but without the redshift bias inherent in SLACS. The present proposed survey promises to do just that.

Flux-density limited large-area surveys conducted at sub-mm wavelengths have two distinct advantages [41, 296, 313]: (i) sources selected in the sub-mm waveband are typically at high redshift ($z \gtrsim 1$) and so the probability of suitable alignment (lensing optical depth) with a foreground object is correspondingly higher; (ii) sub-mm galaxies exhibit strong cosmological evolution that translates into very steep source counts, i.e. intrinsically bright sub-mm galaxies are rare. Between 20 and 100 mJy, the slope of the measured number counts, $\beta$, takes on $2.53 \pm 0.16$, $2.99 \pm 0.51$ and $2.66 \pm 0.24$ at 250, 350 and 500 microns respectively. These values for $\beta$ clearly emphasize the steep decrease in sub-mm number counts with flux density. For a flux-density limited survey, magnification bias leads to an enhanced lensing cross section that scales with magnification, $\mu$, as $\mu^{(\beta-1)}$ for point sources. Assuming this holds for extended sources also, we get a magnification bias of $\mu^{1.5}$ to $\mu^{2}$ – this is the largest magnification bias of any source population in astronomy!

Thus, highly amplified galaxies show up in the tail of the source counts where the unlensed galaxies are rare. Other contributors to the brightest sub-mm counts, namely low-redshift ($z \leq 0.1$) spiral and starburst galaxies [361] and higher redshift radio-bright Active Galactic Nuclei (AGNs) [117] can be easily identified in relatively shallow optical and radio surveys. Therefore flux-density limited sub-mm surveys can provide a sample of lens candidates from which contaminants can be readily removed, leaving an almost pure sample of gravitational lens systems.

This expectation has been validated [297] by the first wide area sub-mm surveys with *Herschel*-SPIRE, H-ATLAS and HerMES. With a coverage of $4000\,\mathrm{deg}^2$, the HSLS should provide a sample of about $1400 \pm 600$ strongly lensed sources with $S_{500} \geq 100$ mJy, almost a factor of 10 larger than H-ATLAS and Hermes will produce together.

This large sample will allow detailed statistical analyses that extract information about the structure of lenses (galaxies), the evolution of their number density, and cosmology. In order to take full advantage of this information, it is essential that we have a complete sample of background sources. In early work, the emphasis was on deducing cosmology with the assumption that early-type galaxies did not have significant evolution out to $z \simeq 0.5$ (e.g., [72, 73, 80, 223]). More recent analysis with the Sloan Quasar Lens Survey [302] and re-analysis of the CLASS sample including evolution [284] has obtained results consistent with previous analyses using CLASS lenses, which is very encouraging.

It is clear that cosmological models may be constrained with strong lensing statistics, but large samples of lensed systems will also help to better understand the connection between dark matter halos and the baryons they host. This expectation has come to fruition with the SLACS sample analyses (with a fixed cosmological model) [16, 48, 164, 229, 390]. The SLACS analyses have revealed that the lensing galaxies follow the general early-type fundamental plane relation and reveal clear evidence for dark matter within the stellar effective radius [48]. The *Herschel* sample will provide a data set about 20 times larger than SLACS and over a wider range of redshifts to explore the connection between the stellar content of a galaxy and its dark matter halo.

The HSLS will also make an important contribution to the study of high redshift star forming galaxies. Follow-up of HSLS lensed sources will allow for a detailed study of the morphology, sizes and luminosity of the sub-mm sources at high redshift that would not be accessible otherwise. A particularly exciting prospect is the possibility of getting information about regions of extremely high star formation on scales comparable to the sizes of local giant molecular clouds [384].

More than three decades after the discovery of gravitational lenses, there is no statistically complete, large sample of lenses over a wide range of redshifts. The HSLS will fill this void and facilitate a wide range of science. The science goals that can be achieved with such a unique sample are described in this section.

## 4.1 Dark Energy

Lensing by a compact mass distribution ('thin lens') in general relativity may be written down in terms of the apparent position of the source $\vec{\theta}$, its true position $\vec{\theta}_S$, and the surface mass density of the lens $\Sigma(\vec{x})$ at position $\vec{x} = D_L\theta$ as

$$\vec{\theta} - \vec{\theta}_S = \frac{D_{LS}}{D_S}\frac{4G}{c^2}\int d^2x'\Sigma(\vec{x}')\frac{\vec{x}-\vec{x}'}{|\vec{x}-\vec{x}'|^2}\,. \tag{22}$$

If the surface mass density can be determined by other observations of the lenses (such as velocity dispersion measurements), then the splitting between images may be used to infer the ratio of the angular diameter distances $D_{\rm LS}/D_{\rm S}$, where $D_{\rm L}$ is the angular diameter distance to the lens, $D_{\rm S}$ the distance to the source, and $D_{\rm LS}$ the distance between lens and source.

The distances to the sources and lenses depend on the matter and energy content of the universe, and provide information on dark energy [160]. This technique for constraining cosmological parameters will come of age with the large data sets now becoming available and the radio follow-ups that will be possible in the near future. It provides complementary information to that obtained from CMB, weak lensing and SNIa measurements with different systematics, especially given the number of sources at high redshift that may be expected.

One method to infer the ratio of distances is to combine velocity dispersion measurements of the lensing galaxy with the measured image positions [178]. Recent analyses [46, 352] along these lines using SLACS lenses have yielded exciting results. The most recent analysis of 53 SLACS lenses [352] finds constraints at the 20% level on $\Omega_\Lambda$, consistent with concordance cosmology. The HSLS offers an opportunity for dramatic improvement in such constraints because of the higher redshift coverage and about 20-fold increase in sample size.

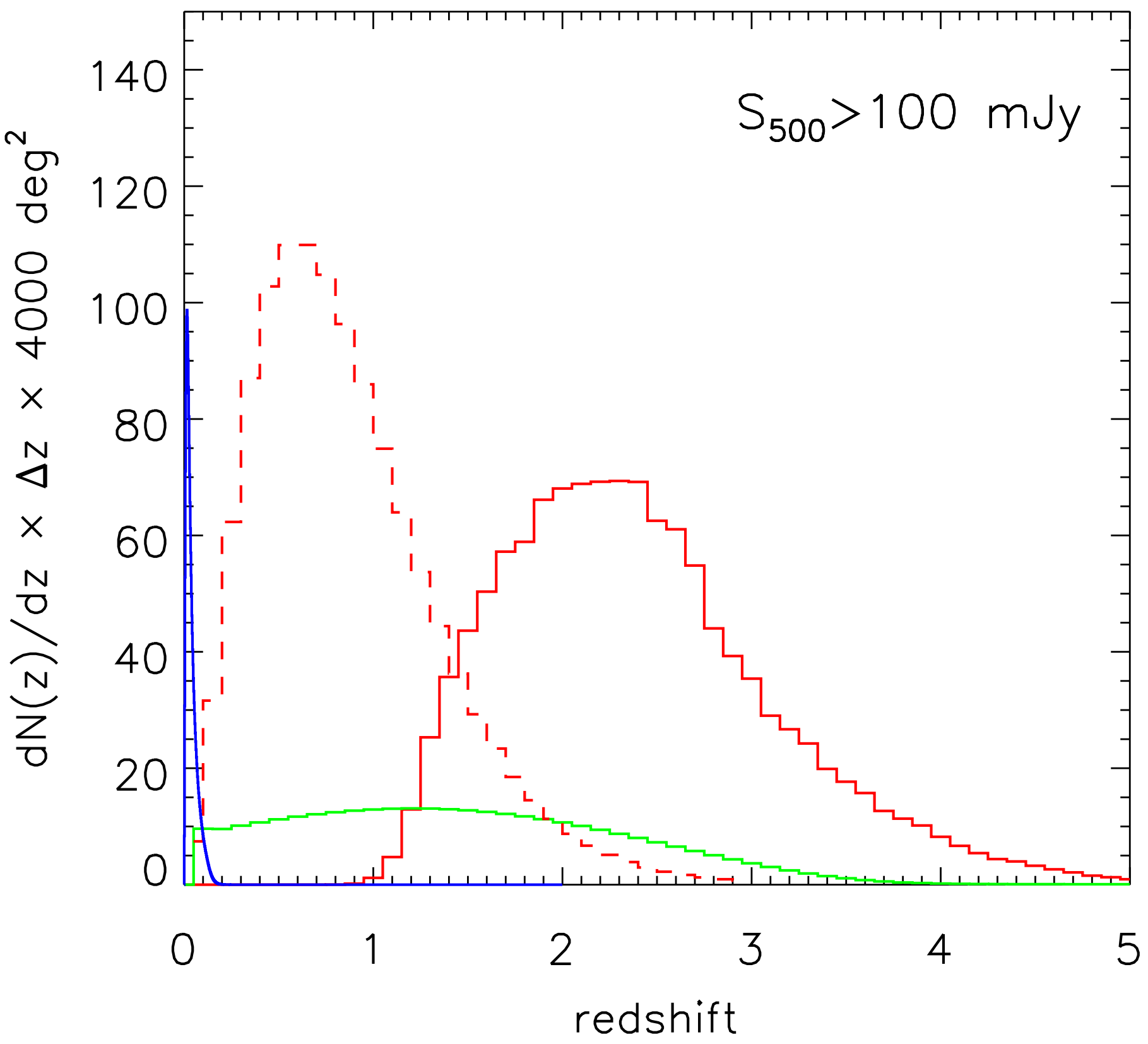

Figure 25: The redshift distribution of $S_{500} > 100$mJy lensed sources (1200 in total in 4000 deg$^2$) is given by the solid red curve, the foreground lenses by the dashed (red) curve, blazars by the solid green curve and low-$z$ spiral galaxies with same flux by the solid blue curve.

In order to estimate the effect of both these improvements, we consider a set-up similar to previous analyses of SLACS lenses [46, 178, 352]. We consider 1000 sources distributed according to Figure 25 and put the lenses at half the angular diameter distance. For each lens we assume that the distance ratio $D_{LS}/D_S$ may be measured to about 10% consistent with the SLACS analysis [352]. In this analysis, it is crucial to account for the unknown slope of the density profile of the lens: $\rho(r) \propto r^{-\alpha}$. An analysis of the SLACS lenses [228] showed that the values of $\alpha$ are consistent with being drawn from a distribution centered around 2 and with width 0.08. We adopt this for our study here with the implicit assumption that detailed follow up of about 50 lenses (the same as SLACS) or more will allow us to place this prior on $\alpha$ for the lenses found in HSLS, which will be at higher redshift. We are also assuming implicitly that $\alpha$ does not show systematic variation with the mass of the lens or its redshift. We note that with $> 1000$ lenses and detailed follow-up observations we should be able to test for the effects of these systematics and correct for them.

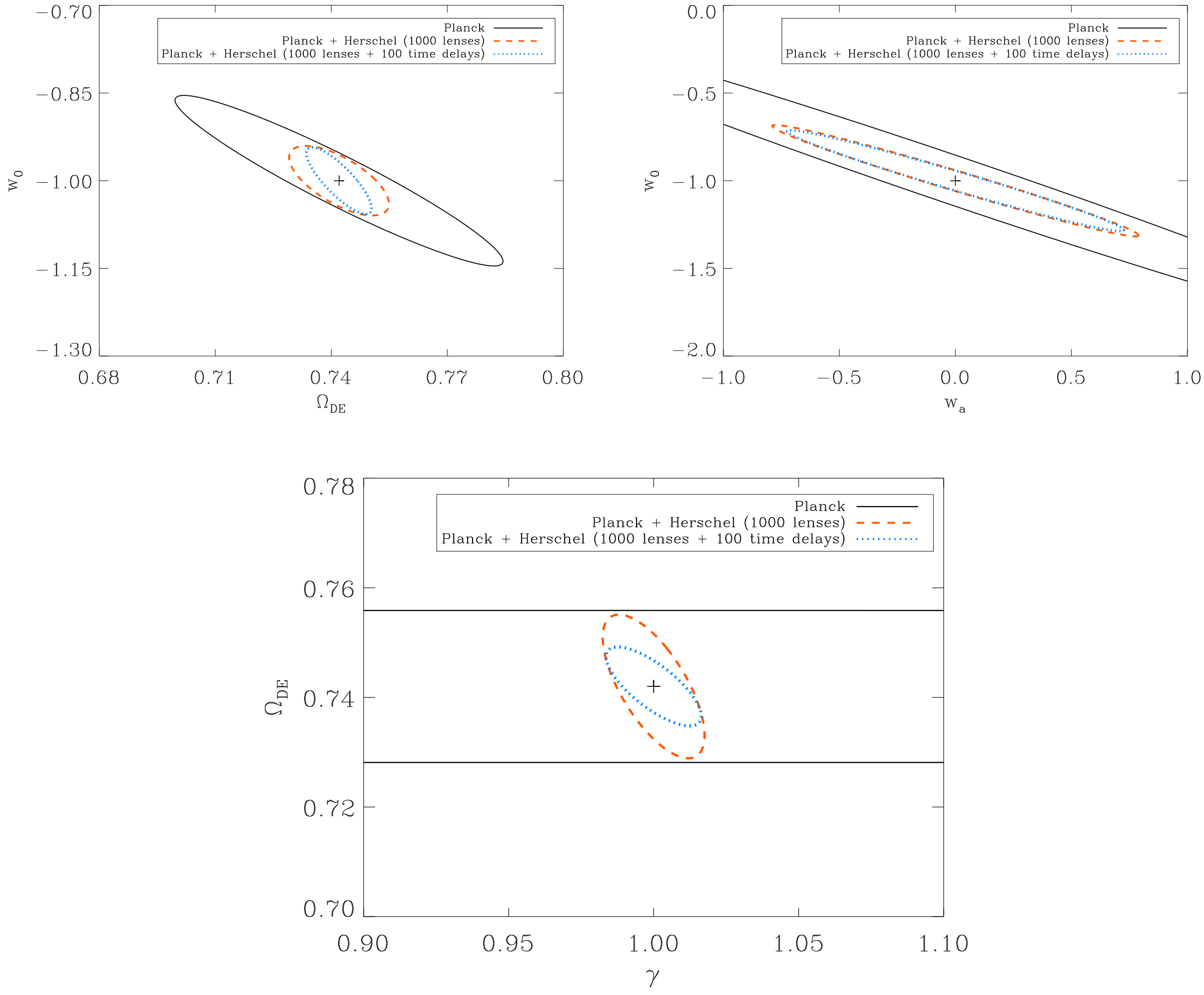

Figure 26: The top panel shows constraints on dark energy parameters assuming General Relativity. The left plot shows the constraints that may be obtained when image positions from 1000 lenses are combined with *Planck* constraints on the dark energy density $\Omega_{\rm DE}$ and the equation of state $w_0$. The right plot shows the constraints on equation of state parameters $w_0$ and $w_a$, where $w(a) = w_0 + w_a(1-a)$. The inner most contours show what happens when we add time delay information for 100 lenses. The bottom panel shows the statistical power of strong lensing to constrain the deviations from General Relativity in terms of the ratio of the gravitational potentials ($\gamma$).

Another avenue to constrain $\alpha$ is offered by the possibility of multiple sources behind one lens. Here we may also exploit the sensitivity offered by JWST to do observations in the near-IR of the lenses. Extrapolating from the counts in NICMOS [409], we estimate that there will be $10^6$ galaxies per square degree down to an H(AB) magnitude of 26. With a typical Einstein radius of $1''$, and making the conservative assumption that roughly half of the near-IR sources are at higher redshifts than the lens, there should be a 12% chance of finding a second lensed image in the near-IR, producing approximately 150 lenses with two lensed sources (one in the near-IR and one in the sub-mm). The double sources will break the degeneracy with cosmology and allow us to characterize the lensing mass profile well.

Some of these double sources will result in double Einstein rings. Using a typical source half-light radius of about $0.25''$, we estimate that there will be about 10–20 such systems. These systems by themselves provide useful cosmological information (e.g., [165]) but the redshifts of the source and lens are crucial for this analysis. We have not included the information from such systems in our analysis here.

Getting back to the distance ratios for 1000 lenses, we find that cosmological parameters may be constrained extremely well. The results are encapsulated in Figure 26. We consider a combination of the strong lensing data set with *Planck*, for which we have an extended cosmological model space with 10 parameters. We assume flatness and include information from the lensing of the CMB. The results in the figure also show the effect of including information from time delays for 100 systems, which is proportional to $D_{\rm LS}/D_{\rm S}/D_{\rm L}$. Detailed studies of time delays between images can break degeneracies between the dark energy parameters [386]. The use of 100 time delay measurements is a rough estimate based on the existing 10% of known lensed systems with time-delay estimates in the literature. This estimate should be used with care since the actual number will depend on the frequency of AGNs in the sub-mm galaxies and their variability at radio wavelengths. Furthermore, we have assumed that the time delays can be measured to about 10% accuracy. This addition improves the constraints at a level between 10 and 50%.

Our results indicate that the statistical power of this sample is sufficient to constrain $\Omega_{DE}$ to about 0.01, while the equation of state of dark energy will be constrained to about 0.06. If we allow for variation in the equation of state parameter, we may also put limits on this variation, although not as well as the average equation of state. To illustrate the statistical power in the strong lensing data set to constrain such variation, we parameterize the equation of state as $w(a) = w_0 + w_a(1 - a)$ and find that $\sigma(w_0) = 0.3$ and $\sigma(w_a) = 0.7$. These constraints improve considerably if $\Omega_{\rm DE}$ can be constrained independently to about 1%, as may be possible with cosmic shear measurements. This will allow for a consistency check with variation in the equation of state measured or bounded from other observations.

An interesting alternative application of such a large sample of lenses is to use them to test the predictions of General Relativity. This is related to dark energy because one of the possibilities often considered in the literature is that dark energy is a manifestation of deviations from General Relativity. Strong lensing (and gravitational lensing in general) is sensitive to these deviations because in modified gravity theories the effective gravitational constant, or the ratio of the metric potentials, could be scale dependent and different from the predictions of GR (e.g., [413]). Here we will consider the sensitivity of a large sample of lenses to the ratio of the metric potentials $\gamma = |\delta g_{00}/\delta g_{ii}|$ [46]. If we fix the cosmological parameters from other probes (say SNe), then it becomes possible, in a statistical sense, to ask whether the Newtonian potential probed by velocity dispersion measurements is consistent with the lensing potential recovered from analysing the lensed images. Analyses of the SLACS lenses have found constraints at the 10% level on $\gamma$ consistent with GR [46, 352]. These are presently the strongest constraints on deviations from GR on galactic scales.

Given the larger lens sample and the promise of better handle on the systematics involved in such a measurement, we expect significantly better constraints from HSLS. This is supported by a Fisher matrix analysis that considers the statistical power of 1000 lenses. We follow the same procedure as that used previously to predict the sensitivity to dark energy parameters. In this case, however, we restrict our attention to $\Omega_{DE}$ and $\gamma$. We find that the ratio of the potentials may be measured to about 2%, providing perhaps the strongest constraint on deviations from GR and complementary to the constraints expected from weak lensing and 21 cm power spectrum measurements [413].

There is, of course, more information in the statistics of strong lenses [231, 350, 391] than what we have used above. We have not considered the probability of finding lensed systems, which depends sensitively on cosmology through the mass function, distances and the concentrations of the lens mass distribution. These statistics are also sensitive to certain assumptions in our models relating to galaxy formation and evolution. One aspect of this is the luminosity-mass relation and its scatter. Further it is important that we understand the redshift distribution and luminosity function of the sources well. There is no doubt that a complete sample like that provided by HSLS will provide the best avenue for this sort of detailed analysis.

### 4.2 Dark matter halo properties

Observations of individual galaxy-lens systems allow us to infer the mass distributions of lensing galaxies independent of their luminosity. The statistics of multiply imaged systems can provide information about the dark matter halos of the lenses and the connection to their luminous component. The major advance in recent years has come from the SLACS survey [16, 48, 164, 229, 390], which selects for objects (usually extended) lensed by luminous red galaxies (LRGs) in the SDSS, and yields images with a lens-plane well covered by lensed structure and hence good model constraints.

Early-type galaxies have been shown to cluster surprisingly tightly around an isothermal distribution in the 1–20 kpc range of radius for which lensing (and dynamical) constraints are available. Indications from weak lensing are that this slope persists out to higher radius [164]. The large sample available through HSLS will allow us to test these results out to higher redshifts and ascertain if the slope of the mass profile changes systematically as a function of the redshift or mass, and luminosity of the lens. These results will inform the discussion about what happens to the dark matter when baryons collapse and form stars on a range of scales from elliptical galaxies to large groups.

With follow-up, these lensed images of sub-mm sources may be separated further into doubles, quads, and cusps. A typical smooth lens has two caustics, an inner astroid caustic and an outer radial caustic. If a source is inside the astroid caustic, it produces five images. The central image is formed close to the center of the lens and is rendered close to unobservable because it gets highly demagnified. Thus, we observe a quad-image configuration.

Sources inside the radial caustic but outside the astroid caustic produce three images, but again one is highly demagnified leading to a double-image configuration. Outside the radial caustic, only one image is produced. The ratio of the doubles to quads is sensitive to the ellipticity of the mass profile and we should be able to use the larger (bigger than $\sim 1''$) separation images to constrain the ellipticity of the underlying dark-matter mass profile at radii comparable to the Einstein radius.

There is also a three-image configuration where all the images are observable. This happens for large lens ellipticities when the astroid caustic becomes large enough in area to pierce through the radial caustic. Sources within the astroid caustic but outside the radial caustic produce three observable images, typically termed as the cusp configuration. The ratio of cusps to doubles and quads at image separations of order $1''$ or smaller is very sensitive to whether the luminous and dark matter components of the lenses are aligned with each other [283]. The alignment and correlation in the shapes of the luminous and dark-matter components in a galaxy are important aspects of galaxy formation. Such a connection between the luminous and dark matter components further lead to a spatial correlation between shapes of galaxies (e.g., [197]), and this is an important systematic effect for cosmic shear measurements that deserves an in-depth study [262]. The large flux-limited sample of lenses made possible by a shallow wide area *Herschel* survey such as HSLS should facilitate just that. An important point to keep in mind in connection with the ratio measurements discussed above is that they are sensitive to the relative magnification (of each image configuration) and hence to the luminosity function. We will need to understand this well to infer dark matter halo properties.

We have also estimated (as described previously) that HSLS, with follow-up using JWST, will provide about 150 lenses with two sources at different redshifts. Such a large sample of double sources offers a unique opportunity to study the mass profile of lenses in greater detail [165].

### 4.3 Fundamental plane of early-type galaxies

The SLACS lenses have been shown to define a fundamental plane that is consistent with that of the general population of early-type galaxies [48, 49]. These lenses are in the redshift range $0.1 < z < 0.4$. The new *Herschel* sample will significantly enlarge the redshift range probed and allow for detailed measurements of the evolution of the fundamental plane and its evolution with redshift. The larger sample will also facilitate a comparison of the fundamental plane of lower and higher mass early type galaxies.

A recent interesting result [216] from the AEGIS and DEEP 2 surveys is that over the redshift range from 0.1 to 1, galaxies follow a modified Tully-Fisher relation with the rotation velocity ($V_{\rm rot}$) being replaced by $(0.5V_{\rm rot}^2 + \sigma^2)^{1/2}$, where $\sigma$ is the dispersion of disordered motion of the gas. This is clearly indicative of an underlying relation between the luminosity and mass. The HSLS sample can elucidate the nature of this relationship by directly measuring the masses of the lenses over a wide range of redshifts.

### 4.4 Substructure in dark matter halos

One of the most important questions that gravitational lensing can answer is whether dark-matter substructure exists within galaxies, as predicted by CDM simulations [220, 290]. In principle, this can be detected by flux anomalies in four-image systems of point-like background sources that are not consistent with smooth lens models [81, 102, 222, 264, 280, 281]. Sub-mm sources may provide both resolved (extended) and unresolved (point-like) sources because a large fraction of sub-mm sources are expected to harbor AGN [6]. In addition, the observations at low frequencies where the emitting regions are large help mitigate the effects of microlensing, and the images of the extended regions provide strong constraints on the smooth mass model.

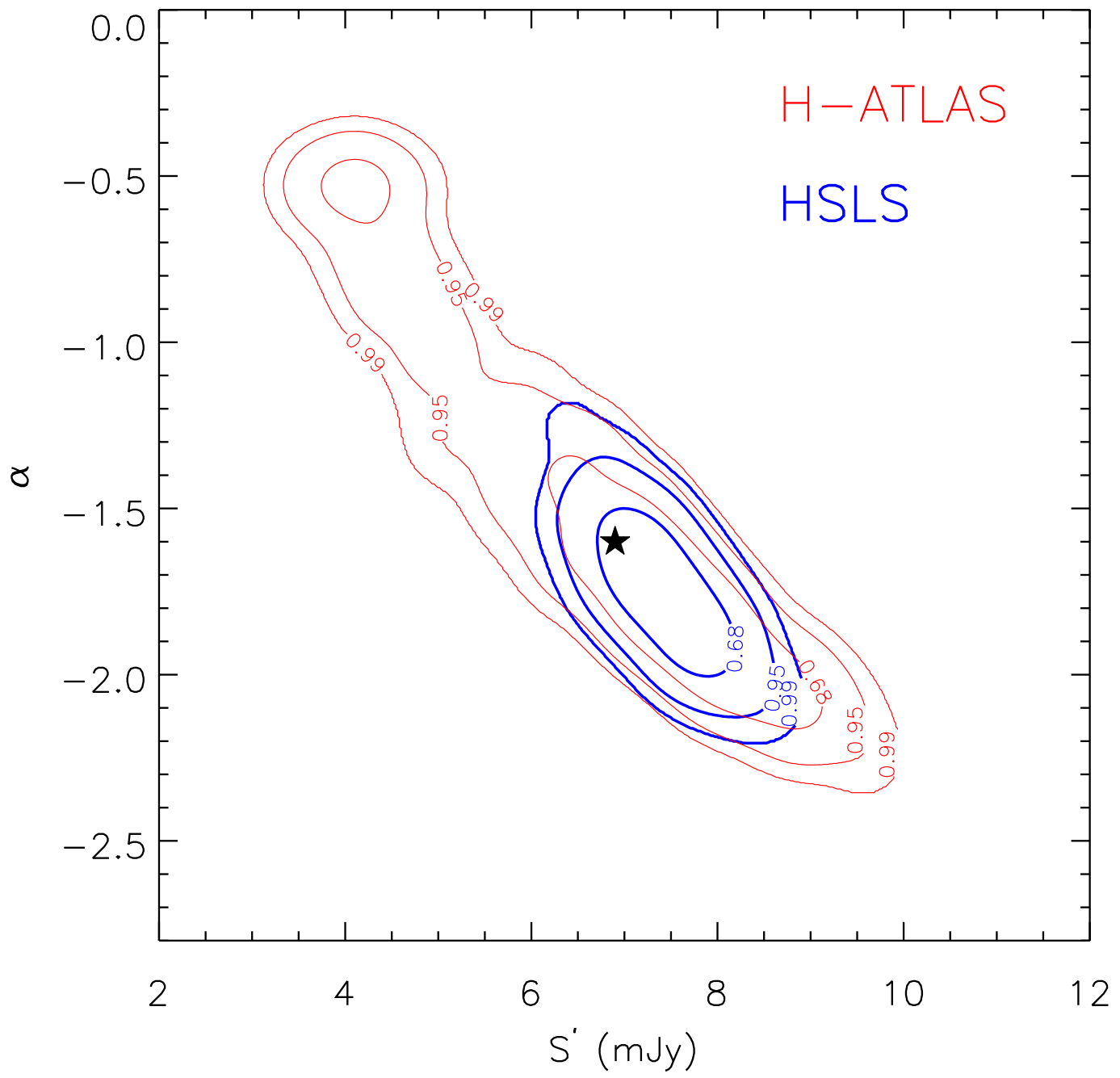

Figure 27: Simulated constraints on slope, $\alpha$, and cut-off flux, $S'$, of the counts of faint sub-mm galaxies as derived from the statistics of lensing events with $S_{500} > 80$ mJy in H-ATLAS (red) and HSLS (blue). The star marks the true values of the parameters used as an input in the simulations.

Alternatively, the presence of dark substructure can be deduced by imaging extended arcs and rings in radio and fitting in a statistical manner for the clumpy lens potential [392]. With these methods, it should be possible to get down to $10^7$–$10^8$ solar mass substructures [393, 394]. To illustrate this method, Figure 28 shows the simulated effect of a $10^8$ solar mass substructure on arcs imaged with resolution of 100 and 50 mas. This simulation shows that low-mass substructure could be observed with the expected resolution and signal-to-noise of ALMA follow-up of Herschel-selected lenses.

Early analyses, using mostly CLASS radio lenses, showed that flux anomalies indicative of dark substructure could be explained by the the predicted LCDM galactic substructure [102, 264], but the situation is far from settled (e.g., [211, 265, 368]). The current sample of four-image lenses with low-frequency flux measurements is small, and the HSLS will provide an opportunity to expand that.

### 4.5 A lensed view of sub-mm galaxies at high redshifts

Because of its large areal coverage, the HSLS will provide tighter constraints on the statistics of strong lensing events than H-ATLAS will achieve. The counts of strongly lensed sub-mm galaxies depends on two factors: (i) the probability distribution of the lensing magnifications;

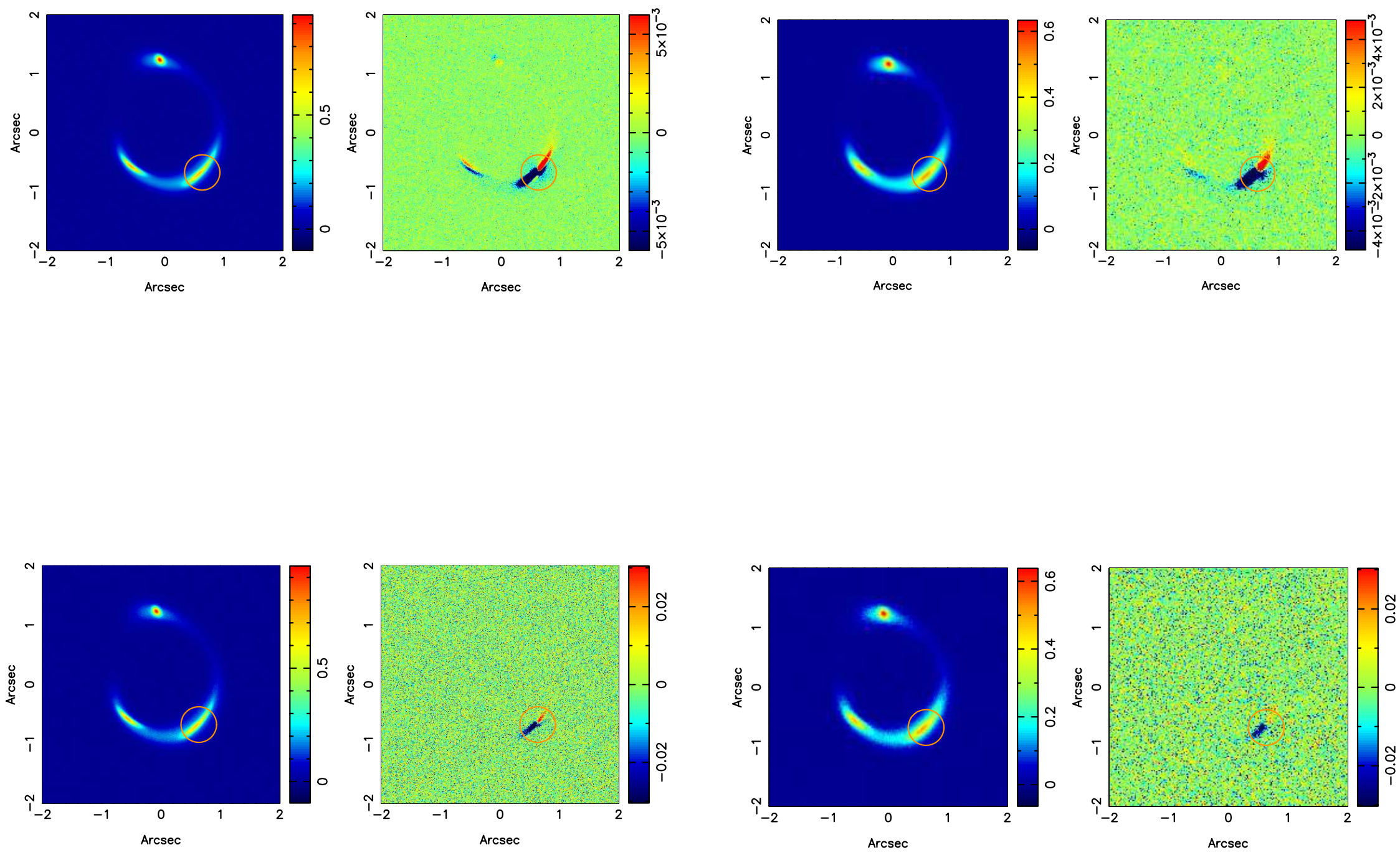

Figure 28: The four panels show the effect of a $10^8$ $M_\odot$ substructure (inside the tidal radius of  1kpc) positioned on the lower right arc of a typical simulated lens system (based ons SLACS). Shown are the lens system observed with a 50mas (left column) and 100mas (right column) FWHM PSF, comparable to ALMA follow-up observations. The top row shows the system observed at a noise level of 0.001 and the bottom row at a noise level of 0.01. The left figure for each panel shows the observed lens system and the right figure shows the difference between the lens system with and without the substructure. These panels illustrate that low-mass substructure could be relatively easily observed with the expected resolution and signal-to-noise of ALMA follow-up of HSLS lenses.

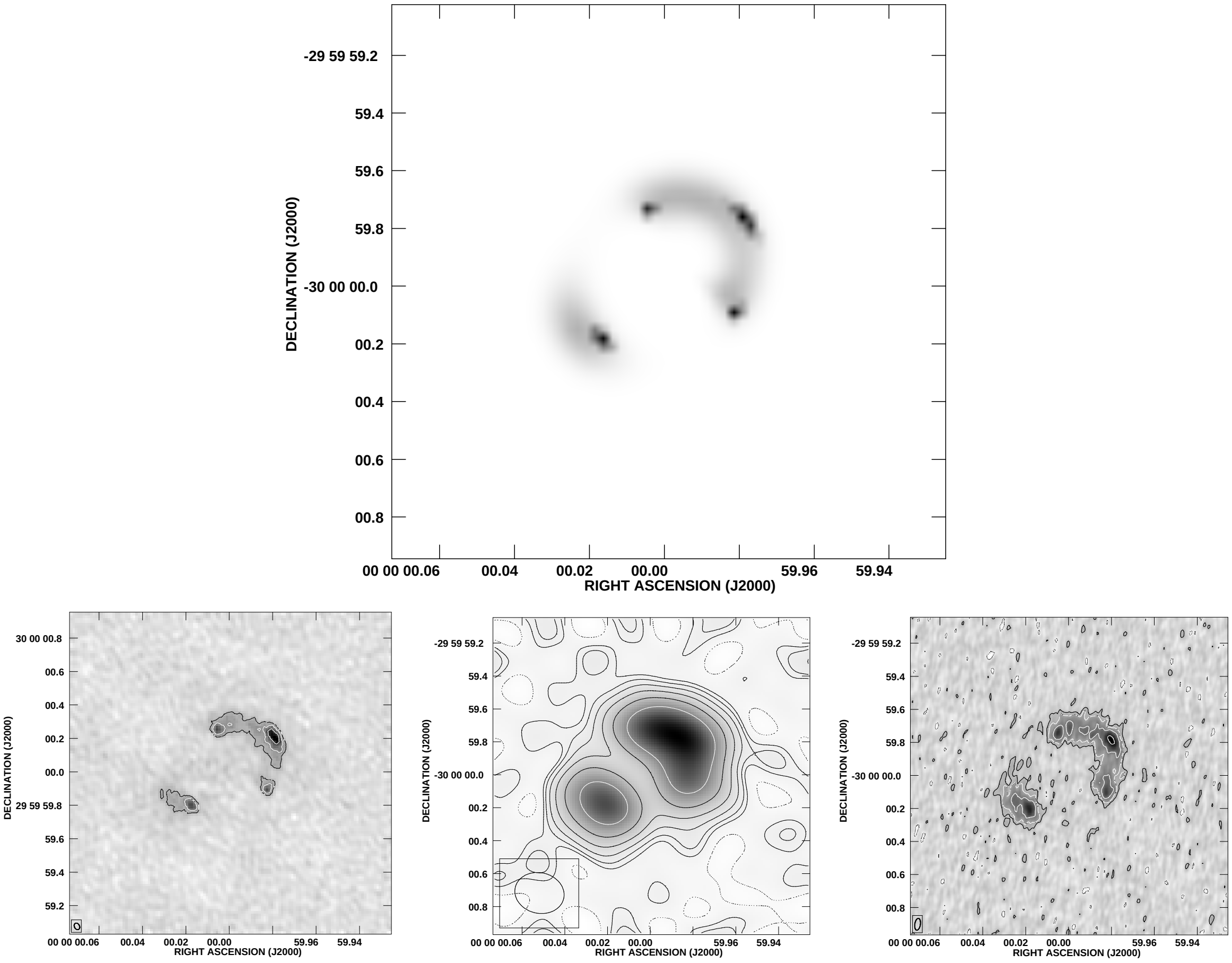

Figure 29: The top panel shows a quad-image system with an Einstein radius of about 0.5″, with total flux density 5 mJy at all frequencies, of which about 1/8 is in four images of the point sources and the rest in a more extended configuration. The source has been put at declination -30 for ALMA and +30 for e-MERLIN. The e-MERLIN simulation at 5 GHz is the left image in the second row and assumes an angular resolution of 50 mas. This system has also been simulated using two configurations of ALMA at 300GHz, from relatively compact (about 250 mas angular resolution) to extended (about the same angular resolution as e-MERLIN) . These follow-up observations will require a few to 10 hours per system, although just getting the fluxes of the point-source images will be relatively quicker.

(ii) the shape and normalization of the unlensed sub-mm counts. The first is determined by the distribution in mass and redshift of the lenses, their mass–density profile (which contributes to defining the functional form of the cross section for strong lensing events), the intrinsic size of the sources that are lensed (the more compact the source the higher the amplification it can experience for a given lens-source configuration) and the redshift distribution of the background sources. The study of the first samples of strongly lensed sub-mm galaxies, discovered with H-ATLAS and HerMES, will allow a direct measurement of the distribution of lensing amplifications. Therefore the study of the abundance of sub-mm bright strong lensing events can be used to obtain indirect constraints on the functional form of the counts of faint (unlensed) sub-mm galaxies. We have performed numerical simulations to test this assertion. We have assumed the differential counts of unlensed sub-mm galaxies at 500 $\mu$m to be described by a Schechter function

$$\frac{dN}{dS}(S_{500}) = \frac{N'}{S'}\left(\frac{S_{500}}{S'}\right)^{1+\alpha} \exp\left(-S_{500}/S'\right) \quad (23)$$

with the flux, $S_{500}$, measured in mJy and the counts expressed in units of mJy$^{-1}$ deg$^{-2}$. We have also assumed the following values for the free parameters in the simulations: $N' = 11267$ deg$^{-2}$, $S' = 6.9$ mJy, and $\alpha = -1.6$, as they produce source counts in agreement with those of high redshift protospheroids, as described by the [176] model and implemented in [296]. We have generated 1000 realisations of the counts from the equation above and used the distribution of amplifications derived from the model of [313] to generate the differential counts of the strongly lensed sources with[6] $S_{500} > 80$ mJy. The resulting source counts were then resampled using Poisson errors according to the simulated area (i.e. 4000 deg$^2$ for HSLS and 550 deg$^2$ for the full H-ATLAS). We fitted the simulated bright counts with the functional form described in Eq. 23 after convolving with the same probability distribution of amplification used in the simulation. The distribution of the amplification is assumed to be constrained to 10% accuracy, and this uncertainty has been taken into account before convolution. In the fitting process we have also used the distribution of amplification as an extra constraint on the predicted value of the background intensity (assuming a 10% uncertainty). Figure 27 shows the results in terms of the constraints achieved on $\alpha$ and $S'$ (after marginalizing over $N'$) derived for H-ATLAS and HSLS. While H-ATLAS will provide reliable information on the shape of the faint sub-mm counts, HSLS will significantly improve these constraints. In addition to providing indirect constraints of the counts of faint sub-mm galaxies, the magnification due to lensing will allow a much easier study of the intrinsic properties of these galaxies (like morphology, extension of the star-forming regions, etc.) on a one-by-one basis, through dedicated follow-up observations (e.g., [384]).

### 4.6 Sub-mm follow-up observations

Unlike some other areas of the proposal, the lens sample extracted from these observations requires further follow-up. As well as accurate redshifts for lens and source, high-resolution observations are necessary in order to determine the morphology of the lensed images, which is vital for the construction of lens mass models. However, the instruments required for this will be available on a short timescale. The most important is ALMA, which will be able to obtain redshifts in a few minutes of integration time for most sources detected in the sub-mm, as well as providing the required resolution to obtain the fluxes and structure for the images (see Figure 29 for a simulated example). Furthermore, parts of the survey at $\delta \geq -20$ are accessible to cm-wave imaging with the new e-MERLIN telescope, at comparable resolution to ALMA. The sensitivity limit of e-MERLIN is about 2–4 $\mu$Jy beam$^{-1}$ at 20 cm in 12 hours, enabling 180-mas resolution maps to be made of sources of a few hundred $\mu$Jy, the likely typical cm-flux density of sources found in this survey. A gravitational lens legacy project is already running on e-MERLIN, with the basic aim of studying existing lenses. A large, flux-limited sample of lenses will provide strong motivation for a systematic follow-up program at e-MERLIN, ALMA and other facilities.

[6] This is the minimum flux threshold above which strongly lensed sources can be easily singled out from the parent sample of bright sub-mm objects.

*We refer the reader to Section 5.3.3 for additional details on what we can learn from followup observations of the HSLS lensed source sample with ALMA.*

## 5 The High-redshift ($z > 1$) Sub-mm Universe

The top-level science goals of high-$z$ studies with HSLS are:

- Approximately 10000 $z > 4$ galaxies leading to a first clustering measurement of SMGs at these highest redshifts.
- At least a few hundred $z > 5$ SMGs to connect the epoch of dusty star formation with the end of reionization and the first galaxies.
- Approximately 200 proto-clusters at $z \sim 2$ that are expected in hierarchical models of structure formation; these are unvirialized structures that are actively forming stars and extend over 10 to 20 Mpc scales.
- Environmental dependence of star-formation in massive, rare clusters at $z > 1$.
- About 1200 strongly lensed high-$z$ sub-mm sources from HSLS that will be fundamental for followup studies of the properties of ULIRGs from $z \sim 1$ to 6 using ALMA.

### 5.1 Overview

Understanding how galaxies form and evolve over cosmological time is a key goal in astrophysics. Over the last decade our understanding of the background cosmology has improved to such an extent [226] that we think we have a reasonable understanding of the structure formation in the underlying dark matter distribution [382]. However, galaxy formation and evolution, involving star formation in the dark matter potential wells, is a more complex process. Observations play a critical role in constraining models of galaxy formation, the evolution of star-formation activity, and the various roles played by galaxy stellar mass, dark matter halo mass, and environment. In some respects, observational evidence may even be ahead of models. For example, observations indicate that present-day big galaxies are older than present-day small galaxies [190], so that there are far more big galaxies at high redshifts than is predicted by the standard model [124, 173, 380]. We expect AGNs to play an important role in the build-up of stellar mass in galaxies, as illustrated by the observed correlation between black hole mass and the mass of the stellar bulge [256].

The central importance of far-infrared (FIR) surveys can be seen from the extragalactic background radiation: roughly half the optical and UV light emitted since the Big Bang by all the objects in the Universe — stars, galaxies, and AGNs — has been absorbed by dust and then re-radiated in the FIR/sub-mm waveband [128, 153, 326]. The optical alone cannot be used to trace this activity, e.g. the brightest sub-mm galaxy in the Hubble Deep Field is not even detected in the optical [97, 125]. *Herschel* is now vastly improving the current state of FIR observations by carrying out unprecedentedly large and coherent surveys in this historically unexplored waveband.

Previous FIR and sub-mm surveys, using SCUBA at 850 $\mu$m [129, 204, 374] and the *Spitzer* surveys at 24 $\mu$m [335], show very strong evolution, particularly of the more luminous FIR galaxies. However, both SCUBA and *Spitzer* typically sample galaxies at wavelengths far from (1) where they peak and (2) the peak of the background radiation. *Herschel's* wavelength and redshift coverage provides a connection between the $z > 2$ Universe, primarily probed by SCUBA, and the lower redshift Universe, sampled by *Spitzer*. SPIRE, which covers the peak of the FIR/sub-mm background, has finally allowed us to investigate directly the sources responsible for the far-IR background and characterise their bolometric emission.

The bulk of submillimeter galaxies (where we define an SMG as a cool, $T_{\rm dust} \sim 35$ K, ULIRG) have been found to lie in a relatively tight redshift range between $z = 1$ and $z = 3$ [77], with very few SMGs spectroscopically confirmed beyond $z > 4$ [93]. The small sub-mm surveys thus far have made it difficult to study the rare population of $z = 3$–6 ULIRGs which trace the build-up of dust and stars in the Universe.

With the formation of the first stars, the Universe began the process of enrichment through feedback from stars and black hole growth. At $z > 6$, the dust extinction in galaxies inferred from the UV is negligible [55] but it becomes increasingly important from $z = 6$ to $z = 3$. In agreement with that, the dust obscured star formation in galaxies as traced by the cosmic infrared background also increases over this period [78].

The key themes of the high-$z$-galaxies part of HSLS are:

(i) the dust-obscured growth of stars and black holes in massive galaxies at early times, including the search for $z > 5$ SMGs to probe the birth places of dust during the era of reionization, and

(ii) the relationship between large scale environment and the most luminous SMGs; specifically, the clustering of $z \sim 3$ to 4 SMGs and their connection to the dark matter distribution at those redshifts, and the connections between proto-clusters and hierarchical structure formation at $z \sim 2$.

The widest-area of the *Herschel*-SPIRE extragalactic programs, H-ATLAS [130], covers 550 deg$^2$ and is expected to find adequate numbers of SMGs to study the key aspects of ULIRGs below $z < 3$. A key motivation for HSLS, with an area of 4000 deg$^2$, comes from SMGs at $z > 3$, where IR galaxies luminous enough to detect are so rare that the combined area of the HerMES and H-ATLAS surveys (620 deg$^2$) limits detailed statistical studies.

Below we first highlight the range of $z > 1$ galaxies expected to be detected by HSLS, and then we explore the scientific drivers that motivate obtaining these samples.

## 5.2 High-$z$ galaxy demographics: What galaxies will be detected by HSLS?

*Herschel* maps are detecting thousands of SMGs per deg$^2$. Figure 30 shows the modeled sub-mm number counts compared to the observations. But what is their redshift distribution, and how are they related to sub-mm and far-IR luminous galaxies we already know about? The average redshift of the sources contributing to the sub-mm background is expected to increase to longer wavelengths [234]; 500-$\mu$m-selected galaxies will, on average, be at higher redshifts than 250-$\mu$m-selected galaxies. In addition, the negative $K$-correction applied to the SPIRE sub-mm observations results in an increasing sensitivity to high redshift sources at the longest wavelength (left panel of Figure 31).

While H-ATLAS [130] was not specifically designed to probe high redshift galaxies — its central theme is improving the understanding of the local universe ($z < 0.5$) — high-redshift sources have already been a great strength of the H-ATLAS science program. In a pilot study using the first 14 deg$^2$ of ATLAS, a third of the $\sim 6000$ galaxies detected (all above the SPIRE confusion limit with a detection in at least one of the channels at $5\sigma$) lie at $z < 0.5$ [9]. However, candidate high redshift HLIRGs ($L_{\rm IR} > 10^{13} L_\odot$), comprising another third of the catalog (selected at 350 $\mu$m) appear to be identified in the $z \sim 1.5-3$ range with a density of $\sim 150$ deg$^{-2}$ [9]. This redshift distribution and number density is similar to that of the bright SCUBA-SMGs [77, 230, 321, 403]. It is likely that there is substantial overlap between the SCUBA and *Herschel* selected galaxies; although the latter might span a larger (hotter) range of SED types (dust temperatures) than probed by the 850-$\mu$m surveys.

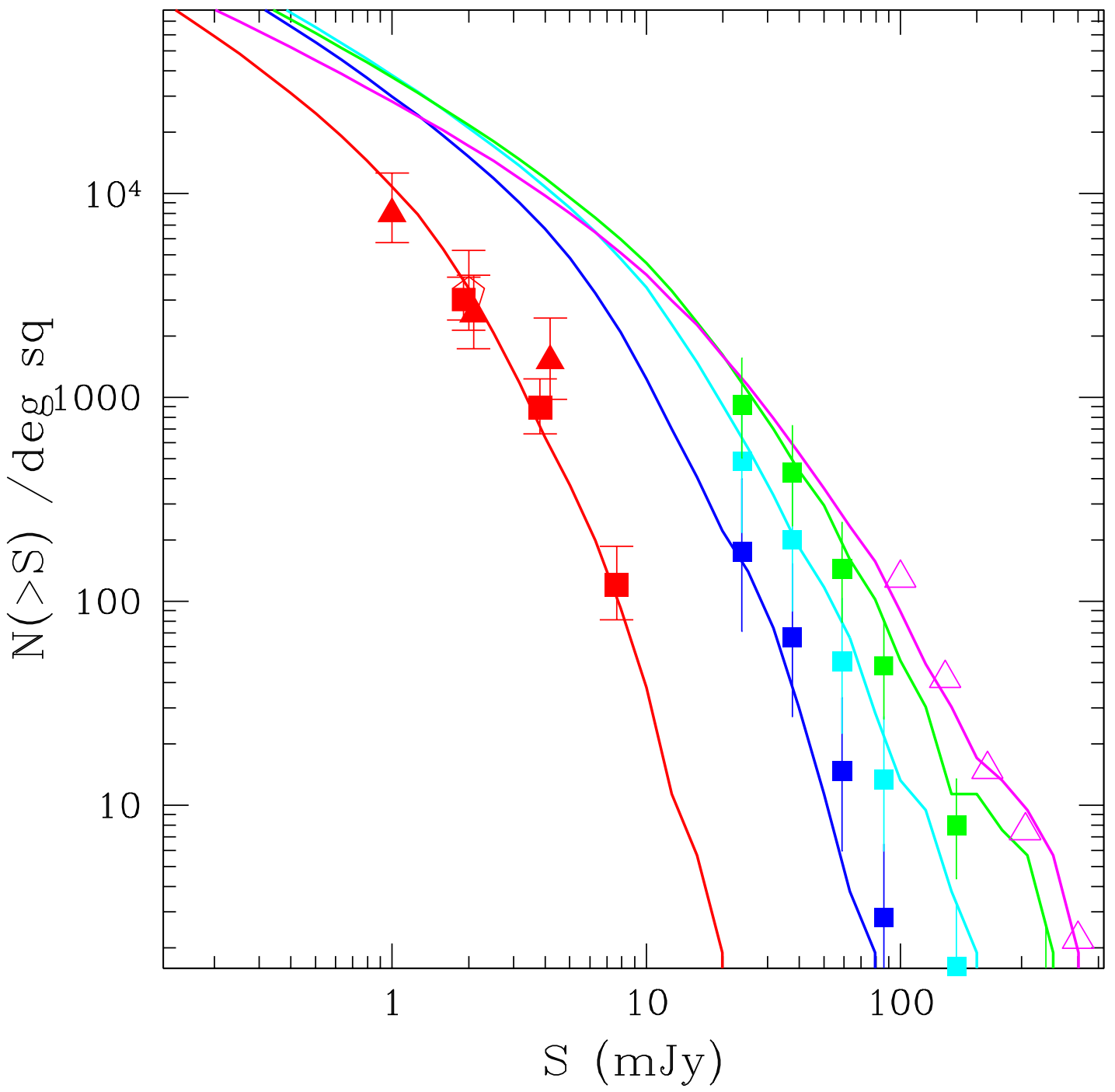

Figure 30: The predicted *Herschel* counts (250, 350, 500-$\mu$m as green, cyan, blue lines) along with SCUBA 850-$\mu$m (red) and *Spitzer* 160-$\mu$m (magenta) in the bivariate luminosity function model of [76], which adopts a similar dust temperature distribution of the sources to the *Herschel*-measured properties (e.g. [9], [140]). The measured counts from [304] are overlaid for comparison, in good agreement with the model.

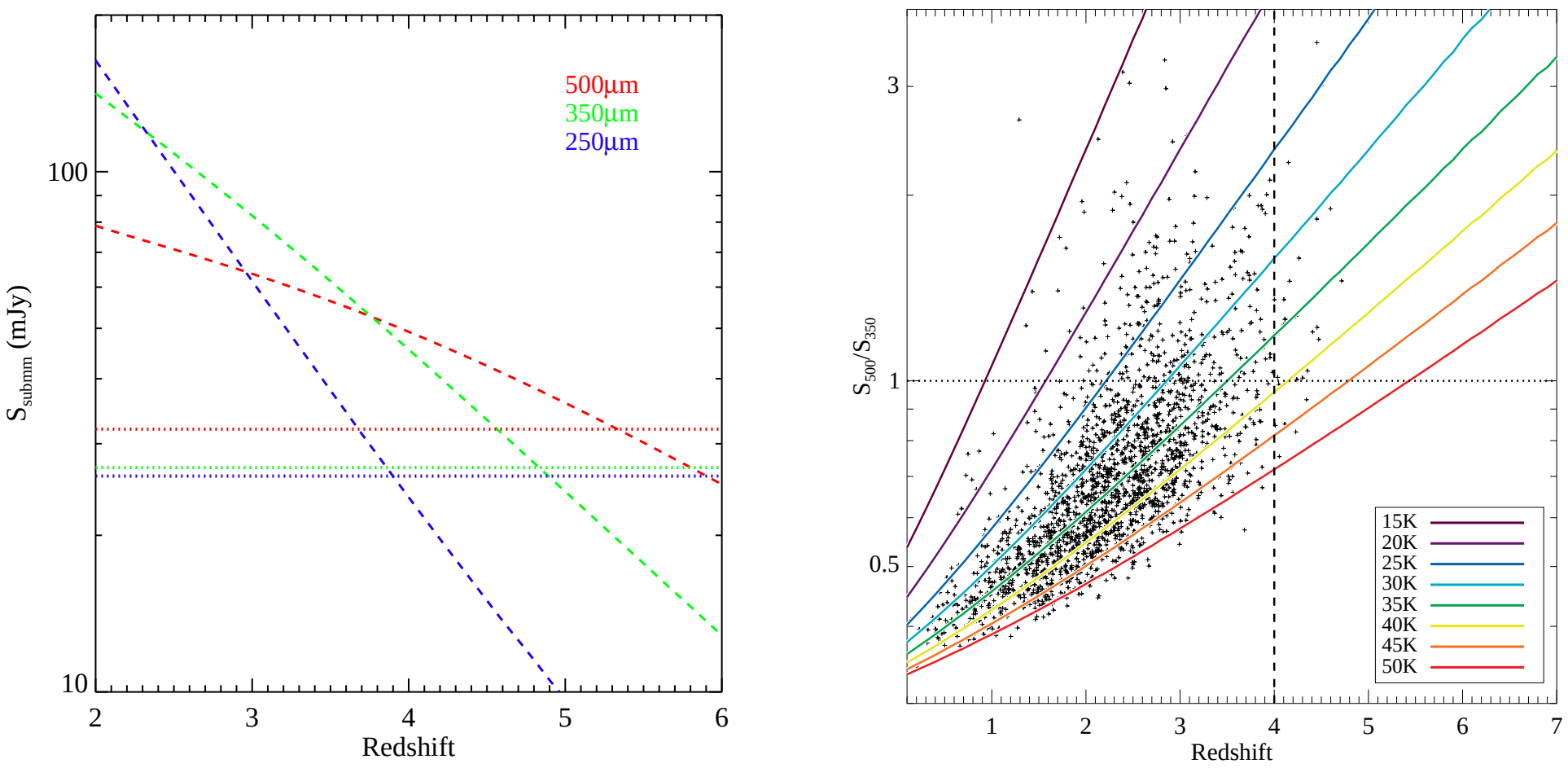

Figure 31: The highest redshift galaxies in HSLS. (*left:*) Submillimeter flux as a function of redshift for a typical SMG SED scaled to $L_{\rm IR} = 2 \times 10^{13}\, L_\odot$. The dotted curves are the 50% completeness limits of HSLS demonstrating that we can detect HLIRGs out to $z \sim 6$. (*right:*) Selecting the highest redshift sources as 500-$\mu$m peakers [320]. The curves show the $S_{500}/S_{350}$ flux density ratio of an SED (single temperature modified blackbody with $\beta = 1.5$) for a given $\rm T_{dust}$ (see color legend) as a function of redshift. The horizontal dotted line shows the criteria used to define a source that peaks at 500 $\mu$m. The data points show a Monte Carlo simulation of 2000 galaxies following a dust temperature distribution of $35 \pm 7$ K and a redshift distribution $z = 2.2 \pm 0.8$. Assuming this conservative scenario, 10% of 500-$\mu$m peakers are at $z > 4$. Scaling from existing results in H-ATLAS, we expect to find $\sim 10^4$ $z > 4$ galaxies.

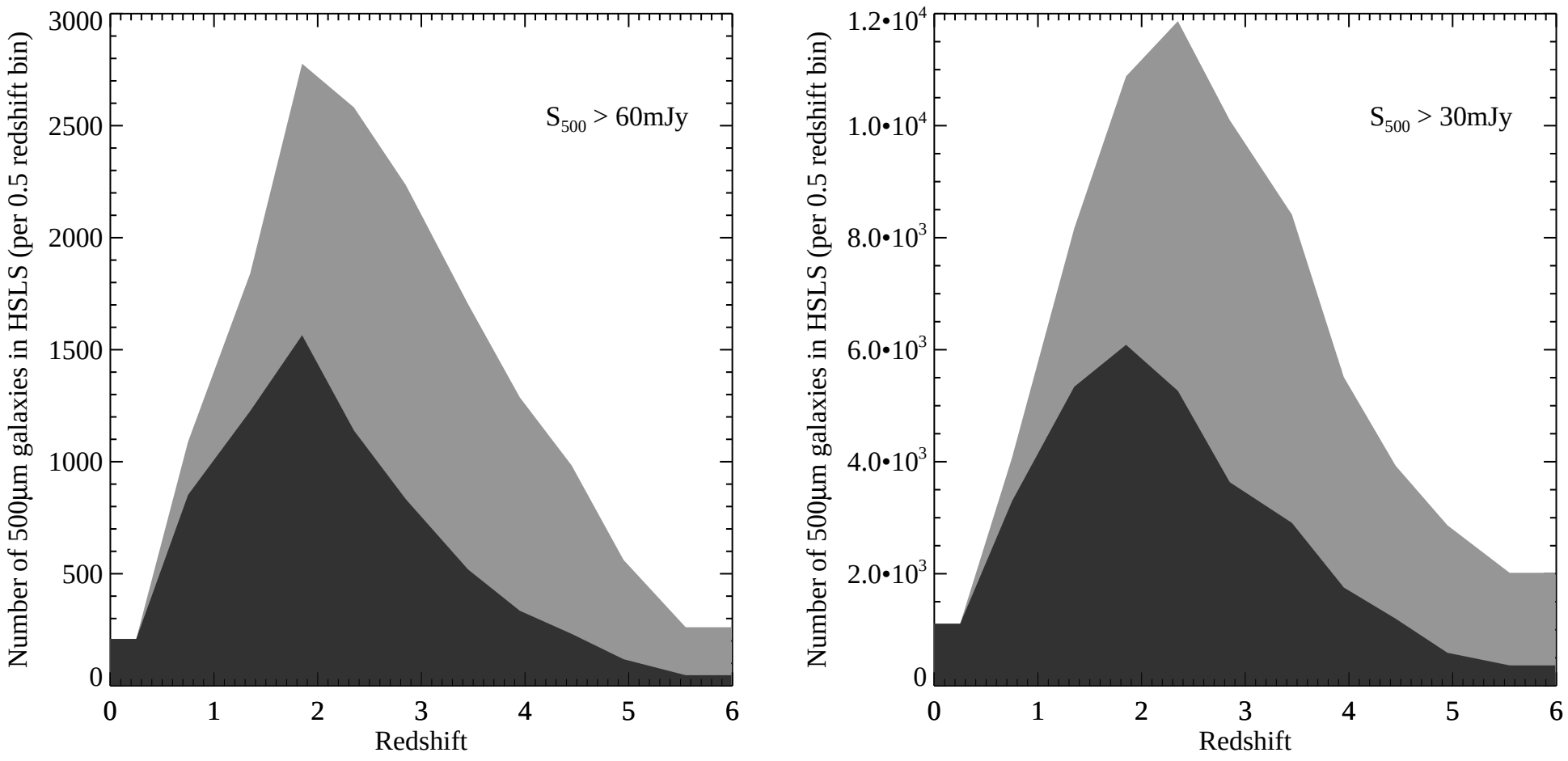

Figure 32: Estimated redshift distribution of blank-field, 500-$\mu$m-selected galaxies from the models of [78]. Note that these estimates do not include the contribution from lensed galaxy populations (see Section 4). These figures show the number of galaxies expected in 4000 deg$^2$ selected above 60 mJy (where HSLS is 100% complete, left panel) and 30 mJy (where HSLS is 50% complete, right panel). The two shaded regions denote different evolutions of the luminosity function at $z > 1$ which are consistent with the total extragalactic background light (see [78] for more details). We anticipate finding thousands of galaxies above $z > 4$ with which we will be able to map out the clustering and environments of the first IR luminous galaxies.

The uniqueness of the HSLS area is its potential to obtain statistical samples of the highest redshift ($z > 4$) SMGs. The development of HLIRG activity at the earliest epochs ($z > 3$) is admittedly not well probed by the current SMG spectroscopic sample, partly due to the radio identification bias (radio detection is required for precise astrometric counterpart identification). The $z > 3$ sources are also not well probed by existing deep surveys such as HerMES, due to source confusion and the low volume covered to observe the few rare objects that are bright enough to be clearly detected above the confusion limit. The 350-$\mu$m selected SMGs from H-ATLAS show no signs of a significant high-$z$ tail beyond $z > 3$ ($< 5\%$ at $z = 3 - 4$, and $< 1\%$ at $z = 4 - 5$). Various models [78, 234, 240, 298] suggest a more significant $z > 3$ component than is inferred from this 14 deg$^2$ field. Part of this discrepancy is likely due to selection effects; the highest redshift sources in SPIRE surveys will be selected at 500 $\mu$m, and may even be undetected at 250 $\mu$m and 350 $\mu$m. The right-hand panel of Figure 31 shows that the majority of $z > 4$ sources have SEDs which peak at 500 $\mu$m or beyond. A simple SPIRE color criterion can thus be imposed on the HSLS catalog to isolate potential $z > 4$ sources and aggressive follow-up efforts at other wavelengths will be employed to confirm candidates and refine this technique. Based on the H-ATLAS results, the HSLS should contain $\simeq 10^5$ of these '500-$\mu$m peakers'; at least $\simeq 10^4$ of these should be at $z > 4$ (Figure 32). Of the sample of six H-ATLAS sources which have had redshifts measured with CO spectrometers, only one is a 500-$\mu$m peaker, and it has a redshift of 4.3.

With the 4000 $\deg^2$ HSLS survey, thousands of distant HLIRGs will be detected. Figure 32 shows the predicted redshift distribution (normalized to the HSLS area) for sources selected at 500 $\mu$m. While the models are largely unconstrained at the highest redshifts (and this in fact is an important science question to be addressed with HSLS) we anticipate detecting $\sim 10,000$ sources above $z > 4$ and at least several hunded at $z > 5$.

Something of a reality check can be obtained by reference to the emerging high-$z$ tail in the millimeter surveys. [93] have published the highest redshift SMG yet at $z = 4.76$, an HLIRG selected from a 0.25 $\deg^2$ field. [100] and [82] have detected galaxies above $z > 4$ with a similar density. Scaling the observed lower limit on the number density of $z > 4$ SMGs to the sensitivity of HSLS, we expect $> 4000$, consistent with the model predictions. The actual number of $z > 4$ SMG that can be studied in details in HSLS will depend on successful deblending of confused sources and on the outcome of detailed followup.

Detailed followup of Herschel-selected high$z$ galaxies will put strong constraints on the highest$z$ specimens, yet there is little question that neither of the two *Herschel* extragalactic programs to date have the area to explore the $z > 5$ HLIRG population in great numbers; nor can they find sources at $z > 4$ with sufficient quantities to establish basic statistical properties such as their spatial distribution and clustering. HSLS will provide sufficiently large numbers of $z > 4$ SMGs to understand the details of enrichment in dusty galaxies at early times, and the large scale structures in which dusty starbursts occur.

### 5.2.1 Lensed SMGs as a Probe of Faint $z > 5$ sources

The strongly lensed SMG populations are emerging as an important result in H-ATLAS [297]. While this is largely its own science case in the HSLS (see Section 4), it is worth noting here that a substantial portion of the high-$z$ HLIRGs discussed above are likely to be lensed by various factors [298]. HSLS is expected to find a very large number of lensed SMGs ($\sim$1200 with $S_{500} > 100$ mJy); models indicate that at least 20 of these sources will be at $z > 5$ [298]. At these bright flux levels, we expect all $z > 5$ sources to be gravitationally lensed. Unlike HLIRGs, such lensed sources allow us to study the fainter samples that will otherwise remain undetectable because of *Herschel* source confusion (see Section 5.3.3 for details related to ALMA followup).

### 5.2.2 How can the redshifts of SMGs be efficiently identified?

Will we be able to calibrate the 250, 350, and 500 $\mu$m color-color diagram to use it as a photo-$z$ estimator, and how well can we do this to establish the sub-mm photometric redshift of an individual galaxy? In H-ATLAS, predictions from [14] for small numbers of sources have come within an average of $\sigma_z = 0.3$, at the 68% confidence level, of the exact CO-line redshift for those sources.

These photometric redshifts were determined using a Monte-Carlo algorithm that yields the probability of producing the fluxes and colours of the sub-mm galaxies at any redshift from mock catalogs of the (sub)-mm sky that adopt the SEDs of a sample of 20 local starbursts, AGN and ULIRGS with a wide variety of spectral shapes [13, 203]. The accuracy of the technique has been experimentally tested for galaxies detected in at least 3 bands in the rest-frame radio to FIR interval, yielding a typical r.m.s. $= \langle (z_{\rm phot}/z_{\rm spec})^2 \rangle^{1/2} \approx 0.25$ over the $0.5 < z < 4.0$ regime [14]. New measurements using only the SPIRE bands for sources with robust spectroscopic redshifts, either CO or IRS *Spitzer* [208, 297], yield an r.m.s. of 0.32 (Figure 33 left panel). Additionally, the inclusion of longer-wavelength (sub-)mm data in the range of 850 $\mu$m to 1.2 mm, radio data, and shorter wavelength (100 and 160 $\mu$m) PACS data can further improve the precision obtained from radio to FIR photo-$z$ to an r.m.s. of $\approx 0.25$ (Figure 33 right panel).

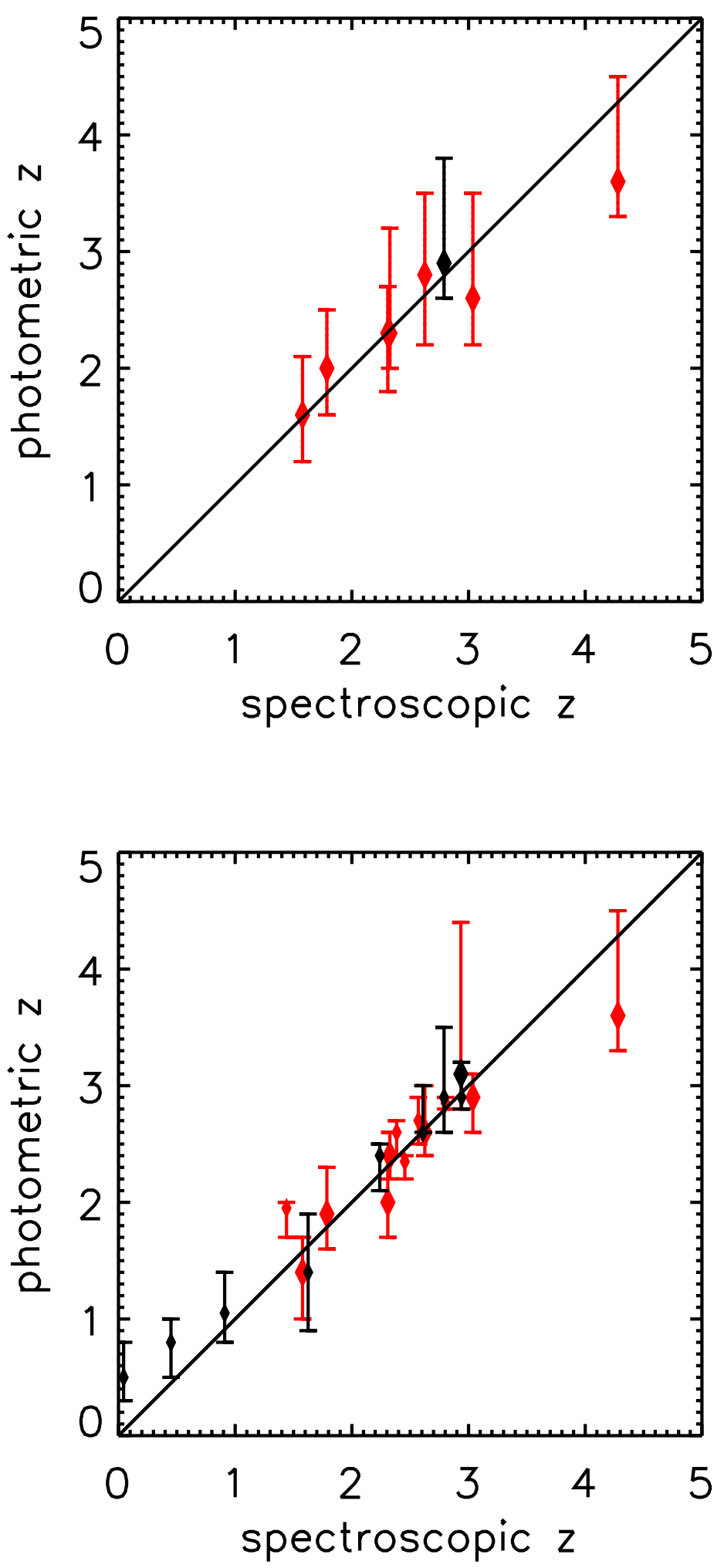

Figure 33: *Top:*Comparison of spectroscopic and photometric redshifts derived from the FIR peak as traced by the SPIRE bands for a sample of 8 galaxies detected by *Herschel*. The sample has robust CO (in red) or *Spitzer* IRS (black) redshifts. The error bars represent 68% confidence intervals in the determination of the redshift. The relationship has an r.m.s. of 0.32. *Bottom:* Comparison of spectroscopic and photometric redshifts derived from the full radio-FIR SED for a sample of galaxies detected by different (sub-)mm telescopes. The sample has robust CO (in red) or optical/IR (in black) redshifts derived from robustly associated counterparts. The error bars represent 68% confidence intervals in the determination of the redshift. The relationship has an r.m.s. of 0.22. Larger symbols correspond to the same galaxies represented in the top panel. Their photometric-redshift determination has been further constrained by additional data in the FIR to radio regime.

Many of the sources detected by HSLS should also be seen in the radio given their likely 250 $\mu$m/radio flux ratios [207]. The radio data will come from the EMU project (a southern all-sky survey to 10 $\mu$Jy rms) which is due to start in 2013 on the Australian SKA Pathfinder, ASKAP. Hence, SPIRE together with radio data should be able to give reasonable redshifts with $\sigma(z)/(1+z) < 0.3$. To use this technique effectively we must, however, know the radio-IR correlation and its evolution at high redshifts ($z > 4$). This is an open question that may only emerge for the most luminous HLIRGs from studying the HSLS itself.

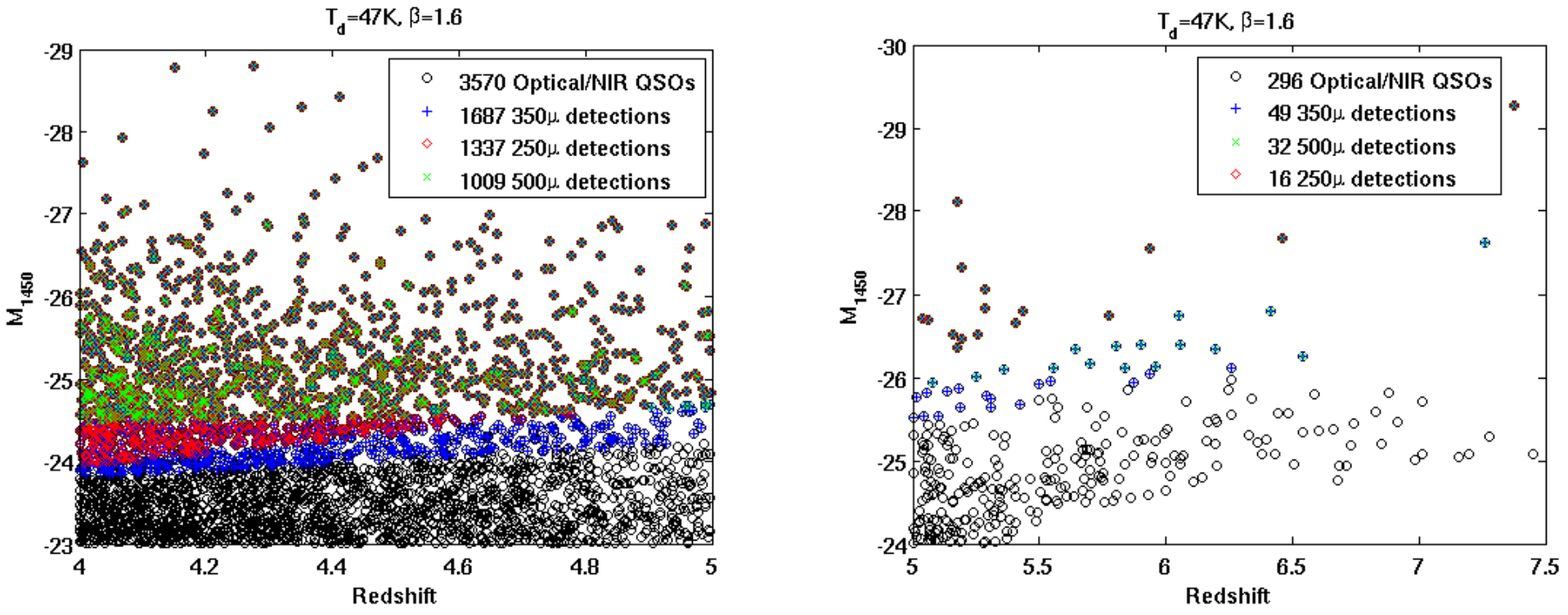

Figure 34: The redshift vs. rest-frame UV magnitude of QSOs in DES/VISTA-VHS with sub-mm detections in HSLS. A best-fit dust model with $T_{dust}$=47K is adopted for the sub-mm SED for $4 < z < 5$ QSOs (left panel) and for the highest-redshift range over $5 < z < 7$ (right panel).

[320] caution that, without priors on the dust temperature distribution, redshifts from SPIRE data alone are difficult due to the degeneracy between redshift and average dust temperature [39, 40]. However, existing *Herschel* surveys (HerMES and H-ATLAS) place statistical constraints on the way in which dust temperature varies as a function of redshift and luminosity, and these can in principle be used to obtain redshift distributions for the HSLS sources. An additional complication may be a significant proportion of QSOs, perhaps especially at the highest redshifts, which may change the SED and strongly affect the small percentages of the highest-$z$ SMGs predicted from our models. We will be able to identify these sources with the available optical data.

We will clearly have the redshift resolution to identify very high-$z$ candidates (with a photometric redshift accuracy of order 0.3 at $z \sim 2$) but we will pursue an aggressive follow-up campaign to get spectroscopy (mainly CO and C+) to confirm the photometric redshifts, at least for a few thousand sources.

### 5.2.3 High-$z$ QSOs

SMGs and QSOs are thought to follow one another in the evolutionary path which terminates with a massive elliptical galaxy. A small fraction of SMGs ($\sim 3\%$), even in deep surveys, are in fact QSOs with lingering star formation allowing them to be detected in the sub-mm [77]. While the most luminous QSOs are rare ($\sim$2 deg$^{-2}$), a 4000 deg$^2$ survey will provide of order 8000 QSOs, of which 25% are likely to be at $z > 3$ (see Section 7 for further discussions). With samples of several thousand sources containing both star forming SMGs and QSOs, the evolutionary relationship between SMGs and QSOs in the early Universe can be further tested and the relative timescales of each phase can be better constrained.

The optical Dark Energy Survey (DES), together with the VISTA Hemisphere Survey (VHS) in the near infra-red, will lead to the detection of unprecedented numbers of bright quasars at high redshifts. Based on current best estimates of the quasar luminosity function at redshifts between 4 and 6, we expect DES and VHS to find $\sim$ 3600 bright quasars ($M_{1450} < -23$) between redshifts 4 and 5 and $\sim$ 300 bright quasars ($M_{1450} < -24$) between redshifts 5 and 7.5 over the total HSLS area of 4000 deg$^2$. This assumes a pure density evolution of the observed $z \sim 6$ luminosity function up to $z \sim 7.5$. Using our current knowledge of the high-$z$ bright QSO SED, we expect a significant fraction of these objects to be bright enough to be detected in all three *Herschel* SPIRE bands in the HSLS.

Figure 34 illustrates the fraction of optically detected quasars between redshifts 4 and 5, and at $z > 5$ these should also be detected in the *Herschel* bands using a FIR QSO SED inferred from observations ($T_{\rm dust} = 47$ K and $\beta = 1.6$). About half the objects ($\sim 1800$) are expected to be detected at 350 $\mu$m, $\sim$ 40% at 250 $\mu$m and $\sim$ 30% at 500 $\mu$m for $z = 4$–5. At redshifts above 5, of the order of 50 bright quasars with $M_{1450} < -24$ will be visible in the HSLS FIR bands; $\sim$ 17% will be detectable at 350 $\mu$m, 11% at 500 $\mu$m and 5% at 250 $\mu$m. The DES, VHS and HSLS data sets together will therefore amass a sample of several thousand quasars with multiwavelength coverage between redshifts 4 and 5. Not only will this allow the study of the QSO SED in detail at these redshifts, but the samples will be large enough to enable statistical studies, for example of the clustering of quasars at these redshifts. These observations will allow us for the first time to constrain the SED of quasars at redshifts above 6 all the way from the optical to FIR and sub-mm wavelengths using samples large enough to draw statistically significant conclusions about the evolution of such objects. The total FIR luminosity of these objects will allow us to constrain star formation rates of the quasar hosts providing us with a direct test of the black-hole mass stellar-mass correlation in the very early Universe. The FIR bright high-$z$ quasars will also serve as ideal candidates for detailed follow-up studies with next-generation telescopes like ALMA.

Radio bright QSOs are also interesting: if they are radio bright ($> 20$ mJy at 150 MHz), they could be used for 21-cm absorption line studies. While such radio sources at $z > 5.5$ are yet to be found, active searches for such sources are ongoing and a handful of sources could easily reveal significant details about the state of the IGM and its reionization history. HSLS may very well provide sources for the low-frequency SKA and other arrays, such as LOFAR and MWA, which aim to study the era of reionization. Some of these radio QSOs, if lensed and are strong enough at mm- or cm-wavelengths, might also be used for molecular absorption line studies in the foreground lens galaxy. Only a very few such systems are currently known and highly studied. HSLS provides adequate volume to find a few more examples of very rare events.

## 5.3 HSLS High-$z$ Science Drivers

### 5.3.1 Environmental dependence of SFR

The environment on various scales plays an important role in the process of galaxy formation. Perhaps the most striking observational evidence for this is that clusters today have a much higher fraction of early-type galaxies than is found in the field. Likewise, the successful physical models of galaxy formation predict a very strong co-evolution of galaxies and dark-matter halos.

There are many ways of determining the role of environment observationally: one can directly examine the galaxy populations (e.g. the SFR distribution functions) in different environments; one can explore the environments of galaxies in different luminosity classes; one can use the clustering of particular galaxy populations to infer the source bias, and thus the mass of the dark matter halos in which they are located through approaches such as the halo model [91], to relate these to their present-day descendants; or one can use fluctuation analysis to constrain such models. All these give us the same basic requirement, a volume sufficiently large to sample enough of the environments of interest, and sufficiently deep to constrain the populations of interest. Simulations show that we can discriminate between hosts of different halo masses for different sub-classes of galaxies and compare the clustering of the FIR galaxies with quasars from optical studies. Our measurements will surpass all previous FIR and sub-mm observations.

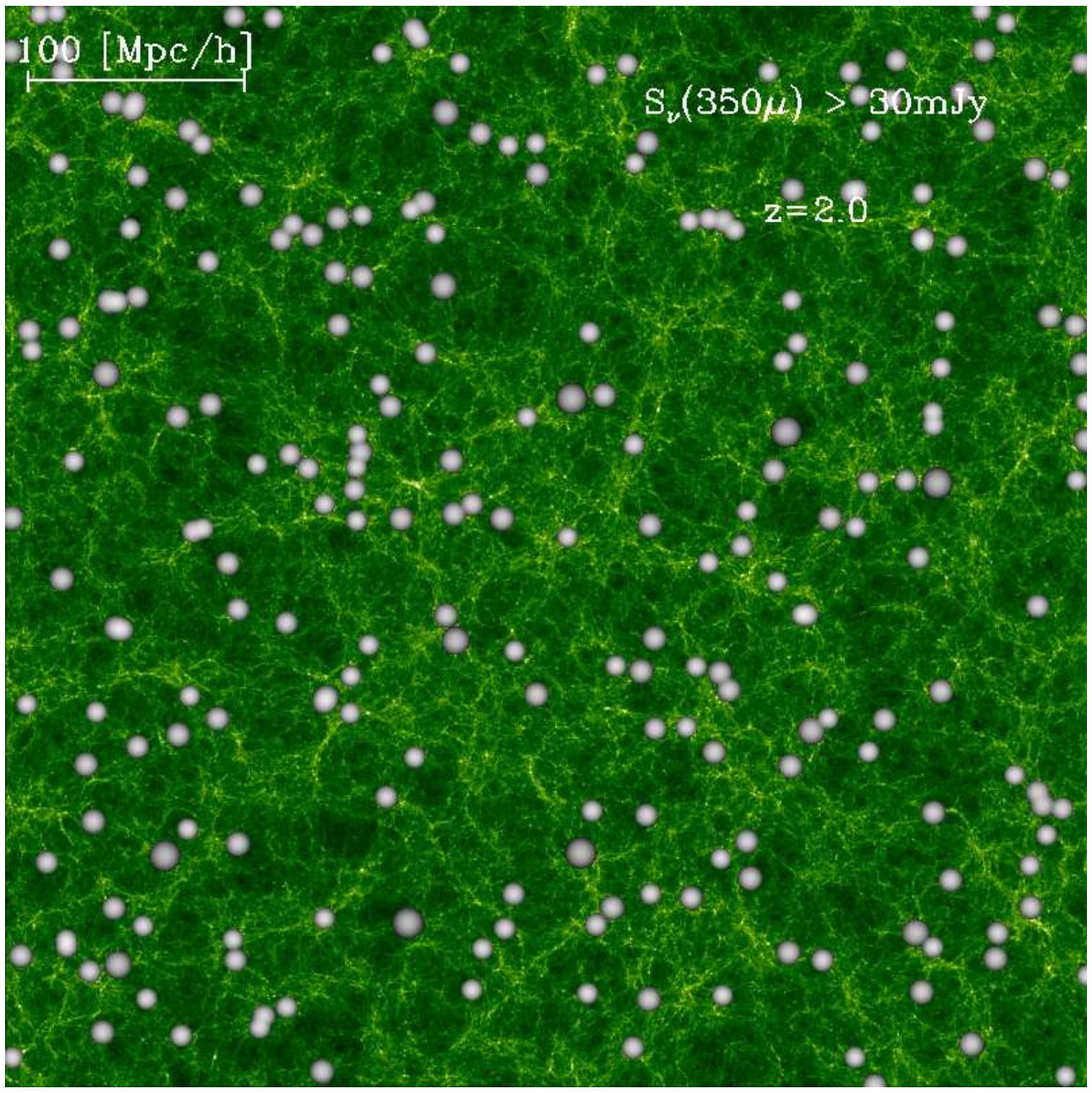

Figure 35: The spatial distribution of 350 $\mu$m sources and dark matter at $z = 2$, taken from the semi-analytic galaxy formation model GALFORM combined with the Millennium $N$-body simulation. Galaxies with $S_{350} > 30$ mJy are shown in white, and dark matter is shown in green. This is a slice of the simulation box which is 500Mpc/h on each side and 50Mpc/h deep.

### 5.3.2 Galaxy clusters and proto-clusters

Studies to date have suggested that the most FIR-luminous, distant galaxies are also clustered very strongly [40, 91, 148] (see Figure 35 for a simulation). Taken at face value, this observation would indicate that they occupy the largest dark-matter halos ($\sim 10^{13} M_{\odot}$). However selection of the most FIR-luminous galaxies may be prone to finding the brief periods when systems of lower matter overdensity are rapidly evolving [158, 344]. Indeed, burst timescales as short as 10 Myr have been estimated for some SMGs [372]. Blain et al. [40] calculated that, complex biases aside, SMGs have a clustering length that is consistent with a form of evolution that ensures that their properties subsequently match the clustering length typical of evolved red galaxies at $z \sim 1$ and finally of rich clusters of galaxies at the present epoch. Chapman et al. [74] suggested that SMGs may also often reside in less overdense environments (and that ULIRG clustering could be dominated by the coordination of highly active periods across modest mass environments).

The *Herschel*-SPIRE surveys HERMES and H-ATLAS are now providing a census of the dust obscured star formation and its evolution from $z = 0$–3. Beyond $z > 3$, the area of these surveys limits the statistical study of the most luminous ($L_{\rm IR} > 10^{13} L_{\odot}$) SMGs. These massive star-forming systems are expected to trace the most massive halos and therefore their distribution should trace the underlying large scale structure. With a large enough area/volume the distribution and clustering of SMGs at $z > 3$ can be used to investigate the role and effect of environment on the formation of massive galaxies at early times. The expected numbers of sources in HSLS will allow for the first time statistical studies of the clustering of SMGs beyond $z > 3$ to investigate the correlation between the most massive star-forming galaxies and the layout of the underlying large scale structure.

Even at $z < 3$, however, HSLS has novel applications not feasible with studies in H-ATLAS. These involve the role of SMGs and massive clusters and proto-clusters. [64] recently claimed a detection of a $10^{15} M_{\odot}$ cluster with the SZ effect at $z = 1.1$, the only one of such a mass at $z \sim 1$ in the first 200 deg$^2$ of the SPT survey. In HSLS, estimates are such that we should see anywhere between a few to 20 such clusters, assuming the standard $\Lambda$CDM cosmological model at $z > 1$. At $M > 5 \times 10^{14} M_{\odot}$, we expect 80 to 150 clusters in HSLS areas. These numbers are sufficiently large that one can study the environmental dependence of SFR from large voids at low-redshifts to $10^{15}$ $\mathrm{M}_{\odot}$ clusters at $z > 1$, a range spanning ten or more orders of magnitude in local density.

Another possibility is to search for $z = 2$ SMG overdensities or proto-clusters. At $z > 1.5$, we predict that the HSLS will find $\sim 200$ 'proto-clusters', unvirialized over-densities extending on scales of 10–20 Mpc which will eventually become the virialized clusters we see around us today. These proto-clusters have so far only been found as clusters of sources around high-redshift radio galaxies and quasars. By observing these systems in other wavebands, we will be able to learn much about the evolution of these structures and their populations. The Dark Energy Survey will trace 3D structure in the Universe out to z $\sim$ 1.5. By determining where the HSLS galaxies fall within the structure traced by DES, we will learn more about the interaction between environment and galaxy evolution. In Stripe 82, where we cover 700 deg$^2$, the required optical data is provided by SDSS.

### 5.3.3 Using lensed SMGs for deep, detailed studies with ALMA

The sample of $\sim 1200$ strongly lensed high-$z$ sub-mm sources from HSLS will be fundamental to follow-up studies of the properties of ULIRGs from $z \sim 1$–6 using ALMA. The lensing magnification by a factor 10 or more will allow for comprehensive studies of the major starbursts, their evolution and their contribution to the growth of galaxies during the peak epoch of buildup. The enormous effective gain in telescope time due to lensing, a factor $\sim$100 or more, will permit studies with ALMA and other facilities which are out of reach for unlensed high-$z$ sub-mm sources.

Even samples of strongly lensed SMGs from H-ATLAS and HerMES as large as $\sim 150-200$ will not provide the statistically significant samples for the various classes of high-$z$ U/HLIRGs that are required for systematic studies of their properties and evolution with redshift. Important rare subclasses will also begin to be uncovered by HSLS (e.g., the highest redshifts, $H_2O$ mega-masers, strongest magnifications).

The example of similarly bright, strongly-lensed sub-mm QSOs (APM08279, the Cloverleaf, FIRAS10214+4724; see e.g. [379]) shows the impact of strong gravitational magnification for high-$z$ studies of sub-mm populations. These three sources were found from the all-sky *IRAS* survey, and before *Herschel* they were still unique, except for the serendipitously discovered 'Cosmic Eyelash' [384] — a cluster-lensed star forming ULIRG at $z = 2.3$. These sources played an important role in the development of high-$z$ sub-mm/mm astronomy through pioneering detections of CO lines, and high sensitivity studies allowed by the strong magnification, such as high-angular-resolution observations of CO lines down to the 100-pc scale, and the first detections at $z > 1$ of various molecular lines much fainter than CO, such as $H_2O$, HCO+, HCN, HNC, CN, and $^{13}$CO.

The order-of-magnitude increase in sensitivity with ALMA, focused on these *Herschel* lensed-SMGs, will allow for detection of large sets of weak molecular lines of several tens of molecules, including deuterated species. This will allow detailed comparisons with local galaxies of the interstellar chemistry in high-$z$ ULIRGs, including XDR and shock chemistry. Maps of the strongest lines will probe spatial variations of the chemistry. Studies of a number of isotopologues such as $^{13}$CO, C$^{17}$O, C$^{18}$O, H$^{18}$O, will allow a check on nucleosynthesis and its evolution, and in addition will permit the study of radiative transfer effects when lines of the main isotopologue are optically thick. High quality line-profile maps (e.g., CO, C+, $H_2O$, CI) will be constructed to probe details of the dynamics in mergers, rotating disk-like components, and galactic or AGN outflows. Very high angular resolution CO and C+

mapping of starbursts will be performed to resolve their cores, as well as to probe the possible contribution of the AGN molecular torus.

### 5.3.4 Dust properties at high $z$

The mechanisms that dominate the production of dust at early times are not well understood [161]. The SPIRE bands alone cannot constrain the dust mass or properties very well, but with detailed follow up studies (IR/sub-mm spectroscopy) of $z > 4$ SMGs, we can probe its composition and put constraints on the mechanisms capable of producing the bulk of the dust.

### 5.3.5 Luminosity function evolution

A primary goal is to determine the total or bolometric FIR luminosity function of galaxies over the redshift range $z < 3$. For this analysis we will use galaxies detected in *Herschel* images, and use either sub-mm or multiwavelength-based photometric redshifts. Monochromatic luminosity functions will be a first result, extending existing studies out to $z \sim 1$ with HerMES and H-ATLAS with enough source statistics all the way to $z \sim 4$, to be followed with modelling of the SED and estimation of bolometric luminosity functions. Later analysis will model the relative contribution of AGN and star formation to the bolometric emission, to investigate processes as a function of galaxy types (with types being determined both from *Herschel* classifications and optical/NIR data in our fields as available from SDSS and in future with DES).

## 6 The Local ($z < 1$) Sub-mm Universe

The top-level science goals of low-$z$ studies with HSLS are:

- The environmental dependence of dusty star-formation for large (150 Mpc) voids to massive ($M > 10^{15} M_{\odot}$) galaxy clusters
- The search for $\sim 500$ cold, luminous galaxies at low redshifts
- The identification of *Planck* extragalactic point sources

### 6.1 Introduction

It is now established that the epoch spanning $z < 1$ witnessed the most dramatic transformation in the processes driving stellar and black hole mass assembly in galaxies since reionization. During this epoch, the comoving star formation rate density declined by over an order of magnitude [32, 237], and the number of bright sub-mm sources per square degree declined by at least four orders of magnitude. These declines were mirrored by a dramatic rise in the number of galaxies with masses $> 10^{11.5} M_{\odot}$, and the emergence of a prominent 'red sequence' of passively evolving galaxies spanning a wide mass range [31, 141].

It seems clear therefore that this epoch witnessed both a strong decline in the comoving density of stellar assembly and BH mass accretion, as well as a shifting of modes, from one in which intense bursts of star formation and near Eddington limited accretion contribute significantly at $z > 1$, to one in which they contribute negligibly at $z = 0$. Such a 'paradigm shift' in the processes driving galaxy evolution is reflected in the latest generation of N-body and semi-analytic models for the formation of galaxies and large-scale structures, which suggest that the most dramatic changes in the density contrast of the underlying dark matter distribution also occur over $z < 1$.

### 6.2 Multiwavelength Surveys of the $z < 1$ Universe

Why did this decline in galaxy assembly processes occur? What drove the shift from obscured starbursts to quiescent star formation? What is the relationship between baryonic and 'dark' matter since $z = 1$? To obtain a complete picture of the evolving Universe over the last half of its history, we need answers to these questions. And while we have made immense progress in obtaining such answers over the last few decades, this progress has slowed in the last few years, due mainly to a serious imbalance in our capacity to survey the Universe at different wavelengths. Surveys in the optical/near-IR ($0.3 < \lambda(\mu\text{m}) < 1.0$), sampling emission from passively evolving stellar populations, unobscured star formation and QSOs, can be performed straightforwardly from the ground in multiple bands, and benefit from relatively high spatial and spectral resolution. Optical surveys can thus detect large samples of passively evolving systems with stellar masses substantially below $M^*$ out to $z = 1$, and distinguish morphologies, stellar ages and metallicities. By contrast, far-IR surveys, spanning $50 < \lambda(\mu\text{m}) < 1000$ and sampling emission from dusty star formation and AGN activity, must in general be performed from orbit, and as a result are usually of much lower effective sensitivity and accuracy, with limited spectroscopic complementary data. To cite three examples;

- (1) The *IRAS* 60 $\mu$m surveys could detect moderately luminous obscured AGN up to

$z = 0.5$, but were relatively insensitive to starbursts above $z = 0.2$.

- (2) Luminosity and dust temperature estimates from ground-based 850 $\mu$m point are uncertain by up to a factor of five, even if the redshift is known, and
- (3) *IRAS* only detected a few hundred sources over the entire sky at $z > 0.2$.

Immediate deliverables from IR surveys are thus cruder than those from optical surveys[7]. Indeed, we currently stand in the rather odd position of knowing more about the far-IR Universe at $1 < z < 3$ than we do at $0.4 < z < 1$!

This discontinuity in our knowledge is best illustrated by a comparison of recent results on galaxy assembly at $z < 1$ from optical and far-IR surveys. From optical surveys we obtain a reasonably good understanding of the passively evolving Universe. We can track the emergence of the red sequence as a function of redshift and environment, and determine the fraction of stars that have assembled onto the red sequence since $z = 1$ [31]. We can also monitor the relative number densities of star forming galaxies vs. passively evolving galaxies, and so determine what fraction of stars in massive red galaxies assembled via *in situ* star formation, and what fraction assembled via minor mergers [141, 218]. We can approximately track the dependence of galaxy assembly processes on environment; examples include constraining separately the halo mass evolution of red sequence and blue cloud galaxies[317, 369], and showing that major starbursts are triggered as galaxies enter the outskirts of clusters, but then are 'quenched' as galaxies move from the outskirts of clusters to their central regions [258, 377].

Conversely, our understanding of the far-IR Universe at $z < 1$ is relatively coarse. We know there is strong positive evolution in the IR luminosity function with increasing redshift, apparent even from the IRAS surveys over $z < 0.2$ [343, 401], and confirmed by ISO [139, 337] and *Spitzer* [17, 237, 255, 312] surveys up to at least $z = 1$. We also know that this evolution is strongest at the bright end of the LF, and *may* be accompanied by a systematic change in far-IR SED shape [140, 364]. We do not however have any real constraints on the environment or mass dependence of the evolution of the IR LF with redshift, since *IRAS* was relatively insensitive to LIRGs at $z > 0.2$ and since the *ISO* and *Spitzer* surveys were performed over relatively small areas of a few tens of square degrees. So, for example, we have only crude halo mass constraints for a few samples of LIRGs scattered over $z < 2$ [40, 148, 252, 410], which, although they clearly show there must have been a major change in environmental richness of LIRGs over $z < 1$, set no constraints on how this change occurred.

Current or planned far-IR surveys will do little to alleviate this imbalance. The SCUBA-2 survey program, which includes projects of order 1000 square degrees, will be an important resource for the bright end of the ULIRG LF at $z > 1$, due to the favorable negative $K$-correction at sub-mm wavelengths, but will do little to advance our knowledge of the $z < 1$ Universe due to the relative insensitivity of 850 $\mu$m surveys to dusty starbursts in this wavelength range. Neither will the *AKARI* [219, 292] far-IR survey program be of much help; *AKARI* has much better spatial resolution than *IRAS* but offers no improvement in sensitivity, meaning that the core problem of insensitivity to LIRGs at $z > 0.2$ remains. The existing *Herschel* imaging surveys do go some way to helping, but suffer the critical problem of not covering enough sky area. HerMES and H-ATLAS *put together* contain approximately 200 (2100) DM haloes of mass $10^{14}M_\odot$ ($10^{12}M_\odot$) or greater [288][8] in the $0 < z < 1$ volume. meaning that, although they can crudely track environmental trends with redshift, they are limited to doing so in redshift slices of width $\Delta z \simeq 0.1$, and with no subdivision by (e.g.) galaxy mass or far-IR luminosity. Put another way, optical surveys such as SDSS, DES and BOSS, which are expected to provide the first measurements of BAO and the dark energy EoS, cover of order several thousand square degrees. As of now, we have no far-IR survey of comparable size to these new optical efforts.

[7]We focus here on the dichotomy in efficacy between optical and far-IR surveys, though of course similar comparisons could be made at all wavelengths from $\gamma$-ray to long-wavelength radio.

### 6.3 The Power of a 4000 Square Degree Survey with Herschel

The proposed 4000 square degree *Herschel*-SPIRE Legacy Survey will be the 'paradigm shift' capable of constraining the duty cycle of dusty activity in assembling galaxies, determining why it diminishes with redshift and if it changes with environment, galaxy mass and luminosity. It is no exaggeration to say that such a survey will be revolutionary: just as *IRAS* transformed our view of the infrared universe by unveiling a huge new population of IR-luminous galaxies whose luminosity function showed the strongest evolution with redshift that had been seen until that point, HSLS will transform it yet again, by placing our understanding of the cosmological evolution of IR-luminous galaxies on the same detailed footing as has been obtained for passively evolving galaxies from optical surveys. This transformation will be achieved by the enormous increase in areal coverage: the survey will be nearly an order of magnitude larger than any existing single *Herschel* survey, and a factor of six bigger than all current *Herschel* imaging surveys put together. It will also be more than twice the size of any planned ground-based sub-mm survey. HSLS will be the perfect complement to the next generation of optical surveys such as DES/BOSS/KIDS in the optical, and in the radio with SKA. This project will be the culmination of the *Herschel* survey program, the final, lowest tier in the classical 'wedding-cake' survey design strategy, and will provide a legacy for studies of the $z < 1$ Universe that will last for several decades.

The specific issues such a survey will address for the $z < 1$ Universe include: (1) the environmental, mass and luminosity dependence of the IR luminosity function over $0 < z < 1$, (2) an extremely robust measurement of the local IR luminosity function, (3) a search for 'cold' luminous galaxies that previous surveys may have missed, and (4) revealing the nature of the *Planck* extragalactic point source population.

#### 6.3.1 Obscured Star Formation & Galaxy Assembly over $0 < z < 1$

The core goal of a 4000 square degree *Herschel*-SPIRE survey will be to investigate the causes behind the dramatic decline in the role of dusty star formation in assembling galaxies between $z = 1$ and $z = 0$. Existing *Herschel* surveys such as HerMES and H-ATLAS are setting precise constraints on the fraction of stellar mass that is assembled by obscured starbursts, 'quiescent' star formation and dry mergers as a function of redshift. A 4000 square degree survey will be the next logical step; further subdividing these constraints by environment, galaxy mass, far-IR luminosity and far-IR SED shape. Put another way, it will produce a mapping of far-IR colours and luminosities onto all regions of the Red Sequence,

[8]These halo masses correspond approximately to those of a moderately rich Abell cluster, and the Milky Way, respectively.

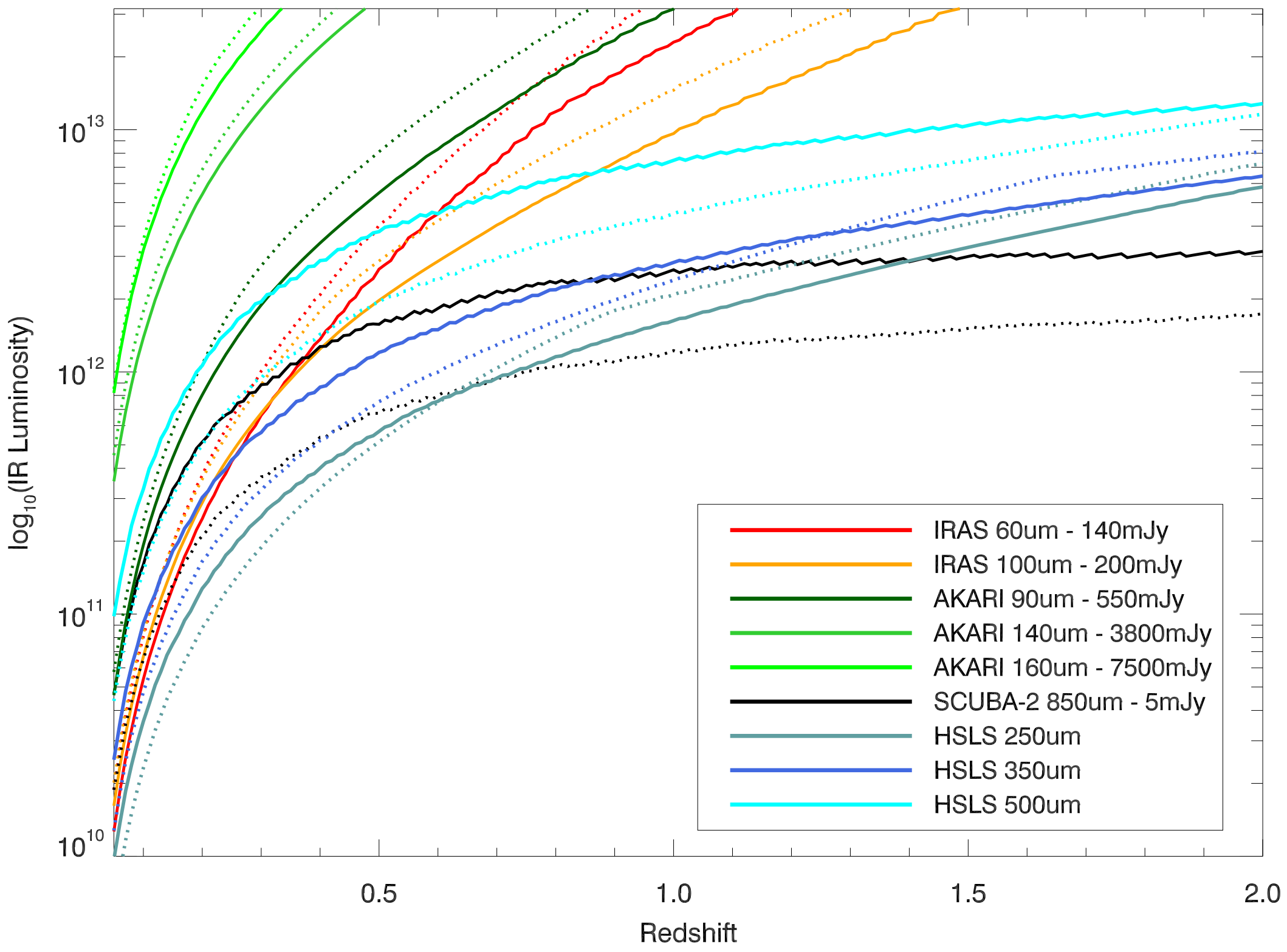

Figure 36: Limiting IR luminosity as a function of redshift for *IRAS*, *AKARI*, SCUBA-2 and HSLS, assuming dust heated purely by star formation. The solid and dotted lines show two different geometries; solid lines assume a 'compact' model [132], while dotted lines are for 'extended' star formation [133]. HSLS is sensitive to an order of magnitude or more fainter star formation than *AKARI* in at least two bands over $0 < z < 1$ for any mode of star formation. Compared to *IRAS*, HSLS can detect at least 0.4 dex fainter starbursts at $z > 0.3$ at 250 $\mu$m, and at both 250 $\mu$m and 350 $\mu$m at $z > 0.5$. Compared to SCUBA-2, HSLS is more sensitive to 'extended' star formation at $z < 0.5$, in two bands, while being comparably sensitive at $z > 0.5$. For 'compact' star formation, HSLS clearly outperforms SCUBA-2 over $0 < z < 1$. HSLS will thus dramatically advance our understanding of all modes of dusty star formation at $z < 1$.

Blue Cloud, and Green Valley as a function of environment and redshift.

We will probe dusty star formation to an order of magnitude fainter than any non-*Herschel* survey at $z < 1$, detecting all (starburst dominated) ULIRGs at $z < 1$ and all LIRGs at $z < 0.5$, the epochs at which their respective luminosity functions show the largest changes in their contributions to the comoving IR luminosity density. In comparison, *IRAS* cannot detect LIRGs above $z = 0.2$, and all existing or planned ground-based sub-mm surveys cannot detect LIRGs below $z = 1$. Our sensitivity to the evolving role of environment at $z < 1$ is exquisite; a 4000 square degree survey will contain approximately 1800 (200,000) DM haloes of mass $10^{14}M_{\odot}$ ($10^{12}M_{\odot}$) or greater [288][9]. The resulting sensitivity to environmental effects with redshift is unprecedented. HSLS will thus contain enough $10^{14}M_{\odot}$ haloes to track their evolution in $\Delta z \simeq 0.02$ bins (the same width as used in COMBO-17 [317]), and with enough $10^{12}M_{\odot}$ subhaloes to examine the far-IR luminosity evolution of individual sources *as a function of morphological type* within each redshift bin. We will thus determine when the transition of LIRGs from field to clustered environments occurs, and whether there is a redshift evolution in the dependence of starburst triggering on cluster infall. At the highest mass end, HSLS will contain approximately 500 $10^{15}M_{\odot}$ DM haloes at $z < 1$; these correspond to the hosts of the most massive Coma/Perseus class clusters seen in the local Universe. While existing *Herschel* surveys will contain small numbers of such objects, HSLS will be the first survey capable of tracking their evolution with redshift in a meaningful way. It is in this mass regime that observational constraints are potentially of the greatest discriminatory power; ref. [288] suggest that there is little evolution in the comoving space density of the most massive DM haloes with redshift, while observations indirectly suggest that major starbursts are still triggered in such haloes down to at least $z = 0.5$ (based on stellar ages in lower redshift clusters [375]). With a 4000 sq deg survey we will test this directly; subdividing into $\Delta z = 0.05$ thick slices gives $\sim 25$ $10^{15}M_{\odot}$ DM haloes per slice, sufficient to test exactly when the most massive DM haloes stop hosting obscured starbursts.

### 6.3.2 The Local Far-IR Luminosity Function

A secondary goal will be to make an exhaustive measure of the morphological, environmental and luminosity dependence of the $z < 0.1$ far-IR luminosity function. This will serve as a benchmark reference to quantify subsequent evolution with redshift. We expect to detect and provide UV to sub-mm SEDs for of order 100,000 sources at $z < 0.2$, an order of magnitude increase over H-ATLAS. Morphological information is readily available for these sources, either already published (e.g. [217]), or via a planned extension of the 'Galaxy Zoo' project to the southern hemisphere.

This provides two key gains. First, we can subdivide efficiently by environment; a 4000 square degree survey will contain of order 450 clusters of Abell class 1 or richer, providing an outstanding benchmark of the low redshift overdense Universe. Just as important is the *under*dense Universe; modern $N$-body simulations suggest that large (> 200Mpc diameter) voids are common at low redshifts, and since a single such void subtends ∼30 degrees on the sky, a survey of several thousand square degrees is needed to study variations in their properties. Second, we will quantify the bright end ($> 10^{11.5}L_{\odot}$) of the far-IR luminosity function for the first time. Objects with such high luminosities are intrinsically rare, with (for example) only 0.026 LIRGs per sq. deg in the latest IRAS redshift surveys [401]. Just 14 such objects are thus expected to be present in the H-ATLAS survey's 550 sq. deg., a number which is clearly inadequate to constrain even the slope of the bright end of the far-IR LF. A 4000 sq. deg. survey, in contrast, will include ∼100 such objects, allowing the shape of the LF to be constrained, and thus direct comparisons with the high redshift sub-mm galaxy population.

[9] These halo masses correspond approximately to those of a moderately rich Abell cluster, and the Milky Way, respectively.

### 6.3.3 Cold, Luminous Galaxies at Low Redshift

There is emerging evidence for 'new' populations of star-forming galaxies at high redshift, characterized by one or more of colder sub-mm colours, higher fractions of neutral polycyclic aromatic hydrocarbons (PAHs), and more extended radio emission compared to low-redshift objects of comparable total IR luminosity [75, 149]. This has been interpreted as implying a softer, less intense ISRF and a colder characteristic dust temperature, which themselves imply that star formation may be more extended in these objects compared to local examples. If confirmed, then the existence of these 'cold' starburst galaxies would have interesting consequences [83]. For example, it would translate to a lower star formation rate per unit sub-mm luminosity, alleviating the need for semi-analytic models to propose exotic solutions such as 'top-heavy' IMFs to explain the observed number of distant sub-mm sources without overpredicting the number of stars formed [25].

The alleged 'newness' of cold, luminous galaxies (CLGs) at high redshift is however based on the absence in the *IRAS* catalogs of a significant population of such sources at $z < 0.3$. *IRAS* however operated between 12 $\mu$m and 100 $\mu$m, wavelengths at which CLGs are expected to be at best only averagely bright. So, as can be seen from Figure 6.3, *IRAS* would be incapable of detecting them. Neither would we expect such systems to be detected in the *AKARI* all-sky surveys at any wavelength, for similar reasons. Since we expect systems with $L_{IR} > 10^{11.5} L_\odot$ to be intrinsically rare at low redshifts (with at most $\sim 1$ per 10 square degrees at $z < 0.4$), irrespective of their dust temperature, we would not expect CLGs to appear in existing sub-mm surveys either, as they typically span areas of much less than a square degree. Upcoming 10-100 square degree ground-based 850 $\mu$m surveys, such as S2CLS, will detect such sources but will be incapable of weeding them out from the general LIRG population as they lack wavelengths near the peak of the IR SED. Finally, H-ATLAS lacks the area to find more than a handful of them, unless they are very numerous.

A very large area, shallow far-IR survey with SPIRE is the perfect way to confirm or refute the existence of CLGs at low redshifts. The potential for such a survey to make a landmark discovery of a hitherto unknown low-$z$ population is neatly demonstrated by results from targeted sub-mm observations of local galaxies — the SLUGS survey [85, 126] found four objects at $z < 0.1$ with sub-mm luminosities comparable to the most luminous low-redshift ULIRGs, but with total IR luminosities well below the ULIRG limit, at around $L_{IR} = 10^{11.2} L_\odot$. These properties imply very large masses of cold dust — $M_{dust} = 10^{8.5} M_\odot$ and $T_{dust} \simeq 30$ K — presumably heated by the ambient ISRF rather than a compact starburst or AGN. If CLGs are even one tenth as numerous as 'ordinary' LRGs and ULIRGs of comparable total IR luminosity, we may thus expect that a 4000 sq deg survey would detect several hundred CLGs up to $z = 1$ in at least two *Herschel* bands (Figure 6.3).

### 6.3.4 Dust in brightest cluster galaxies (BCGs)

X-ray observations with the *Chandra* and *XMM-Newton* satellites have provided a breakthrough in the study of feedback in massive elliptical galaxies located at the centers of groups and clusters of galaxies [276, 314]. Whereas previous lower resolution X-ray studies of cluster cores provided clear evidence for short gas cooling times in many systems, the large implied cooling rates were found to be reduced by over an order of magnitude with higher resolution *Chandra* and *XMM-Newton* data which agree much better with the star formation rates computed from H$\alpha$ in brightest cluster galaxies (BCGs). Moreover, the high-resolution *Chandra* images have revealed cavities, ripples and other features that are thought to be caused by intermittent AGN activity [38, 142, 156]. This has given rise to the new paradigm where the cooling intracluster medium (ICM) in cluster cores is largely balanced by episodic heating by the AGN.

To obtain a complete picture of AGN feedback in cluster cores, it is necessary to fully understand how the AGN communicates its energy to the hot ICM. A potentially important channel for AGN feedback is the heating of dusty gas observable in the far-IR. Studies of BCGs with *Spitzer* have shown that some of the most luminous X-ray clusters have significant far-IR excesses implying star formation rates that might be consistent with radiative cooling of the hot cluster gas [120, 134, 327], which is supported by a recent study of three cooling flow clusters with Herschel [131]. A critical step to assess the role of dust in AGN feedback is to determine the origin of the dust; i.e., whether it arises from the stellar population of the BCG or from external sources.

For this purpose it is necessary to obtain far-IR luminosities for a large, representative BCG sample such as will be possible with the proposed survey. The dust mass and star-formation rate (SFR) is then estimated by modeling the far-IR SED [131]. If the inferred SFRs exceed a few 0.1 $M_{\odot}$/yr they probably cannot be attributed to dust-cooled gas at the center of a single galaxy, but must involve some radiative cooling of the cluster gas. The inefficiency of central AGN to completely stop cluster cooling flows can be determined by comparing the central SFR to the uninhibited cooling rate in $M_{\odot}$/yr for the cluster which depends mostly on the cluster mass. This inefficiency factor may vary among clusters of similar mass if SF is interrupted by feedback events. Such a far-IR/dust–selected BCG sample will also prove valuable for selecting targets for detailed follow-up studies essential for mapping the spatial distribution of dust and further clarifying its relationship to the cooling ICM and the AGN [388].

### 6.3.5 The *Planck* Extragalactic Point Source Population

While most attention focuses on *Planck* as a probe of the cosmic microwave background radiation, it will also make potentially seminal contributions to our understanding of galaxies at all redshifts. *Planck* is surveying the entire sky in nine bands, spanning 857 GHz to 30 GHz, and is expected to detect $\sim$10000 – 50000 galaxies (*Planck* Scientific Programme, ESA, 2005, see also [84]). The populations detected are expected to be diverse, including (1) intrinsically luminous sources at moderate redshift, (2) strongly lensed sources at moderate to high redshift [42], (3) blazars, (4) clumps of cool dust both in and just outside the Milky Way, and (5) galaxy clusters via the SZ effect ([166, 353], see also the Cosmology section of this document). *Planck* on its own however will find it extremely difficult to determine the nature of its point source population [238], particularly for strongly clustered populations at moderate to high redshifts [295]. The *Planck* beams are comparatively large ($5'$ or more), meaning that even galaxy clusters appear as point sources. Further contamination arises from arcminute scale structures in Galactic cirrus. And while existing *IRAS* & *AKARI* data will go some way to resolving *Planck* sources, they will only do this at $z < 0.1$, where most *IRAS*/*AKARI* sources reside.

The HSLS, covering ∼1/10th of the extragalactic sky in the SPIRE bands with beams as small as 18" at 250 $\mu$m, will dramatically alleviate the problem of point-source confusion for *Planck*. The key strength for HSLS will be the direct identification of *Planck* point sources at $z > 0.1$ in the two SPIRE bands that are matched to the highest frequency *Planck* channels. The 4000 square degrees of HSLS will thus serve a fundamentally important goal: by increasing by nearly an order of magnitude the number of *Planck* point sources with high spatial resolution IDs, from $\sim 1000$ with HerMES/H-ATLAS combined, to $\sim 8000$, it will determine precisely the level of cirrus contamination in the *Planck* extragalactic catalogs.

# 7 The Search for Rare Objects

The top-level science drivers of AGN/rare source studies are

- The buildup of the Magorrian relation and constraints on quasar feedback mechanisms
- The effects of hot-mode and cold-mode black hole accretion on the stellar mass assembly of galaxies
- Tests of unified models of radio-loud AGN
- The first large sample of sub-mm-selected blazars
- New galaxy populations: is there a limit to the specific star formation rate of a galaxy?
- Are there undiscovered planets in the outer Kuiper belt or inner Oort cloud?

## 7.1 Active Galactic Nuclei and rare objects

Our wide-area Herschel survey offers an unparallelled opportunity for the discovery and/or detection of rare object populations in the sub-mm, made possible by the fast mapping speed of SPIRE and a sensitivity that exceeds pre-launch expectations. This is an extremely rich seam with a very wide range of rare populations.

In terms of the active galactic nucleus population, key science targets include very rare phenomena, such as Fanaroff-Riley class II radio sources [145] which are radio galaxies whose source density is roughly six per square degree [e.g. 408], whilst others include rarer sub-populations within larger samples, such as the most luminous quasars found in the Sloan Digital Sky Survey [e.g. 351] or the most distant (and also most luminous) $z > 6$ quasars [e.g. 144].

There are, however, other celestial objects which will only be picked up in the largest area surveys, such as asteroids in our own Solar System and rare halo stars (see Section 8). Here we highlight the 'AGN and rare object' science case for the *Herschel*-SPIRE Legacy Survey.

### 7.1.1 Tracing the build up of the Magorrian relation through optical QSOs

One of the principal areas of uncertainty in the understanding of galaxy and quasar evolution is the nature of the feedback processes that regulated stellar mass assembly and black hole growth. The discovery of an astonishingly tight correlation between central supermassive black hole masses and the bulge properties of their hosts, the so-called Magorrian relation [257], clearly requires that star formation and black hole accretion must have been closely linked.

At the present day, this relation can be investigated directly with high-spatial-resolution spectroscopy of local galaxies where the resolution of, e.g., *HST* or VLBI is needed to resolve the sphere of influence of the black holes and thus measure the dynamics of the gas or stars orbiting the supermassive black hole. However, many models of quasar feedback predict that this relationship should vary with cosmic time [e.g. 99]. By combining indirect methods of black hole mass estimation with various tracers of stellar mass and star formation rate, benefiting from large statistically complete AGN samples, we will be be able to probe the evolution of the Magorrian relation and its time derivative and provide a key constraint on feedback models which will be possible only with this survey.

One way of estimating black-hole masses of high-redshift AGN is the use of the virial black-hole mass estimator [e.g. 400]. This assumes the broad emission line region gas in AGN has a Keplerian motion which is related to the mass of the central supermassive black hole. Given a velocity $v_{\rm BLR}$ and a radius $R_{\rm BLR}$ of the gas, the mass of the black hole can me estimated via

$$M_{\rm BH} = f\frac{R_{\rm BLR}v_{\rm BLR}^2}{G}, \tag{24}$$

where $f$ is a factor which describes the geometry of the broad-line region. For a spherically symmetric BLR $f = 3/4$ [315], while $f = 1/(4\sin^2\theta)$ for a disk BLR with an inclination angle of $\theta$ to the observer [274]. The velocity of the gas can be measured by the width of various elements within the broad-line region material. The most commonly used are H$\beta$ [400], MgII [275] and CIV [397], which together open up the full redshift range, through to the epoch of reionization [e.g. 407] for which we can measure the velocity of broad-line region gas. The radius of the BLR is calculated assuming the empirical relation between the continuum quasar luminosity and the radius determined through reverberation mapping [215]. Thus with a single quasar spectrum we are able to estimate black-hole mass. However, as shown by a number of groups, the intrinsic scatter in determining black-hole masses via the virial estimator means that large samples are required to obtain a statistically significant result and to overcome inherent biases from optical and/or radio wavelength selection[150, 151, 212, 213, 273].

In 4000 deg$^2$ we expect to have optical data to an $i-$band magnitude of $i_{\rm AB}$ =24–25 from the Dark Energy Survey and the VST ATLAS surveys. This is $\sim$ 2–3 magnitudes deeper than is currently used for the SDSS quasar survey with $i = 19.1$. This in turn allows us to define a much larger sample of quasars to much fainter magnitudes than is possible with the SDSS. Thus, the depth of the optical data in our chosen regions allows us, for the first time, to define a quasar sample which contains objects from the Seyfert-1 luminosity régime through to the most powerful quasars in the Universe at all redshifts up to the so-called 'quasar epoch' at $z \sim 2 - 3$ [e.g. 98, 332].

*Herschel* allows us to trace the star-formation rate in the host galaxies of these quasars, and thus enable us to determine the relationship between the black-hole mass and accretion rate of the quasars and the build-up of stellar mass in the host galaxy. This can then be considered in the framework of downsizing models, where the most massive galaxies appear to be in place before less-massive present-day systems [96]. Evidence is accumulating that black hole masses also follow a similar downsizing evolution [18]. The measured correlation between CO luminosity to far-IR luminosity in SMGs means that high-$z$ HSLS QSOs will be strong CO emitters. ALMA will then be able to rapidly provide CO maps of such sources. These will then allow a direct determination of the total dynamical mass of the galaxies for a significant subsample of these high-$z$ QSOs.

Crucially, the 4000 deg$^2$ survey with *Herschel* provides a large enough volume of the Universe to allow measurements constraining this relationship from the least luminous quasars at $z < 3$ through to the most luminous quasars at all redshifts. Such a sample will allow us to determine whether star formation rate changes with accretion rate of the AGN, with black hole mass or with redshift and to delineate these relations.

Early results using the limited area from the *Herschel*-ATLAS survey [130] suggest that the star-formation rate in quasar host galaxies is strongly related to the redshift [e.g. 51, 362] but with large dispersion on this relation, and a weaker but significant ($> 4\sigma$) relation between the quasar accretion luminosity and the star formation rate in the quasar host galaxy, traced by the *Herschel* data [51]. Supplementing the *Herschel* data with data from other facilities improves the significance of these relations [e.g. 362] but at the cost of using a heterogeneously-selected quasar sample with heterogeneous rest-frame photometry. Our *Herschel* survey will benefit from homogeneous photometry of large AGN samples with well-defined (i.e. statistically complete) selection criteria.

Furthermore, the new radio surveys which will be undertaken in the southern hemisphere by both of the Square Kilometre Array (SKA) precursor experiments, Australian SKA Pathfinder and MeerKAT telescopes, will allow us to test how far-infrared luminosity may be related to the issue of radio loudness in the quasar population. Recent work suggests that the radio emission may provide an indirect method of determining black-hole spin; thus, if major mergers spin up black holes then we would expect the radio-loud population to have enhanced star formation when compared to the radio-quiet population. The data has not been good enough to test this scenario thus far, due to the fact that it is almost impossible to determine the star-formation rate in quasars from UV, optical or near-infrared due to the dominant emission from the quasar itself. With *Herschel* we no longer have this problem and the combination of sensitivity and survey speed will allow us to investigate such model on a sound statistical footing.

### 7.1.2 Hot and Cold mode accretion and AGN driven feedback

In recent years there has been a growing consensus that both positive and negative feedback can occur due to the presence of an AGN. In the most recent semi-analytic models this feedback is in the form of a negative feedback which works to truncate star formation in the host galaxy in order for the semi-analytic models to match the bright end of the local luminosity function and also to explain the space density of massive galaxies at the highest redshift [e.g. 56, 99].

Feedback process may have two distinct modes [99], which are known as 'hot-mode' (or 'radio-mode') and 'cold-mode' (or 'quasar mode') feedback. These are physically associated with the efficient accretion of cold gas (the quasar mode) and less efficient accretion of hot gas from the interstellar medium ('radio mode'), see also [188]. The former is thought to be the dominant accretion mechanism at the bright end of the AGN luminosity function, i.e. in the quasar régime which also includes most of the more powerful FRII-type radio sources. The main influx of gas in this scenario is usually assumed to originate from galaxy mergers, although accretion via cold flows could also play an important role [e.g. 112, 118].

There is now observational evidence that powerful AGN show indirect signatures of the presence of cold gas. Herbert et al. [195] showed that, in a sample of radio galaxies, the less powerful FRI-type radio galaxies showed no evidence for star formation in their host galaxy, whereas the more powerful FRIIs showed clear evidence for such residual star formation. This is also in line with the work by Baldi & Capetti [20], who found a similar result using

| $z$ | Number of FRIIs degree$^{-2}$ | Area for $5\sigma$ stacked SPIRE detection in degrees$^2$ |
|---|---|---|
| 0.5 - 0.75 | 0.03 | 74 |
| 0.75 - 1.0 | 0.11 | 49 |
| 1.00 - 1.25 | 0.25 | 41 |
| 1.25 - 1.50 | 0.43 | 40 |
| 1.50 - 1.75 | 0.63 | 45 |
| 1.75 - 2.00 | 0.71 | 61 |
| 2.00 - 2.25 | 0.71 | 92 |
| 2.25 - 2.50 | 0.67 | 146 |
| 2.50 - 2.75 | 0.59 | 250 |
| 2.75 - 3.00 | 0.50 | 427 |
| 3.00 - 3.25 | 0.43 | 714 |
| 3.25 - 3.50 | 0.33 | 1195 |
| 3.50 - 3.75 | 0.27 | 1825 |
| 3.75 - 4.00 | 0.20 | 3207 |
| 4.00 - 4.25 | 0.13 | 6318 |
| 4.25 - 4.50 | 0.09 | 11465 |
| 4.50 - 4.75 | 0.06 | 21553 |
| 4.75 - 5.00 | 0.04 | all sky |

Table 7: The areal coverage required to achieve a $5\sigma$ stacked detection of FRII galaxies in *at least one* SPIRE band, for a series of redshift bins, assuming an M82 starburst SED in the far-infrared and a star formation rate of $100\,M_\odot\,\mathrm{yr}^{-1}$ in each AGN. The inevitable conclusion is that to investigate the evolution in star formation rates in FRII galaxies at $z > 3$, a wider-area survey than H-ATLAS is essential, even if using broader redshift bins than tabulated here.

*HST* UV data.

With a $4000\,\mathrm{deg}^2$ survey with *Herschel* we can explore this further and finally pin down whether the star formation in the host galaxies of radio-loud AGN is linked to the jet power and also the accretion mode. It is only with radio sources that we can do this efficiently, as it is extremely difficult to find the hot-mode (or radio mode) accreting AGN via any other observations.

FRII radio galaxies are intrinsically very rare objects, with roughly six per square degree peaking around a redshift of $2 < z < 3$ with tails towards lower and higher redshifts. Thus in order to gain a complete census of the star formation in the host galaxies of FRIIs we require sufficiently large areas covered by *Herschel*. Furthermore, given the necessity that large area also implies relatively shallow depth, many of the key results will rely on stacking of data at the position of the radio sources or optical identification. In Table 7 we show how many square degrees are needed per $\Delta z = 0.25$ to obtain a $5\sigma$ detection of a galaxy with $100\,\mathrm{M}_\odot\ \mathrm{yr}^{-1}$ for redshifts $z = 0 \rightarrow 5$. One can immediately see that the 4000 square degree survey is ideal for tracing the star formation in the hosts of powerful radio sources over $\sim 90$ per cent of cosmic time.

To understand the feedback mechanism, however, we also need to determine the star-formation rate in the lower-radio luminosity FRI sources. This has been attempted with the SDP data of the *Herschel*-ATLAS survey: initial findings are that these low-luminosity sources have far-infrared luminosities, and therefore star-formation rates, consistent with the the non-active galaxy population spanning the same galaxy stellar masses [187]. However, the wider area of the present survey will allow stacking of a larger number of sources which will in turn allow us to probe to much deeper limits in star-formation rate.

### 7.1.3 High-redshift quasars

High-redshift quasars are natural targets for rare object science in our survey. In the SDSS DR5 quasar catalogue alone there are 111 quasars per $1000\,\mathrm{deg}^2$. Among these, 4.5 per $1000\,\mathrm{deg}^2$ are at an interesting redshift range of $5 < z < 5.5$ with I-band absolute magnitudes brighter than $-26.5$. Where individual objects are not detected, stacking analyses can yield constraints on their mean submm properties. This is a clear driver for large areas since this requires large numbers of quasars, given that the stacked signal-to-noise ratio scales as the square root of the number of stacked targets.

The mean $850\,\mu$m stacked fluxes of quasars are fairly constant throughout $1 < z < 5$ at $\sim 2\,$mJy ([323]; $500\,\mu$m equivalent $> 5\,$mJy assuming an M82-like SED). However, about one in six have luminosities 5–10 times this level ($17 \pm 6\%$ in [323] at $850\,\mu$m), and 1/3 of the $z > 6$ quasars observed so far are strongly far-IR luminous [334].

### 7.1.4 Other rare AGN populations

Can we find youthful quasars in the act of shutting down star formation in their hosts? Broad Absorption Line QSOs (BALQSOs) form $\simeq 10\%$ of the QSO population, and it has been been proposed that they are youthful AGN in a brief vigorous outflow phase [e.g. 26, 60]. Are BALQSOs systems in which radiative or kinetic energy input has started shutting down star formation? Alternative models suggest that BALQSOs are normal quasars seen along atypical lines of sight [e.g. 87, 175]. If BALQSOs have the same far-IR luminosity as normal quasars for a given (unabsorbed) optical continuum, it makes it harder to maintain that they are seen at an unusual point in a quasar's star formation history.

Other candidate youthful quasars are low-ionization BALQSOs [LoBALs, e.g. 184] or iron-rich LoBALs (FeLoBALs) or nitrogen-enriched quasars. There are only 46 LoBALS per $1000\,\mathrm{deg}^2$ in SDSS, and 33 FeLoBALs per $1000\,\mathrm{deg}^2$ so the study of these objects clearly benefits from wide areas. Curiously there are a few famous far-IR-luminous LoBALs (e.g. the Cloverleaf) and there is evidence for $> 100\,$pc winds in FeLoBALs [113]. There are also tentative links between submm luminosity and CIV BAL equivalent width [324]. Our very large quasar sample will enable us to test whether these or some other AGN subset are more closely connected with starburst activity [e.g. nitrogen-enriched quasars 34]. Our wide-field survey is ideal for the detection and discovery of rare transition populations, such as starbursts occurring along the merger sequence in major galaxy-galaxy mergers [e.g. 30], or youthful quasars still in the process of shutting down star formation in their hosts (see above). We will attempt to identify rare populations by searching for colour space outliers (e.g. SDSS+PACS+SPIRE+NVSS/FIRST).

### 7.1.5 Radio-loud AGN unification

Decimetric observations of radio-loud AGN tend to have radio luminosities dominated by orientation-independent optically-thin synchrotron emission. The accessibility of our fields to LOFAR and KAT will make radio spectral classification simple in the medium term. We will thus be able to investigate whether the hosts of the most luminous radio AGN are distinguishable from the hosts of more modest AGN, or whether the latter AGN are simply more evolved versions of the former. The key test of radio-loud unification schemes is to select radio galaxies and quasars on one orientation-independent quantity (optically-thin synchrotron radio lobes) and compare another (host galaxy star formation rates). With the a wide-area survey we will obtain far-infrared photometry of radio-loud AGN with a wide range of lobe luminosities at any fixed redshift; again, stacking analyses will constrain the mean fluxes of undetected populations.

### 7.1.6 Blazar physics

The submm wavelength range is crucial for the physics of blazars, comprising BL Lac objects and flat-spectrum radio quasars. These objects allow the investigation of physics at the highest energies and in the most extreme conditions. They exhibit rapid variability and apparent superluminal motion and have SEDs that peak at $\gamma$-rays and radio-wavelengths. The physical interpretation of these extreme objects is that they are accreting black holes where the relativistic jet is aligned close to the observer's line of sight. The emission is a combination of synchrotron emission, peaking at far-infrared–radio wavelengths, and inverse Compton scattering which is thought to explain the $\gamma$-ray peak.

The far-infrared and submm waveband is essentially unexplored for the selection of blazars. This is a particular handicap for the study of the low-frequency synchrotron peaking blazars (LSP; with peak frequency $\nu_p^s < 10^{14}$ Hz), which have their synchrotron peak close to the *Herschel* bands. This peak frequency contains crucial information about the physical parameters which govern the system, such as the Lorentz factor of the emitting electrons, the Doppler factor and the strength of the magnetic field.

The results from [174] suggests that the H-ATLAS sample will set strong constraints on the abundance of blazars with low values of the synchrotron peak frequency ($\nu_p^s \lesssim 10^{13}$ Hz) which, in the blazar sequence scenario, are those with more powerful jets, more luminous accretion disks and higher black hole masses [167]. However, a $4000\,\mathrm{deg}^2$ survey to a depth similar to the H-ATLAS one should yield a complete, homogeneously selected sample of $\sim 600$ blazars, a substantial fraction of which are expected to be also detected by *Fermi*-LAT. Multi-frequency radio-microwave data for sources in the southern hemisphere are provided by the AT20G survey. Such sample would allow us to investigate similarities and differences between different blazar sub-populations (BL-Lacs, FSRQs, LSP, ISP, HSP, ...). This cannot be done with the sample of $\sim 80$ blazars provided by H-ATLAS.

### 7.1.7 Kuiper belt objects

No new populations of main belt asteroids are expected to be detected, but there is an outside possibility of making new detections of populations of outer solar system objects. At a distance of 60 AU a 1300 km radius planet with an albedo of 0.05 would have a 250 $\mu$m flux of 41 mJy, i.e. at the survey limit. Such objects need not be confined to the ecliptic plane, as the discovery of Sedna (inclination 12°) and Eris (44°) have demonstrated. It is still possible that there are large scattered disk Kuiper belt objects waiting to be discovered.

With only one SPIRE observation, simultaneous optical imaging would be needed to confirm their solar system origin. Eris (radius 1300 km, albedo 0.86) has $V = 18.6$ at its current distance of 97 AU. The proper motions of these objects are typically $< 1''$/hour. If Eris had an albedo of 0.05, its apparent magnitude would be $V = 21.6$, but this might still be detectable from a comparison with e.g. digitized sky survey data.

A more exotic and speculative possibility is a Mars or Earth-sized planet at $> 100$ AU, which has been argued to explain the fine structure of the Kuiper belt [247]. At 100 AU an object with the radius of Mars and an albedo of 0.2 would have a 250 $\mu$m flux of 52 mJy. An even more exotic outside possibility is a gas giant in the inner Oort cloud, which cannot be ruled out on dynamical arguments. At 3000 AU, an 0.5 $M_{\rm Jupiter}$ planet with an age of 4.6 Gyr (using the models of [67]) would have a 250 $\mu$m flux of 151 mJy. Such an object would be unfeasibly faint in the optical, but might be uncovered by unexpected parallax in sub-mm follow-ups of e.g. optically-unidentified sources with warm colour temperatures.

### 7.1.8 Discovery space for new rare object populations

Which are the most luminous galaxies in the Hubble Volume? By surveying $\sim 10\%$ of the sky, and thanks to the favourable sub-mm K-corrections, the statistical expectation is that our survey will detect galaxies among the ten most luminous in the $z < 6$ observable Universe. Taken in conjunction with the fainter Herschel surveys (e.g. HerMES) and the de-magnified lensed populations from this survey and H-ATLAS, this makes it possible to sample *the complete range of sub-mm galaxy luminosities in the $z < 6$ observable Universe.*

Is there a limit to the specific star formation rate of a galaxy? For example, does AGN/supernova-driven feedback suppress the specific star formation rates in the most massive high-$z$ galaxies, or are the gravitational potential wells too great to prevent gas expulsion? Do all massive galaxies form at the putative $z \simeq 2$ peak seen in SCUBA surveys, or does the downsizing trend extend to higher-redshift and more massive systems?

The few empirical insights on feedback processes do not give sufficient constraints for semi-analytic models of galaxy evolution to provide unequivocal answers to these questions. Could there be $10^{14-15} L_\odot$ populations waiting to be discovered? No survey to date has had the depth and area capable of identifying a $z > 3$, luminous galaxies with $5 \times 10^{14} L_\odot$ galaxy (though they may exist, undiscovered and unidentified, in deep decimetric radio surveys). There is a clear pedigree for surprising populations of extreme starbursts challenging models of galaxy evolution: for example, the discovery of the hyperluminous galaxy IRAS FSC 10214+4724 [336], or the discovery of the sub-mm-luminous population itself [202, 373].

At redshifts $z > 0.5$, the comoving volume sampled by our 4000 deg$^2$ survey is 12$\times$ larger than the *entire* $z < 0.5$ observable Universe. We therefore have the unprecedented opportunity to sample galaxies too rare to have local counterparts. These extreme starbursts and active galaxies will more clearly illustrate the feedback processes that are only weakly apparent in fainter sub-mm galaxy populations.

# 8 Galactic Dust and Stellar Sources

The successful Hi-GAL survey has shown that the Galactic Plane is a crowded region with high levels of filamentary cirrus emission that pose their own particular challenges to the discovery of faint objects [289]. A large area survey like the HSLS will allow the investigation of High Galactic Latitude and local star formation, molecular cloud formation and other Galactic phenomena along sight-lines that are less confused and less affected by the high background of the Galactic Plane.

Here, we explore the main science drivers for a Galactic science component to the *Herschel*-SPIRE Legacy Survey, which may be summarised as:

- An unbiased volume limited census of high galactic latitude star formation
- The properties of dust in the diffuse high galactic latitude interstellar medium, and its role in forming cold condensations and core
- Determination the origin of field T Tauri stars & solving the distributed T Tau problem
- A census of dust producing stars; the detection of Debris Disks with unbiased statistics on frequency and mass as a function of star, and the properties of cold dust around White Dwarf stars

## 8.1 High galactic latitude star formation

Most studies of star formation have been confined to large molecular clouds near the Galactic plane. In the Galactic disk, clusters of young stars are invariably associated with giant molecular clouds (GMCs). The disk includes more than 90% of all known young open clusters and even a larger fraction of GMCs. Therefore, if field stars are born in some type of cluster and clusters are formed out of clouds, recent star formation is mainly expected to occur within the disk. However, all-sky surveys with IRAS and AKARI have made possible the discovery and study of more tenuous molecular material. Although molecular clouds are ubiquitous throughout the crowded and boisterous Galactic Plane, there are advantages if it proves possible to study star formation at high latitude above the molecular scale height where sightlines are particularly simple, and where the clouds are generally nearby.

Our understanding is also that most star formation appears linked to the spiral structure of our Galaxy, and/or on filamentary or sheet-like structures. There are three distinct populations of pre-main-sequence stars seen at high Galactic latitudes. These are those still found in association with their parent molecular clouds, those found in isolation far from any present molecular cloud, and those formed in extraplanar environments high above the Galactic plane. Large-scale shock fronts induced by spiral arms are widely regarded as the dominant, primary triggering mechanism in the disk. There is, however, a growing body of observational evidence indicating that high Galactic latitude star formation, even if proceeding at a slower rate, is happening now. But, far from the regions in which the usual driving forces of triggered star formation can act efficiently, by what mechanism could star clusters form out of high Galactic altitude clouds? Possible formation scenarios have included supernovae from first generation stars [268] and Open Clusters [115].

The Galaxy is known to harbour a population of high Galactic altitude clouds which are, in some cases, similar to those in the disk. About 3% of open clusters younger than 100 Myr are located at least 200 pc away from the disk; in the outer Galaxy, some embedded clusters are found at 500 pc. But, by what mechanism could star clusters form far from the regions in which the usual driving forces of triggered star formation can act efficiently? One interesting area is to examine whether passing pre-existing star clusters can induce tidal forces able to trigger star formation in such clouds. Another interesting line of study will be to correlate the results of the proposed *Herschel*-SPIRE Legacy Survey with the distributions of T Tauri stars, particularly those located far away from molecular clouds [59].

Typical galactic plane GMCs have masses of a few 1000 to 10's of thousands of solar masses, diameters of $\sim$ 45 pc, and internal velocity dispersions of $\sim$2 km s$^{-1}$. They are found along the Galactic equator with a small vertical scale height of 75 pc [254]. Molecular clouds located off the Galactic plane appear somewhat different, and can be classified in two different categories: low mass, transient clouds [155, 259] and larger clouds with properties similar to those in the solar circle. These clouds may be primordial but they could also form in Heiles-type shells-supershells [191] or by the supernova-driven Galactic fountain mechanism [253]. The high-latitude clouds are assumed to be close to the Sun or high above the Galactic plane, and are predominantly diffuse or translucent and are thus difficult to detect on photographic surveys. The Gould Belt and the Taurus-Auriga complex are the most well known, and local, high galactic latitude star formation region [272]. However, whilst the majority of high latitude clouds are diffuse or translucent and not thought to be capable of forming stars, there are a number of growing examples that high latitude clouds are not entirely devoid of star formation. Dense molecular cores have been detected even in Galactic cirrus, e.g. MCLD 123.5+24.9 in the Polaris Flare [193], which is also gravitationally stable and with in falling motions similar to known prestellar cores [192]. T Tauri stars are found to be associated with the MBM 12, MBM 33 and MBM 37 high latitude clouds [267]. Understanding the nature of the star formation in high latitude clouds

From early studies with the *Herschel*-ATLAS survey [130] we are beginning to find a population of small, compact molecular clouds with *Herschel* that could not be efficiently discovered by any other means. Figure 37 shows a compact molecular cloud identified in the first science demonstration image from *Herschel*-ATLAS (Thompson et al, in prep). It is extended on the scale of the SPIRE beam, dark in the visible and near-infrared (down to K$\simeq$18, from the UKIDSS Large Area Survey), but associated with strong J=2–1 CO emission at an LSR velocity of $-20$ km s$^{-1}$ and with a mass between 0.01 and 0.5 M$_\odot$. Current efforts are to determine the distance of the cloud via a photometric technique, however the observed CO line width of the cloud suggests that it may be actively forming a star or substellar object. Figure 37 also shows the SED of the cloud compared to the detection limits of the IRAS Faint Source Catalogue, although it is undetected in the IRAS FSC. Large-beam CO surveys at high latitude [103, 104] will also not detect these compact clouds due to the extreme levels of beam dilution involved. Large area surveys with *Herschel* are the only way in which to discover these clouds, study them at high angular resolution and determine the peak of their SED.

Taking the number of these clouds that have been discovered so far (1 in the H-ATLAS science demonstration data and another in the HeVICS survey field, see [95]) implies that HSLS could detect on the order of $\sim$200 molecular clouds — compiling the first statistical sample of these clouds. These clouds are important to star formation studies because they are likely to be very close to the Sun, extremely low-mass (∼0.1 solar mass) and trace the uncertain formation processes for substellar objects and the formation of small molecular clouds. It will also be possible to compare the properties of high latitude star formation to those within more well-known Galactic Plane clouds to test whether the star formation mode varies with Galactic environment, in particular the role of triggering versus gravitational collapse in such low density high latitude clouds. Indeed, the discovery of star forming clouds at high latitude has the significant scope to discover the closest star forming clouds to the Sun (currently MBM 20 or Taurus Auriga) which would enable the study of the physics of star formation in unprecedented detail.

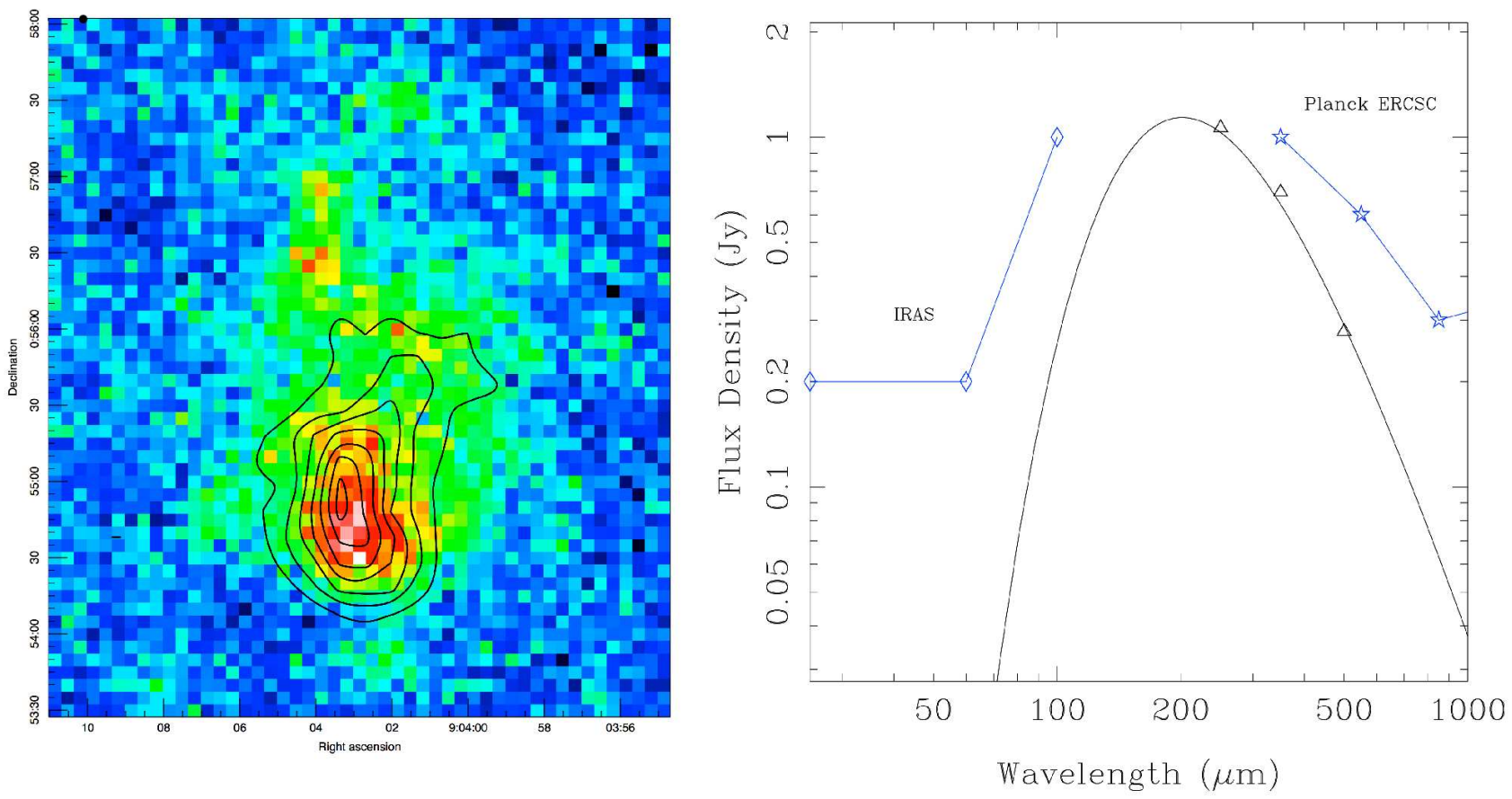

Figure 37: A small and isolated molecular cloud discovered during the science demonstration observations for the *Herschel*-ATLAS survey. *Left:* SPIRE 250 $\mu$m image overlaid with contours of CO J=2–1 integrated emission. The CO emission peaks at a $V_{LSR}$ of $-20$ km s$^{-1}$, confirming the Galactic origin and molecular nature of this cloud. *Right:* The SED of the cloud, with a greybody model fitted to the H-ATLAS SPIRE fluxes. The best fitting greybody has a temperature of 15 K and $\beta$ of 1.8. The blue lines show the detection limits of the IRAS FSC, indicating that clouds of this type are essentially below the ERCSC and IRAS FSC detection thresholds.

## 8.2 The properties of dust in the diffuse High Galactic latitude ISM

Dust is the most robust tracer for the 'Galactic ecology' - the cycling of material from dying stars to the ionised, atomic, and molecular phases of the Interstellar medium (ISM), into star forming cloud cores, and back into stars. Atoms, ions, and molecules are imperfect tracers because they undergo complex phase changes, chemical processing, depletions onto grains, and are subject to complex excitation conditions. In contrast, dust is stable in most phases of the ISM; it is optically thin in the Far Infrared (FIR) over most of the Galaxy, so that its emission and absorption simply depend on emissivity, column density and temperature. Cold dust in particular (10K $<$ T $<$ 40K) traces the bulk of non-stellar baryonic mass in all of the the Galactic ecosystem.

Observations of high-latitude clouds provide an unbiased search for star formation as a function of galactic latitude. The Far-Infrared (FIR) emission of the Galaxy is dominated by emission from the largest dust grains (BG: Big Grains) in the ISM with sizes of the order of 0.1 $\mu$m. In the diffuse ISM, these grains radiate in thermal equilibrium and their temperature is set by the balance between cooling through Infrared emission and heating through the absorption of the VIS and UV photons of the Interstellar Radiation Field (ISRF). sub-parsec resolution is a critical observable needed to formulate a global predictive model of the Galactic ISM"star formation cycling process which drives the Galactic ecology in normal spirals and is a cornerstone for the unveiling of the formation and evolution of galaxies throughout the cosmos. The properties of the dust have recently been modelled to include several new grain emission mechanisms, including the presence of the so-called 'anomalous' spinning dust, or electric-dipole radiation from spinning very small grains. Previous FIR observations of the diffuse ISM at high and intermediate galactic latitude with the IRAS, DIRBE and FIRAS instruments have established that the average dust equilibrium temperature in the solar neighbourhood, corresponding to G0 = 1, is about 17.5 K [235] assuming a dust emissivity index of $\beta = 2$.

The Polaris flare is an example of a high Galactic latitude cirrus cloud, located at a distance $\sim$ 150 pc [143]. It has significant CO emission [143, 194, 282] and colder dust grains than typical diffuse clouds [35, 235]. Although the Polaris flare does not show strong signs of star formation activity, it is the archetype of the initial phase of molecular cloud formation, and studies of objects like it will allow us to study the formation of low- to intermediate-mass stars in remnant molecular clouds at unusually large heights above the galactic plane, reflecting the transport of molecular gas to such heights by expanding superbubbles.

For the ISM, there are three main drivers which together form an unique combination:

- **Studies of dust properties on sub-arcmin scales** Combining SPIRE 250, 350 and 500 um data with, for instance, AKARI/FIS data [406], we would provide full coverage of the ISM SED (predicting a peak, in most regions, around $\sim$150 $\mu$m). Hence, we could study colour variations (i.e. variations of Big Grain abundances, etc.) as well as temperature and spectral emissivity index variations. Many CMB experiments (ARCHEOPS, BOOMERANG, PLANCK) have recently pointed out an anti-correlation between these two parameters, which cannot be explained with classical dust models like the Finkbeiner et al [152], but rather through two level systems (TLS models [279]). These variations have been seen to occur on arc minute scales, and *Herschel* has started reporting similar variations [11, 36, 305]. However, these studies have thus far been limited to either Galactic Plane regions or to nearby clouds

- **Constraining spinning dust models.** One of the problems of spinning dust studies is to constrain the possible contribution of the grey-body tail due to cold dust. In fact, using only IRAS bands is not enough, since IRAS 100 $\mu$m does not constrain the peak of the grey-body. The issue has been shown for sources like Rho Oph [70] or M78 (Casassus et al in press) detected (by CBI) anomalous emission - when longer wavelengths data are available, for instance from SCUBA, they are usually limited by the fact that, due to beam switching, they miss the extended emission - hence they cannot be used. SPIRE data would provide this information and compared to PLANCK, would have the advantage of having angular resolution comparable to Spitzer IRAC/MIPS data, i.e. templates of, respectively, PAH and VSG emission, so that they could be used as spatial templates of Big Grain emission.

- **Cold Cores and cold ISM structures in all the environments of the high galactic latitude ecosystem.** The direct detection of cold (i.e. T≤20 K) dust possibly the quiescent counterparts to traditional molecular clouds, has been difficult [235, 329, 378] either because of insufficient wavelength coverage (e.g. IRAS) or inadequate spatial resolution (DIRBE, FIRAS). CO observations have been problematic due to molecular freeze-out onto grains [154], or photochemical effects in low-metallicity environments [54]. The recent detection of very cold clumps in the GP with Archeops [108] has confirmed that the FIR/submm continuum as the best tool to directly trace cold ISM components. Notable examples are Infrared Dark Clouds (IRDCs) and HI Self-Absorption (HISA) clouds. Our programme will search for cold cores located at intermediate/high latitudes, providing an alternative to those which we are beginning to recognise close to the galactic plane.

### 8.3 Dust and the formation of molecular clouds at high latitude

In this era of deep FIR extragalactic surveys, it has become obvious that a better grasp of the contribution from the high galactic latitude foreground is necessary. For example the radio astronomers of the Planck Deep Field (PDF) international consortia (Martin et al. in preparation) have embarked, through the use of the DRAO telescopes (interferometric 21 cm HI line mapping at 1' resolution), in a quest for understanding the Milky Way's ISM as a foreground to the Planck cosmological observations. The PDF covers approximately 40 square degrees of a vast cirrus cloud taking the shape of an intricate network of connected HI filaments. A similar area has also been observed in shallower depth but this time as a large number of unconnected smaller fields. Cirrus clouds of varying morphology and column densities distributed across the northern sky have been sampled. All these observations of high galactic latitude fields open the way to understanding the kinematics and dynamics of relatively low HI gas column density areas where in principle only the interstellar radiation field (ISRF), turbulence and magnetic fields dominate the physics. Therefore in the next paragraphs we will concentrate on Galactic physics.

High latitude clouds are obvious targets of studies. They do not suffer from velocity crowding or self-absorption effects and very few show evidence of star formation. Hence they provide the opportunity to study the initial steps of molecule formation. Since cirrus clouds span the range of column densities – from a few $10^{19}$ to a few $10^{21}$ H cm$^{-2}$ – over which UV absorption studies show that the conversion from HI to $H_2$ occurs [328], the presence of $H_2$ is probably common. Only a small fraction of the cirrus have been detected in CO surveys [254] but this is not unexpected since the density required for CO excitation ($> 500$ cm$^{-3}$) is one order of magnitude larger than either the mean density one gets by dividing column densities by cloud sizes or those inferred from the $H_2$/HI column density ratio in UV studies [179]. Furthermore, self-shielding of the $H_2$ molecule is much more efficient than that of the CO molecule [239]. The CO emission is thus tracing dense structures filling a small fraction of the cloud volume. The more diffuse and wide-spread $H_2$ gas can be traced by far-IR dust observations since the far-IR emission from cirrus is a density-independent tracer of total hydrogen column density.

Studies of high latitude clouds abound. For example, Refs. [214, 285, 286] have demon-strated the richness of the information which one can extract on the physics of the ISM. Barriault et al [22], using the IR excess method, have found two molecular clouds in the making at two different stages of evolution in the PDF. Two of their major findings are that dynamical effects (velocity shears) have a strong impact on the formation of CO and that OH may be a better tracer of $H_2$ than CO at least at high latitude. The interstellar radiation field also seems to have a say in the early phase of molecular cloud formation. Understanding the formation of molecular clouds is a fundamental issue in Galactic physics. One only has to mention that stars form from such objects to be convinced! Therefore in order to make significant progress in our understanding of star formation, it is clear that we require knowledge on the pristine conditions a molecular cloud offers. Do they vary depending on the conditions of molecule formation? Hence the need to observe transition regions (HI to $H_2$) at different stages of evolution and in different environments. Joncas and his collaborators are pursuing these investigations using DRAO, the Green Bank Telescope, the Onsala Space Observatory and the Planck observatory.

Building on this expertise, it is now time to learn more on the contribution of the dust component to the evolutionary scenario of molecular clouds. It is a key complementary agent in the physical mechanisms that take place: dust has a role of catalyst in $H_2$ formation and is the main heating agent of the diffuse ISM via the photoelectric effect. Miville-Deschênes et al. [285] have shown from ISO observations that dust properties change in the transition zone between purely atomic and molecular areas while HI and CO observations demonstrated that the magnitude of turbulence also changes. If one accepts that dust grains are kept small through collisions in highly turbulent areas then the variation of dust characteristics can be a major parameter in the efficiency of a given environment in producing molecules. The sensitivity, spatial resolution and spectral multiplicity of *Herschel*-SPIRE make it the ideal instrument for this investigation. Combining the observations with modeling such as provided by the "DUSTEM" software will provide insight on dust composition, size and shape under the varying conditions uniquely provided by the proposed large area survey.

### 8.4 Debris Discs and other stars associated with dust

Debris discs are the dusty remnants of the planet formation around main sequence stars, detected through their emission at infrared wavelengths. The study of these objects improves our understanding of both the frequency and the timescale of planet formation around other stars, and, in resolved disc systems, can be used as an indirect method of planet detection in regions of orbital radius-planet mass space where standard methods (e.g. radial velocity, transits) are insensitive. Although very few older stars show evidence of hot dust, likely due to clearing by planets [28], it is reasonable to believe that Kuiper-Belts are more long lived and many cool debris disks should be detectable at far-IR wavelengths. A significant fraction of the mature FGK stars have cool dusty disks at least an order of magnitudes brighter than the solar system's outer zodiacal light. Since such dusts must be continually replenished, they are generally assumed to be the collisional fragments of residual planetesimals analogous to the Kuiper Belt objects.

The potential correlation of cold dust and planetesimal belts with the formation of exoplanets in these systems is another incentive for larger submm surveys of nearby solar-type main sequence stars. Beichman et al. [27] and Bryden et al. [66] reported Spitzer MIPS observations of a large sample of nearby F, G, and K main-sequence stars, and detected an overall excess of 70 $\mu$m emission toward 13% of the stars. From a larger sample of solar-type stars, spanning ages between 3 Myr and 3 Gyr, studied in the *Spitzer* Legacy program Formation and Evolution of Planetary Systems (FEPS), showed a total 70 $\mu$m excess rate of 7% with high (50 times the photosphere) fractional excesses at ages between 30 and 200 Myr, and significantly lower fractional excess for older stars [69, 196]. An additional advantage in making observations in the far-IR region is that measured integrated fluxes are directly proportional to the temperature and mass of the disk, due to the fact that they (in most cases) sample over, or close to the intensity peak of the Rayleigh-Jeans tail of the spectral energy distribution (SED), and that the disk can be assumed to be optically thin at these wavelengths particularly if the far-IR data can be supplemented with ground based submm observations at $\sim$ 1 mm wavelength.

A natural approach to understand the origin and diversity of planetary systems is to study the birth sites of planetary systems under varying environmental conditions. The disk geometry may change from a flaring to a more flattened structure, gaps may develop under the gravitational influence of protoplanets, and eventually the disk will dissipate, terminating the planet formation process. Dedicated debris disc search programmes are underway with *Herschel* in the form of the DEBRIS and DUNES Key Projects [135, 271]. In both cases these programmes are targeting with PACS a volume-limited, carefully selected uniform sample of main sequence stars ranging from $\sim$ 400 A –M stars in the case of DEBRIS and $\sim$ 200 FGK stars in the case of DUNES. The advantage of these carefully selected samples is that sensitive PACS observations to photospheric levels can be achieved for the entire sample, allowing disk frequency as a function of spectral type and age to be fully investigated. However, the nature of the sample selection means that only a limited number of main sequence stars with types A–M can be included. These dedicated surveys are thus relatively insensitive to rarities in the debris disc population of which there are likely to be few counterparts in the local population of discs (for example bright and/or massive discs) and cannot be used to investigate debris or primordial dust around other types of stars not on the main sequence (e.g. T Tauris or white dwarfs). Recently Thompson et al [389] showed that large area galaxy surveys such as the *Herschel* ATLAS could be used to search for debris discs (see Figure 38 for an example) by combining *Herschel* and SDSS catalogues using Bayesian Likelihood Ratio techniques originally developed for identifying radio galaxies [376, 385]. The combination of DUNES, DEBRIS & H-ATLAS then forms a classic wedding-cake survey design with H-ATLAS being the lower tier that is uniquely sensitive to bright and massive discs that may not be picked up by the other surveys.

As the deepest and widest unbiased survey since AKARI, the proposed *Herschel*-SPIRE Legacy Survey has the potential to take this to a new level, improving the potential for the discovery rare and extreme objects. The access to supporting high quality optical data from SDSS Stripe 82 and eventually from the Dark Energy Survey allows the straightforward selection of stars on the main sequence locus using the colour selection techniques of Covey et al 2007 and Kimball et al 2009. Whilst SPIRE offers some disadvantages to a debris disc survey, namely the higher confusion limit due to the larger beam, it offers the advantage that typical photospheric fluxes at 250 $\mu$m are $\sim$ $\mu$Jy and hence any detection of a star in this SPIRE band implies that it is likely to host a debris disc. The sensitivity of HSLS (40 mJy 5$\sigma$ at 250 $\mu$m) is sufficient to detect analogues of known debris discs, e.g. HR 4796 and $\beta$ Pictoris, out to $\sim$ 120 pc (at 120 pc we are sensitive to $\sim$ 2 Lunar masses of dust at a temperature of 60 K). The number of stars that could be searched by such a wide area survey is truly staggering. Scaling from the stellar densities observed in the H-ATLAS science demonstration field we expect that the HSLS will allow us to search for debris discs toward $\sim$10,000 main sequence stars within 120 pc — some two orders of magnitude higher than the DEBRIS & DUNES surveys. It is also notable that several of the pre-main-sequence stars found high galactic latitudes are members of local moving groups and stellar associations, and likely that the proposed HSLS fields will also be able to explore the relationship of star formation and Debris Disc frequency to triggering events resulting from the presence of local co-moving groups. The legacy value of this data set is extremely high. Following the completion of the ESA GAIA mission *all* the stars in our search area brighter than $r = 20$ will have accurate spectral types and trigonometric parallaxes determined, allowing precise disk masses to be obtained.

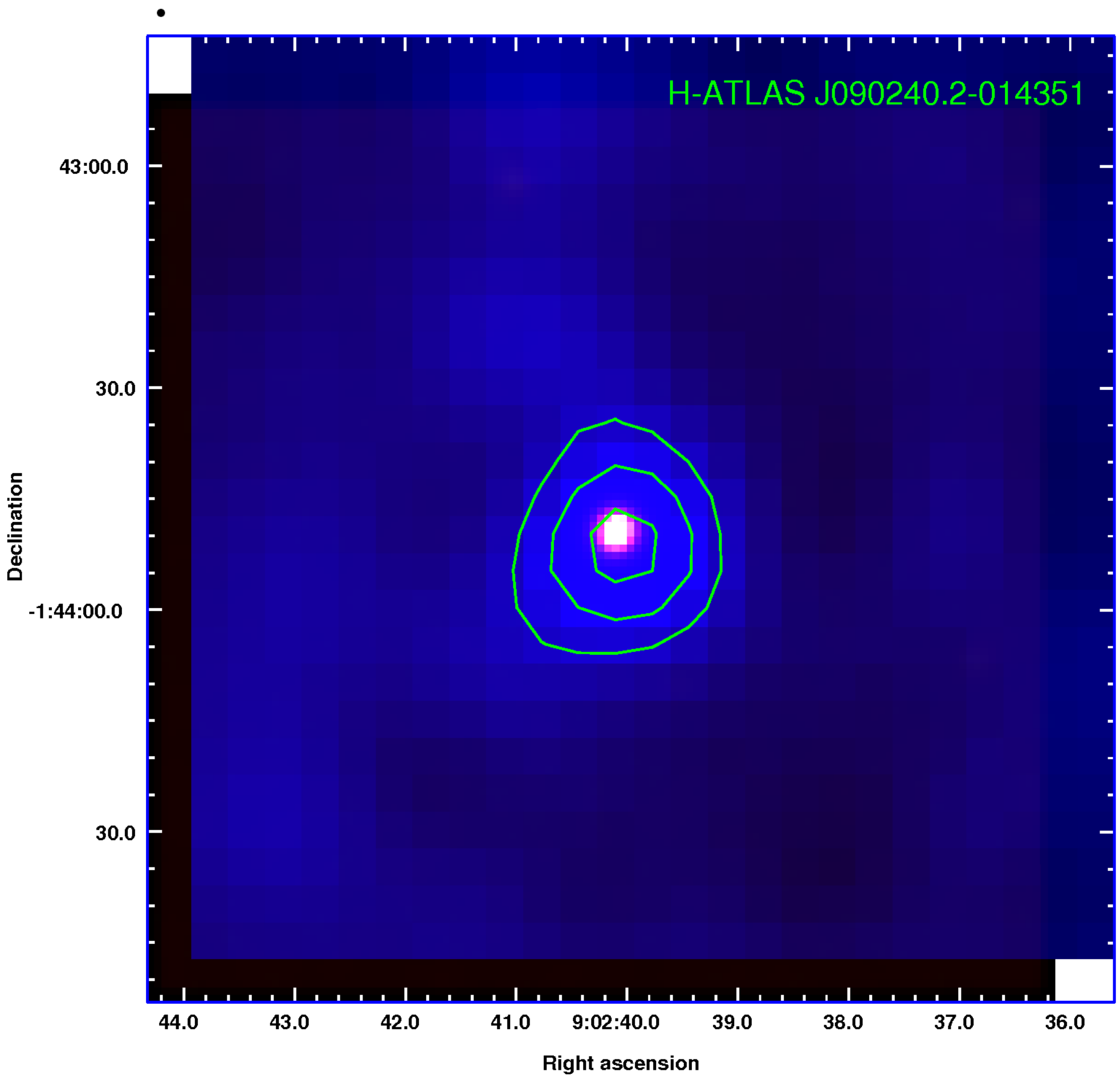

Figure 38: Three colour image of a candidate debris disc identified in the H-ATLAS science demonstration field. Red & green channels are 2MASS J & $K_s$ respectively. The blue channel is SPIRE 250 $\mu$m, smoothed by a 3 pixel Gaussian kernel to increase the signal to noise ratio. Contours are of 250 $\mu$m emission starting at $3\sigma$ and spaced by $2\sigma$. The host star has a spectral type derived from its $g - i$ colour of G5, a photometric distance of 190–290 pc. The candidate disc has a temperature of 60 K and a disk mass of 0.5–1.3 $M_\oplus$.

There is also the scope to examine stars that are not on the main sequence for evidence of dust-producing activity. Within the region of HSLS covered by SDSS there exist a number of spectroscopically determined samples of ultracool dwarfs [404] and White Dwarfs [138]. White dwarfs offer an unique view into the properties of planetary systems. A star can retain much of its planetary system as it leaves the main sequence and evolves through the red giant stage. Although the inner few AU of a planetary system may evaporate within the red giant atmosphere, and planets that were in marginally stable orbits may escape during stellar mass loss, most outer planets and even much of the Oort-Cloud like extremities of planetary systems should persist around white dwarfs [109]. The destiny of planetary systems through the late evolution of their host stars is very uncertain. White dwarfs are the compact end products of stars with masses up to 8 solar masses. An excess infrared flux has been discovered around several white dwarfs [61, 121, 147], providing links to the existence of planetary systems around their main-sequence progenitor stars. The presence of cold dust around these systems may hint at the existence of Kuiper Belt type objects. Beyond these programmes HSLS will provide a wide census of dust producing stars at high galactic latitudes (T Tauri circumstellar disks, giants, AGB/post-AGB, etc).

# A Institutional Addresses

[1] Center for Cosmology and the Department of Physics & Astronomy, University of California, Irvine, CA 92697, USA
[2] Department of Physics and Astronomy, Cardiff University, The Parade, Cardiff Cf24 3AA, UK
[3] Department of Physics & Astronomy, The Open University, Walton Hall, MK7 6AA, Milton Keynes, UK
[4] Astrophysics Group, Physics Department, Blackett Lab, Imperial College, Prince Consort Road, London SW7 2AZ, UK
[5] Jet Propulsion Laboratory, 4800 Oak Grove Drive, Pasadena, CA 91109, USA
[6] Astronomy Centre, Department of Physics & Astronomy, University of Sussex, Brighton BN1 9QH, UK
[7] Astrophysics Branch, NASA/Ames Research Center, MS 245-6, Moffett Field, CA 94035, USA
[8] Centre for Astrophysics Research, Science & Technology Research Institute, University of Hertfordshire, College Lane, Hatfield, AL10 9AB, UK
[9] Dipartimento di Fisica "G. Marconi", University of Rome "Sapienza", 00185, Rome, Italy
[10] Dipartimento di Fisica,Università di Roma "Tor Vergata", Via della Ricerca Scientifica 1, I-00133, Roma, Italy
[11] Department of Physics and Astronomy, University College London, London, WC1E 6BT, UK
[12] Department of Astronomy, University of Massachusetts, Amherst, MA 01003, USA
[13] CENTRA, Departamento de Física, Instituto Superior Técnico, 1049-001 Lisboa, Portugal
[14] Institute of Astronomy, University of Cambridge, Madingley Road, Cambridge CB3 0HA, UK
[15] Space Science & Technology Department, Rutherford Appleton Laboratory, Chilton, Didcot, Oxfordshire OX11 0QX, UK
[16] LUTh, Observatoire de Paris, CNRS-UMR8102 and Université Paris VII, 92195 Meudon Cédex, France
[17] Berkeley Center for Cosmological Physics, Department of Physics, University of California, Berkeley, CA 94720, USA
[18] Argelander-Institut für Astronomie der Universität Bonn, Auf dem Hügel 71, D-53121 Bonn, Germany
[19] SISSA, via Bonomea 265, I-34136, Trieste, Italy
[20] Korea Institute for Advanced Study, Seoul 130-722, Korea
[21] European Southern Observatory, Karl-Schwarzschild-Str. 2, D-85748 Garching bei Munich, Germany
[22] Lawrence Berkeley National Laboratory, Berkeley, CA 94720, USA
[23] South African Astronomical Observatory, 7935 Observatory, Cape Town, South Africa
[24] Physics Department, Stanford, CA, 94305, USA
[25] Institute for Computational Cosmology, Department of Physics, Durham University, South Road, Durham DH1 3LE, UK
[26] Spitzer Science Center, California Institute of Technology, MC 314-6, Pasadena, CA 91125, USA
[27] Department of Astrophysical Sciences, Princeton University, Princeton, NJ 08544, USA
[28] UCLA Physics & Astronomy, Los Angeles, CA 90095, USA
[29] Institute for Space Imaging Science, University of Lethbridge, Lethbridge, Alberta, T1K 3M4, Canada
[30] ISAS, JAXA, Sagamihara, Kanagawa 229-8510, Japan
[31] Centre for Astronomy & Particle Theory, School of Physics and Astronomy, Nottingham University, University Park Campus, Nottingham, NG7 2RD, UK

[32] Department of Physics, Durham University, South Road, Durham , DH1 3LE, UK
[33] Department of Physics, University of Pennsylvania, Philadelphia, PA 19104, USA
[34] Texas Cosmology Center & Department of Astronomy, University of Texas, Austin, TX 78712, USA
[35] Astrophysics, Denys Wilkinson Building, University of Oxford, Keble Road, Oxford OX1 3RH, UK
[36] School of Physics and Astronomy, Nottingham University, University Park Campus, Nottingham, NG7 2RD, UK
[37] Dept. of Physics & Astronomy, University of British Columbia, 6224 Agricultural Road, Vancouver, B.C. V6T 1Z1, Canada
[38] Department of Astronomy and Astrophysics, University of Toronto, Toronto, ON, M5S 3H4, Canada
[39] Department of Astronomy, University of Arizona, Tucson, AZ 85721, USA
[40] Canadian Institute for Theoretical Astrophysics, University of Toronto, Toronto, ON, M5S 3H, Canada
[41] Space Sciences Laboratory, University of California, Berkeley, CA 94720, USA
[42] National Research Council of Canada-NRC, Herzberg Inst of Astrophysics-HIA, CA Victoria BC V9E 2E7, Canada
[43] National Radio Astronomy Observatory, Green Bank, WV 24944, USA
[44] Steward Observatory, University of Arizona, Tucson, AZ 85721, USA
[45] Joint Astronomy Centre, University Park, Hilo, Hawaii, USA
[46] Sydney Institute for Astronomy, School of Physics A28, The University of Sydney, NSW 2006, Australia
[47] Instituto de Astrofísica de Canarias, c/Vía Láctea s/n, La Laguna, Tenerife, Spain
[48] University of Manchester, Jodrell Bank Centre for Astrophysics, Turing Building, Oxford Road, Manchester M13 9PL, UK
[49] Institut d'Astrophysique Spatiale, bât 121 Université Paris Sud XI, F-92405 Orsay Cedex, France
[50] Department of Physics, Princeton University, Princeton, NJ 08544, USA
[51] Laboratoire d'Astrophysique de Marseille, UMR 6110 CNRS, 38 rue F. Joliot-Curie, F-13388 Marseille, France
[52] Departamento de Astrofísica, Facultad de CC. Físicas, Universidad Complutense de Madrid, E-28040 Madrid, Spain
[53] APC/Université Paris 7 Denis Diderot/CNRS, Bâtiment Condorcet, 10, rue Alice Domon et Léonie Duquet, 75205 Paris Cedex 13, France
[54] Institute for Astronomy, University of Edinburgh, Royal Observatory, Blackford Hill, Edinburgh EH9 3HJ, UK
[55] Astrophysics, Cosmology and Gravity Centre (ACGC), Astronomy Department, University of Cape Town, Private Bag X3, 7700 Rondebosch, Republic of South Africa
[56] Center for Astronomy and Astrophysics, Observatório Astroómico de Lisboa, Faculdade de Ciências, Universidade de Lisboa, Tapada da Ajuda, 1349-018 Lisbon, Portugal
[57] Kavli Institute for Astrophysics and Space Research, Massachusetts Institute of Technology, Cambridge, MA 02139, USA
[58] Department of Physics and Astronomy, Rutgers University, Piscataway, NJ 08854-8019, USA
[59] Department of Physics and Astronomy, University of Southern California, Los Angeles, CA 90089, USA
[60] Instituto de Física de Cantabria, UC-CSIC. Avda. Los Castros s/n 39005, Santander, Spain
[61] ICREA (Institució Catalana de Recerca i Estudis Avançats) and Instituto de Ciencias del

Cosmos,(ICC-UB-IEEC) Universidad de Barcelona, Martí i Franqués 1, 08028, Barcelona, Spain
[62] Agenzia Spaziale Italiana, Via G. Galilei snc, I-00040 Frascati, Italy
[63] Astrophysics & Cosmology Research Unit, School of Mathematical Sciences, University of KwaZulu-Natal, Private Bag X54001, Durban 4000, South Africa
[64] Department of Physics, University of Miami, Coral Gables, Florida 33146, USA
[65] Universtité de Toulouse, CNRS, IRAP, CESR, 9 av. du Colonel Roche, BP44346, 31028 Toulouse cedex 4, France
[66] Mullard Space Science Laboratory, University College London, Holmbury St. Mary, Dorking, Surrey RH5 6NT, UK
[67] NASA Herschel Science Center, California Institute of Technology, Pasadena, CA 91125, USA
[68] ALMA, National Radio Astronomy Observatory, Santiago, Chile
[69] Department of Physics and Astronomy, Seoul National University, Seoul 151-742, Korea
[70] Berkeley Lab & University of California, Berkeley, CA 94720, USA
[71] Dipartimento di Astronomia, Universita di Padova, I-35122 Padova, Italy
[72] SETI Institute, 515 N. Whisman Avenue, Mountain View, CA 94043, USA
[73] Physics Department, University of Johannesburg, P.O. Box 524, Auckland Park, 2006, South Africa.
[74] Herschel Science Centre, European Space Astronomy Centre, P.O.Box 78, 28691 Villanueva de la Cañada (Madrid), Spain
[75] Département de physique, de génie physique et d'optique et Centre de recherche en astrophysique du Québec, Université Laval, Québec, QC, G1V 0A6, Canada
[76] Harvard-Smithsonian Center for Astrophysics, 60 Garden Street, Cambridge, MA 02138, USA
[77] Departamento de Astrofisica, Universidad de La Laguna, Tenerife, Spain
[78] INAF-Osservatorio Astronomico di Padova, Vicolo dell'Osservatorio 5, I-35122 Padova, Italy
[79] Institut d'Astrophysique de Paris, CNRS UMR7095 & Univ. Paris 6, 98bis Bd Arago, 75014 Paris, France
[80] Laboratoire AIM, CEA/DSM -CNRS - Université Paris Diderot, DAPNIA/Service d'Astrophysique, CEA Saclay, Orme des Merisiers, F-91191 Gif-sur-Yvette Cedex, France
[81] Kavli Institute for Cosmological Physics, University of Chicago, 5640 South Ellis Avenue, Chicago, IL 60637, USA
[82] Kapteyn Astronomical Institute, University of Groningen, P.O.Box 800, 9700AV, Groningen, The Netherlands
[83] UK Astronomy Technology Centre, Royal Observatory, Blackford Hill, Edinburgh EH9 3HJ, UK
[84] Instituto Nacional de Astrofísica, Óptica y Electrónica, Aptdo. Postal 51 y 216, 72000 Puebla, Pue., Mexico
[85] Sterrenkundig Observatorium, Universiteit Gent, Krijgslaan 281 S9, B-9000 Gent, Belgium
[86] IPAC/Physics/Astronomy, California Institute of Technology, Pasadena, CA 91125, USA
[87] National Astronomical Observatories, Chinese Academy of Science, 20A Datun Road, Chaoyang District, Beijing 100012, China
[88] Excellence Cluster Universe, Technische Universität München, Boltzmannstr. 2, D-85748 Garching b. München, Germany
[89] Dipartimento di Fisica "G. Galilei", Universita' degli Studi di Padova and INFN, Sezione di Padova, via Marzolo 8, I-35131 Padova, Italy
[90] Leiden Observatory, Leiden University, PO Box 9513, NL - 2300 RA Leiden, The Netherlands
[91] Physics Dept., University of the Western Cape, Private Bag X17, Bellville, 7535, South Africa
[92] SOFIA-USRA, NASA Ames Research Center, MS N211-3, Moffett Field, CA 94035, USA
[93] Center for Astrophysics and Space Astronomy, University of Colorado, 593 UCB, Boulder, Co,

80309-0593, USA
[94] Kavli Institute for Particle Astrophysics and Cosmology, Stanford, CA 94305
[95] Department of Physics, University of Crete, 71003 Heraklion, Greece
[96] Department of Astronomy, University of Bologna, Via Ranzani 1, 40139, Bologna, Italy
[97] Max Planck Institute for Extraterrestial Physics, Giessenbachstrasse, 85741 Garching, Germany
[98] Department of Physics, University of California, Davis, One Shields Avenue, Davis, CA 95616, USA
[99] University of New South Wales, Australia
[100] Institute for Astronomy, University of Hawaii, Hawaii, USA
[101] Physics and Astronomy Department, University of California, Riverside, California 92521, USA
[102] Kavli Institute for Cosmology, University of Cambrdige, Madingley Road, Cambridge CB3 0HA, UK
[103] Institute for Advanced Research, Nagoya University, Furo-cho, Chikusa-ku, Nagoya 464-8601, Japan
[104] Department of Astronomy and Astrophysics, University of California Santa Cruz, CA, 95064, USA
[105] Departamento de Astronomía y Astrofísica, Facultad de Física, Pontificia Universidad Católica de Chile, Casilla 306, Santiago 22, Chile
[106] Department of Physics and Astronomy, The Johns Hopkins University, 3400 N. Charles St., Baltimore, MD 21218-2686